\documentclass[fleqn,usenatbib]{mnras}

\usepackage[T1]{fontenc}
\usepackage[svgnames]{xcolor}

\DeclareRobustCommand{\VAN}[3]{#2}
\let\VANthebibliography\thebibliography
\def\thebibliography{\DeclareRobustCommand{\VAN}[3]{##3}\VANthebibliography}

\usepackage{graphicx}	
\usepackage{amsmath}	
\usepackage{amssymb}	
\usepackage{color}

\newcommand{\rd}{\color{red}}

\newcommand{\msun}{{\,\rm M_\odot}}
\newcommand{\dd}{{\rm d}}
\newcommand{\flun}{\, \rm{cm^{-2}\,s^{-1}}}
\def\noterd #1]{{\bf \rd #1]}}

\title[Core overshoot of the solar convective core]{Core overshoot constrained by the absence of a solar convective core and some solar-like stars}

\author[Q.-S. Zhang et al.]{
Qian-Sheng Zhang,$^{1,3,4,5}$\thanks{E-mail: zqs@ynao.ac.cn (QSZ)}
J{\o}rgen Christensen-Dalsgaard$^{2}$
and Yan Li$^{1,3,4,5}$
\\
$^{1}$Yunnan Observatories, Chinese Academy of Sciences, 396 Yangfangwang, Guandu District, Kunming 650216, China\\
$^{2}$Stellar Astrophysics Centre and Department of Physics and Astronomy, Aarhus University, DK-8000 Aarhus C, Denmark\\
$^{3}$Center for Astronomical Mega-Science, Chinese Academy of Sciences, 20A Datun Road, Chaoyang District, Beijing 100012, China\\
$^{4}$Key Laboratory for the Structure and Evolution of Celestial Objects, Chinese Academy of Sciences, 396 Yangfangwang, \\
Guandu District, Kunming 650216, China\\
$^{5}$University of Chinese Academy of Sciences, Beijing 100049, China
}

\date{Accepted XXX. Received YYY; in original form ZZZ}

\pubyear{2020}

\begin{document}
\label{firstpage}
\pagerange{\pageref{firstpage}--\pageref{lastpage}}
\maketitle

\begin{abstract}
	Convective-core overshoot mixing is a significant uncertainty in stellar evolution. Because numerical simulations and turbulent convection models predict exponentially decreasing radial rms turbulent velocity, a popular treatment of the overshoot mixing is to apply a diffusion process with exponentially decreasing diffusion coefficient. It is important to investigate the parameters of the diffusion coefficient because they determine the efficiency of the mixing in the overshoot region. In this paper, we have investigated the effects of the core overshoot mixing on the properties of the core in solar models and have constrained the parameters of the overshoot model by using helioseismic inferences and the observation of the solar $^8$B neutrino flux. For solar-mass stars, the core overshoot mixing helps to prolong the lifetime of the convective core developed at the ZAMS. If the strength of the mixing is sufficiently high, the convective core in a solar model could survive till the present solar age, leading to large deviations of the sound-speed and density profiles comparing with the helioseismic inferences. The $^8$B neutrino flux also favours a radiative solar core. Those provide a constraint on the parameters of the exponential diffusion model of the convective overshoot mixing. A limited asteroseismic investigation of 13 \textsl{Kepler} low-mass stars with $1.0<M/{\rm M_\odot}<1.5$ shows a mass-dependent range of the overshoot parameter. The overshoot mixing processes for different elements are analyzed in detail. It is found that the exponential diffusion overshoot model leads to different effective overshoot mixing lengths for elements with different nuclear equilibrium timescale.
\end{abstract}




\section{Introduction} \label{SecIntro}

There is a convective core in the main-sequence stars with mass higher than $1$ to $1.1 \msun$ (depending on metallicity). The convection mixes nuclear fuels in the core, playing an important role to determine the structure and the lifetime of the star. The presence of convection is controlled by $\nabla_{\rm R} = (\dd \ln T / \dd \ln P)_{\rm R}$, the gradient of temperature $T$ with respect to pressure $P$ required to transport the energy by radiation, and the corresponding adiabatic gradient $\nabla_{\rm ad}$. In the region where $\nabla_{\rm R}\geq\nabla_{\rm{ad}}$, the radiative temperature gradient would lead to convective instability and convection sets in, transporting the excess energy flux beyond the capacity of radiation. However, the details of the motion in the adjacent region with $\nabla_{R}<\nabla_{\rm{ad}}$ are not clear.

In a classical phenomenological point of view, a fluid element is always accelerated in the convection zone; thus it cannot stop at the convective boundary where $\nabla_{R}=\nabla_{\rm{ad}}$. The fluid elements penetrating into the radiative region is called the convective overshoot. In the classical overshoot models \citep[e.g.,][]{shaviv1973,maeder1975,bressan1981,zahn1991}, the focus is on how far a penetrating fluid element can move, which is considered as the extent of the overshoot region, and the whole convection zone and the overshoot region are assumed to be fully mixed. The main properties of those models are as follows \citep{zahn1991}. The radial velocity and the temperature fluctuation in the overshoot region should be strongly correlated. The temperature gradient in the overshoot region should be slightly less than the adiabatic temperature gradient $\nabla_{\rm{ad}}$ due to the high P\'{e}clet number ${\rm Pe}$, i.e., the ratio between the turbulent diffusivity and the radiative thermal diffusivity. The boundary of the overshoot region is located near ${\rm{}Pe}=1$. The temperature gradient quickly changes from $\nabla_{\rm{ad}}$ to $\nabla_{\rm{R}}$.

In the hydrodynamical point of view \citep{zqs13}, the convective overshoot should be regarded as an overshoot of the turbulent kinetic energy rather than the fluid elements, and the extent of the overshoot region should be significantly larger than the penetrating distance of a penetrating fluid element because \emph{the penetrating fluid element should disturb the local fluid elements and result in extended transport of turbulent kinetic energy}. The temperature gradient is determined by the entropy mixing of the fluid elements gaining kinetic energy from the overshoot of kinetic energy. Due to buoyancy braking, the characteristic size of the fluid elements is $\sqrt{k}/N$ with $k$ the turbulent kinetic energy and $N$ the Brunt-V\"{a}is\"{a}l\"{a} frequency, which is not large enough to ensure a high efficiency of entropy mixing. Therefore the temperature gradient in the overshoot region is close to $\nabla_{\rm{R}}$, and the radial velocity and the temperature fluctuation in the overshoot region should be weakly correlated. In this scenario, the overshoot mixing can be regarded as a diffusion process \citep{zqs13}. On the other hand, numerical simulations of stellar convection show more complicated behaviours of the convective overshoot. It is found that the turbulent entrainment exists near the convective boundary, leading to a convective boundary mixing \citep[e.g.,][]{meakin2007,arnett2015,cristini2017,cristini2019}. This convective process cannot be regarded by a diffusion approximation. In the convective mixing model by \citet{zqs13}, spherical symmetry is adopted, leading to zero mean velocity. Therefore the contributions to the variation of abundances and the chemical flux by the mean field including a turbulent entrainment are ignored. The entrainment velocity satisfies a turbulent-entrainment law depending on conditions \citep[e.g.,][]{fernando1991}, leading to varied strength for stellar models with different stellar mass \citep[e.g.,][]{staritsin2013,staritsin2014,scott2021}.

The differences between the classical phenomenological overshoot models and the hydrodynamical turbulent convection models are significant and they should been benchmarked by numerical simulations. The weak correlation between the radial velocity and the temperature fluctuation in the overshoot region has been confirmed by numerical simulations \citep{singh1995,meakin2007}. Numerical simulations of convective overshoot \citep{brummell2002} have shown that the overshoot region is not adiabatically stratified even for ${\rm{}Pe}\gg1$. Recent numerical simulations by \citet{cai2020a,cai2020b,cai2020c,cai2020d} have comprehensively investigated the upward overshoot and shown that the details of the convective status and the structure of the overshoot region predicted by the stellar turbulent convection model \citep{zqsly12a} show overall consistency with the result of numerical simulations, while the classical overshoot model is not. Helioseismic investigation of the layer beneath the base of the solar convection zone has also concluded that the required temperature gradient favours the hydrodynamical turbulent convection models \citep{cd2011}.

For the diffusion model of overshoot mixing, numerical simulations \citep{Freytag96} and turbulent convection models \citep{xiong89,xiong02,zqsly12a,zqs13} have predicted an exponentially decreasing diffusion coefficient. For an exponential diffusion coefficient, there are two parameters, i.e., the initial diffusion coefficient and the index of the exponential function. The initial diffusion coefficient is usually assumed to be the typical value of the diffusion coefficient in the convection zone near the convective boundary. Here we note that the assumption may not hold. In a widely used diffusion model \citep{herwig2000}, the value of the index has been given by the calibration of the width of the main sequence on the HR diagram. On the other hand, overshoot calibrations on eclipsing binary systems \citep[e.g.,][]{ribas2000,claret2016,claret2017,claret2018,claret2019} have shown that the strength of overshoot mixing depends on stellar mass, especially for low-mass stars. Further information about overshoot is provided by asteroseismic investigations \citep[for a review, see][]{deheuvels2019}. Such analysis of low-mass stars has suggested a smaller overshoot region \citep{deheuvels2016}. Therefore it is necessary to investigate the overshoot parameter in low-mass stars. It is well-known that a solar-mass star has a convective core at the ZAMS stage, and the overshoot mixing can prolong the lifetime of the convective core because of the extra mixing brings $^3$He into the core \citep[e.g.,][]{shaviv1971,Roxburgh1985,hd203608,buldgen2019a}. If the convective core survives at the present solar age in a solar model, it was found that the sound-speed deviation in the core is significant and the $^7$Be and $^8$B neutrino fluxes are reduced \citep[e.g.,][]{RV1996,SW1999,shaviv1971,CH1996,DR1998}.

Asteroseismology is an advantageous tool to probe the core overshoot mixing because the oscillation frequencies are directly affected by the stellar interior. \citet{mrv15} investigated the $3 \msun$ star KIC10526294 and found that the diffusive mixing is better than the classical step mixing in fitting the frequencies of the observed oscillation modes.
\citet{Yang2015} found a rather large overshoot region in the asteroseismic investigation of KIC2837475 by using ratios between small and large frequency separations. In contrast, \citet{Wu2020} found very weak mixing outside the convective core in the asteroseismic investigation of KIC8324482. For central helium burning stars, investigations of the oscillation period spacing favour a moderate overshoot region \citep{Bossini2015,Bossini2017}, which is obviously different from the case of the main sequence stars.
\citet{Noll2021} found strong evidence for convective-core overshoot in their asteroseismic investigation of the subgiant KIC10273246,
while diffusive mixing showed no clear improvement over the classical treatment. On the other hand, other mixing mechanisms including rotation mixing and turbulent entrainment could also be required to explain observations \citep[see, e.g.,][]{Saio2021,Pedersen2021,Johnston2021}, leading to systematic uncertainties of probing the core overshoot mixing.
We also note that even in a normal asteroseismic investigation with least-$\chi^2$ method of fitting all frequencies, the details in the numerical calculations such as the time step can significantly affect the $\chi^2$ of frequencies \citep{Wu2016}, and hence the inferences about overshoot.

In this paper, we will investigate the effect of a new formula for the convective core overshoot mixing on solar models and some solar-mass stars. Our results show that the convective core survives in the solar model to the present solar age when the overshoot mixing is sufficiently strong. This leads to signatures in the sound-speed and density profiles and the $^8$B neutrino flux that can be benchmarked by comparing model properties with observations. The range of model parameters for some \textsl{Kepler} solar-mass stars is investigated via a limited asteroseismic investigation by using the frequency separation ratios.

\section{Input physics of solar models} \label{Secinputphy}

\subsection{Standard input physics of solar models} \label{Secinputstdphy}

Solar models are calculated by using the YNEV code \citep{YNEVZ15}. The element abundances are based on the AGSS09 \citep{AGSS09} solar photosphere composition and the $\sim40\%$ upward revision of Ne abundance \citep{Young18, AAG21}. This is denoted as A09Ne composition in this paper, which leads to $(Z/X)_{s}=0.0188\pm0.0012$ \citep{ZLCD2019}. The thermodynamical functions are interpolated from the OPAL equation of state tables \citep{OPALEOS}. The opacities are interpolated from the OPAL tables \citep{OPAL} at high temperature and the \citet{OPLT} opacity tables at low temperature. Nuclear reaction cross sections are based on SFII \citep{SFII}, enhanced by weak screening \citep{SCR}. Molecular diffusion in the screening case \citep{DIFSCRZ17} is taken into account. The temperature gradient in the convection zone is calculated by using the standard mixing-length theory. The K-S relation \citep{KSATM} between temperature and optical depth in the solar atmosphere is adopted.

\subsection{Models of convective core overshoot mixing} \label{Secinputovm}

The only extra physical process outside the framework of standard solar models taken into account is the overshoot mixing outside the convective core, which appears in solar models at the zero-age main sequence (ZAMS). We investigate two kinds of models of the convective-core overshoot mixing.

The first is an exponential diffusion overshoot model (EDOM) with the diffusion coefficient in the overshoot region defined as
\begin{eqnarray} \label{OVMmodel}
D = C D_0 \left(\frac{P}{P_{\rm{cz}}}\right)^{\theta},
\end{eqnarray}
where $C$ and $\theta$ are dimensionless model parameters, $P$ is pressure and $P_{\rm{cz}}$ is the value at the boundary of the convective core, $D_0$ is the typical diffusion coefficient in the convective core near the convective boundary. $D_0$ is calculated as $D_0=u(r_*)l(r_*)/3$, where $l=\alpha H_P$ is the mixing length, $u$ is the mean turbulent speed calculated by the MLT, $H_{P}=P/(\rho g)$ is the local scaleheight of the pressure $P$, $\rho$ is the density, $g$ is the local gravitational acceleration, $r_*=r_{\rm{cz}}-d$, $r_{\rm{cz}}$ is the radius at the convective boundary, and $d$ is a distance ($0.1 H_P$ in default). The reason of adopting $u(r_*)$ as the typical turbulent speed at the boundary is that the local MLT gives $u(r_{\rm{cz}})=0$ and the nonlocal effects of convection should be significant near the convective boundary.

The second is the classical overshoot model (COM), i.e., assuming an overshoot region with the length
\begin{equation}
l_{\rm{ov}}=\alpha_{\rm{ov}} {\rm{min}}(H_{P},r_{cz})
\label{eq:com}
\end{equation}
outside the convective core, where $\alpha_{\rm{ov}}$ is a free parameter. The diffusion coefficient in this model is not assumed to be infinite in the convective core and the overshoot region. The inappropriateness of setting infinite diffusion coefficient will be discussed in the Appendix. The diffusion coefficient is calculated as $D=u(r)l(r)/3$ by using the MLT for $r<r_*$ and assumed to be $D=D(r_*)$ near the convective boundary and in the overshoot region, i.e., for $r_*<r<r_{\rm{cz}}+l_{\rm{ov}}$. In this case, the diffusion coefficient in the convection core and the overshoot region in solar models is about $10^{13}-10^{14}\, {\rm cm^2\,s^{-1}}$, ensuring that the convective core and the overshoot region are efficiently mixed for the great majority of elements. There is another choice in the classical model: setting $l_{\rm{ov}}=\alpha_{\rm{ov}} H_{P}$. However, for solar-mass stars, it is not a good choice. As the convective core retreats and finally vanishes, $H_{P}$ at the convective boundary and $l_{\rm{ov}}$ become larger and larger. That is physically unreasonable.

As discussed in the Appendix, the EDOM diffusive mixing can also be characterized by an effective overshoot mixing length (denoted as $l_{\rm{ov,dif}}$).
A main difference between the COM and the EDOM is that in the EDOM $l_{\rm{ov,dif}}$ increases with the increase of stellar age because $l_{\rm{ov,dif}}^2 \propto D t$, which means that the low-$D$ region far away from the convective boundary will finally lead to significant mixing if the stellar lifetime is long enough, while $l_{\rm{ov}}$ in the classical overshoot model is constant. Another difference is that the COM results in a discontinuity of the abundance profile after the convective core reaches its maximum. In contrast, the EDOM leads to a smooth abundance profile.

For both overshoot models we assume that the temperature gradient is purely radiative in the overshoot region.

\subsection{On the exponential diffusion overshoot model} \label{Secinputovm2}

Theoretical analysis of the convective mixing has shown that the convective/overshoot mixing can be regarded as a macroscopic diffusion process \citep{zqs13}. Numerical simulation \citep{Freytag96} and non-local turbulent convection models \citep[e.g.,][]{xiong89,xiong02,deng06,zqsly12a,li17} have predicted exponentially decreasing turbulent \textit{rms} speed. Therefore it is reasonable to regarded the overshoot mixing as a diffusive process with an exponentially decreasing diffusion coefficient. For an exponential function, there are two free parameters, i.e., the initial value and the e-folding length. Therefore equation~(\ref{OVMmodel}) is a universal formula for an exponentially decreasing diffusion coefficient in a convective overshoot region. Comparing the model (i.e., equation~(\ref{OVMmodel})) with the widely used formula of the diffusion coefficient of overshoot \citep{herwig2000} adopted in the MESA code \citep{mesa}, i.e.,
\begin{eqnarray} \label{herwigovm}
&&D = D_0 \exp \left( - \frac{{2\left| {r - {r_{\rm{cz}}}} \right|}}{{{f_{\rm{ov}}}{H_P}}}\right),
\end{eqnarray}
the parameter $\theta$ in equation~(\ref{OVMmodel}) is related to $f_{\rm{ov}}$ because
\begin{eqnarray} \label{paratheta}
&&\exp \left( - \frac{{2\left| {r - {r_{\rm{cz}}}} \right|}}{{{f_{\rm{ov}}}{H_P}}}\right) \approx \exp \left( - \frac{2}{{{f_{\rm{ov}}}}}\int\limits_{{r_{\rm{cz}}}}^r {\frac{{dr}}{{{H_P}}}} \right) \\ \nonumber
	&&= \exp \left(\frac{2}{{{f_{\rm{ov}}}}}\ln \frac{P}{{{P_{\rm{cz}}}}}\right) = \left(\frac{P}{{{P_{\rm{cz}}}}}\right)^{2/f_{\rm ov}}\; , 
\end{eqnarray}
which shows that the e-folding lengths of the diffusion coefficient in the two models are similar when $\theta=2/f_{{\rm{ov}}}$. The above derivation replaced $H_P$ at the convective boundary by the local $H_P$ as a function of $r$. This approximation generally holds because $f_{\rm ov} \ll 1$ (Herwig recommends $f_{\rm ov} \simeq 0.016$) leads to a quick decrease of $D$ near the convective boundary so that the region in which $D$ is high enough to result in an efficient mixing is narrow and in this narrow region $H_P$ changes little. The difference between equation~(\ref{OVMmodel}) and \citeauthor{herwig2000}'s \citeyearpar{herwig2000} model is the parameter $C$ in equation~(\ref{OVMmodel}). In \citeauthor{herwig2000}'s model, $C$ is set to be 1, i.e., the initial value of the exponentially decreasing diffusion coefficient is set as the typical diffusion coefficient in the convection zone near the convective boundary. However, in equation~(\ref{OVMmodel}), $C$ is not fixed {\it a priori}. In the following we provide a physical justification for the possible variation of $C$ from 1.

The diffusion coefficient depends strongly on the radial characteristic length of convection, which could significantly change near the convective boundary because the stellar convection is driven by buoyancy that changes sign at the convective boundary. In the convectively unstable zone, buoyancy helps radial convection; thus a large radial characteristic length could be expected. Even in a slightly sub-adiabatic overshoot region (if it exists), the buoyancy is weak since it is proportional to $\nabla-\nabla_{\rm{ad}}$. Therefore we can also expect that a large characteristic length could be kept although the buoyancy prevents convective movement. However, in a significantly sub-adiabatic overshoot region in which $\nabla \approx \nabla_{R}$, a strong buoyancy brakes radial convective motion and then significantly reduces the radial characteristic length. Therefore, the radial characteristic length should significantly decrease near the convective boundary (or the boundary of the nearly adiabatic overshoot region).

The relation between convective heat transport and mixing of material has been discussed in \citet{zqs13}. Let us recall the conditions of high P\'{e}clet number ${\rm{Pe}} = \lambda_{\rm{turb}} / \lambda_{T}$ and high diffusivity ratio $\tau = \lambda_{T} / \lambda_{\mu}$, where $\lambda_{\rm{turb}}$ is the turbulent diffusivity, $\lambda_{T}$ is the thermal diffusivity, and $\lambda_{\mu}$ is the compositional diffusivity, respectively. Those conditions hold in most of the overshoot region in the deep stellar interior. In this case, both the timescale of radiative heat transport and the timescale of molecular diffusion are much longer than the timescale of turbulent dissipation so that a fluid element keeps its entropy and abundance ($\delta S=0$ and $\delta X_i=0$) in the convective motion. The convective heat transport can be thought as the result of \emph{mixing of entropy}, which is revealed in the energy conservation equation in the hydrodynamic equations. The convective heat transport and mixing of material are two results of the pure mechanical mixing caused by turbulent dissipation. In the case of both $\delta S$ and $\delta X_i$ are zero, the mixing of entropy and the mixing of material are therefore strongly correlated.

The discussion above indicates that the diffusion coefficient should significantly vary near the convective boundary or the boundary of adiabatic overshoot region (if it exists). Therefore the initial diffusion coefficient of the exponential formula equation~(\ref{OVMmodel}) used in the overshoot region could be much smaller than the typical diffusion coefficient in convection zone, i.e., $C \ll 1$.

Comparing with the widely used model in equation~(\ref{herwigovm}), there is another advantage of the model in equation~(\ref{OVMmodel}). When the convective core shrinks to nothing, $H_P$ becomes very large ($H_P\rightarrow\infty$ for $M_{\rm{cz}}\rightarrow0$), and equation~(\ref{herwigovm}) would suddenly lead to a strong overshoot that significantly refreshes the nuclear fuel in the core and extends the size of the core. This leads to an instability in the calculation of stellar evolution because the diffusion coefficient far away from the convective core is suddenly enhanced when $H_P\rightarrow\infty$. This instability does not exist when the model in equation~(\ref{OVMmodel}) is adopted because it ensures that the diffusion coefficient decreases sufficiently in the region far away from the core, even for $H_P\rightarrow\infty$.

\section{Presence of a convective core in solar models} \label{Secconvcore}

Before we show the results for the solar models, it is necessary to investigate the existence of a convective core and its relationship with convective overshoot mixing. The determining factor is that
\begin{equation}
	\nabla_{\rm R} \propto \overline{\epsilon}=L_r/M_r \; ,
\end{equation}
where $\overline{\epsilon}$ is the mean energy release rate and $L_r$ and $M_r$ are luminosity and enclosed mass at radius $r$. This is determined by the temperature sensitivity of the energy generation rate, $\eta = \partial \ln \epsilon/\partial \ln T$. When $\eta$ is high, energy generation is strongly concentrated towards the centre, leading to a high $\overline{\epsilon}$ and hence a tendency towards convective instability \citep[see also][]{Roxburgh1985}. On the other hand, in normal models of the present Sun, energy generation is dominated by the \textsl{p-p} chain operating in nuclear equilibrium, with $\eta \approx 4$, resulting in convective stability.

\subsection{The convective core in ZAMS solar models} \label{subSeczamsccore}

\begin{figure}
\includegraphics[width=0.98\columnwidth]{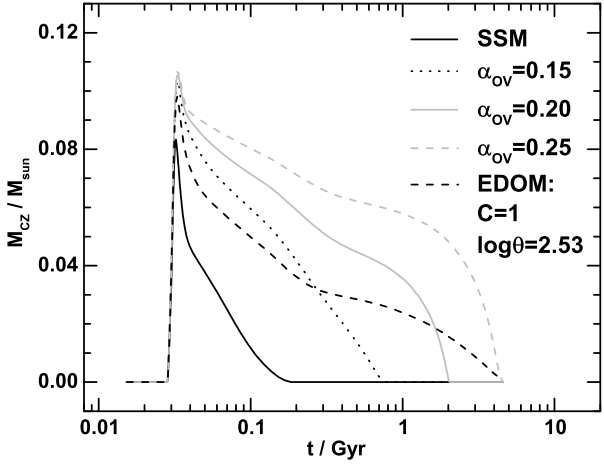}
\caption{Evolution of fraction of convective-core mass for COM overshoot with different $\alpha_{\rm{ov}}$ (cf.\ equation~\ref{eq:com}) and a EDOM solar model with $\log C=0$ and $\log \theta=2.53$.}
\label{resstdmc}
\end{figure}

The reason of the existence of a convective core in solar-mass stars at ZAMS has already been clearly investigated \citep[e.g.,][]{hd203608}. It can be summarized as follows. As a solar-mass PMS star evolves toward the main-sequence stage, the core temperature of the star increases. Near the ZAMS stage, the core temperature is high enough to drive the proton-capture reactions of $^{12}$C and the fusion of $^{3}$He. Since they are more sensitive to temperature than the equilibrium \textsl{p-p} chain, according to the argument given above a convective core appears. However, as the initial $^{12}$C and $^{3}$He have almost depleted and reached its equilibrium abundance in about $ \sim 100 $ Myr, $\eta$ decreases to the value for the equilibrium \textsl{p-p} chains. As a result, the convective core caused by the burning of initial $^{12}$C and $^{3}$He vanishes in a short time in standard solar model, which is shown in Fig.~\ref{resstdmc} with the black solid line.

\subsection{The effects of overshoot mixing on the convective core} \label{subSecoveffects}

As shown in Fig.~\ref{resstdmc}, the convective core is larger and exists for a longer time when the core overshoot mixing is taken into account. Because the main effect of overshoot is an extra mixing and $\nabla_{R}$ is proportional to $\overline{\epsilon}$, the reason of an enlarged convective core has to be that the mixing adds nuclear fuels, i.e., $^{12}$C and $^3$He, disturbing the \textsl{p-p} nuclear equilibrium and hence increases $\eta$. \citet{hd203608} have analyzed the phenomenon and concluded that the existence of a peak of the equilibrium abundance of $^3$He is the reason of the overshoot mixing extending the lifetime of the convective core because it brings $^3$He from the overshoot region into the core.

We have found that a pump cycle mechanism, which has not been noticed, should be the main reason of the phenomenon. When there is no overshoot mixing, the abundance of $^3$He in the layer above the core is in local nuclear equilibrium, increasing with decreasing temperature and hence increasing distance from the centre. However, if the overshoot mixing is taken into account, the abundance of $^3$He in the mixing region (core and the overshoot region) is in a nonlocal nuclear equilibrium as
\begin{eqnarray} \label{dehe3}
0 \approx \frac{{\partial {X_3}}}{{\partial t}} = {R_3} - \frac{{\partial {F_3}}}{{\partial m}},
\end{eqnarray}
where $X_3$ is the abundance of $^3$He, $F_3$ is the radial flux (multiplied by $4 \pi \rho r^2$) of $^3$He, and $R_3$ is local generation rate of the abundance of $^3$He due to nuclear reactions. The flux $F_3$ is dominated by the overshoot mixing. The physical boundary condition of the equation above is that $F_3$ at the boundary of the overshoot region and at the centre is zero. A key point is $F_3$ at the convective core boundary. Because the mixing brings $^3$He from the overshoot region into the core, $F_3$ at the surface of the core must be negative. This leads to a positive $ \partial  F_3 / \partial m > 0 $ in the overshoot region and a negative $ \partial  F_3 / \partial m < 0 $ in the core. Therefore the nonlocal dynamical equilibrium results in a pure $^3$He generation $R_3>0$ in the overshoot region and a pure $^3$He consumption $R_3<0$ in the core.

That is a $^3$He pump cycle mechanism driven by the convective mixing and the nuclear reactions. The total effect is that $^3$He is reproduced in the overshoot region, transported into the core by the mixing, and consumed in the core. Because the contribution to $\epsilon$ of the reaction $^3{\rm{He}}+ {}^3{\rm{He}}= {}^4{\rm{He}}+2p+12.86{\rm{Mev}}$ and its temperature sensitivity are high enough to drive the convection when the abundance of $^3$He keeps a high level, the pump cycle mechanism can significantly extend the life time of the convective core. The main difference between this mechanism and that of \citet{hd203608} is that we emphasize the \emph{consecutive reproduction} of $^3$He. If $^3$He in the overshoot region is mixed into the core once only, it should be depleted on its nuclear equilibrium timescale and cannot maintain the convective core over a much longer lifetime, e.g., $\sim1$Gyr.

\section{Core properties of solar models with COM} \label{SecFullymix}

\begin{figure*}
	\includegraphics[width=0.66\columnwidth]{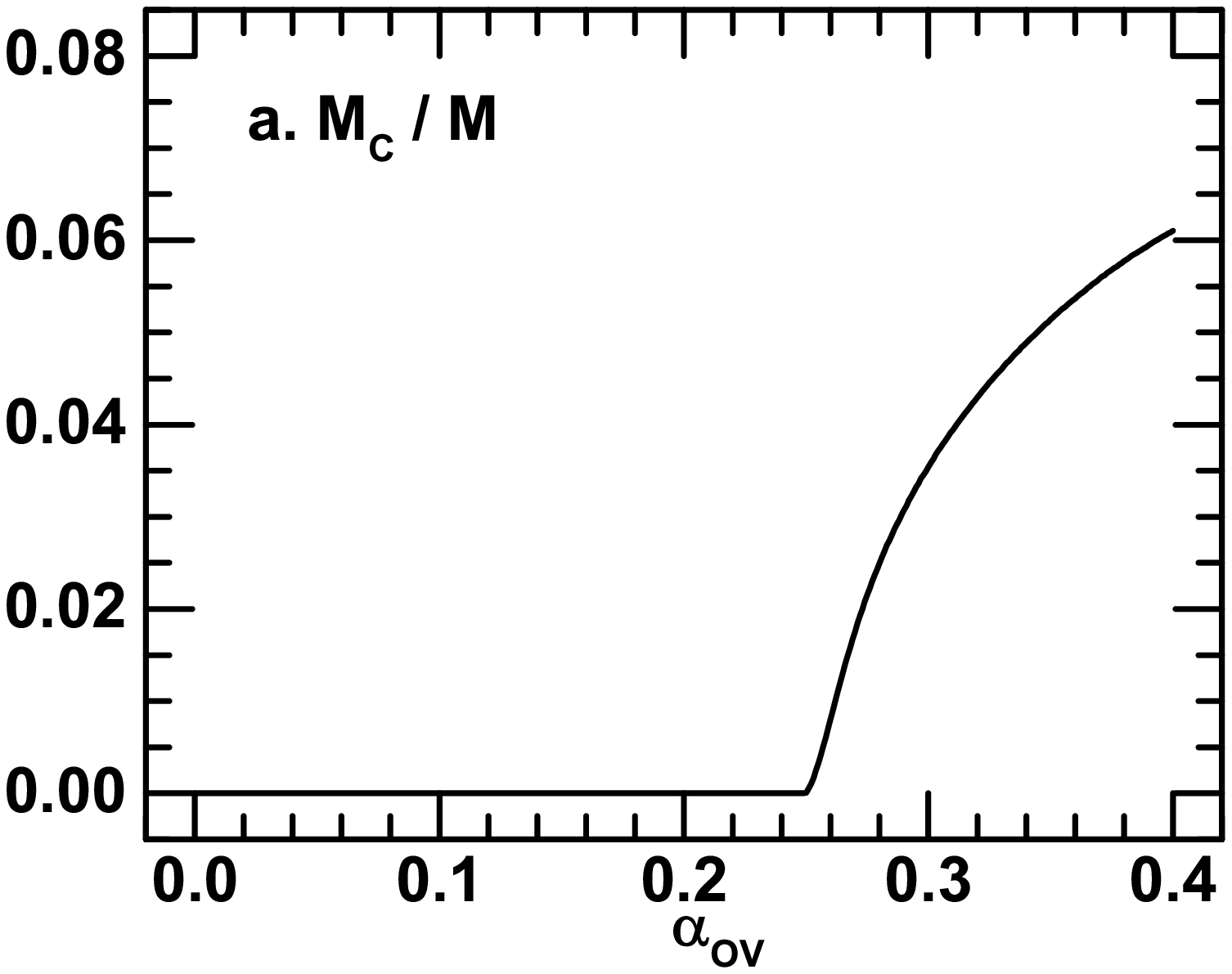}
	\includegraphics[width=0.66\columnwidth]{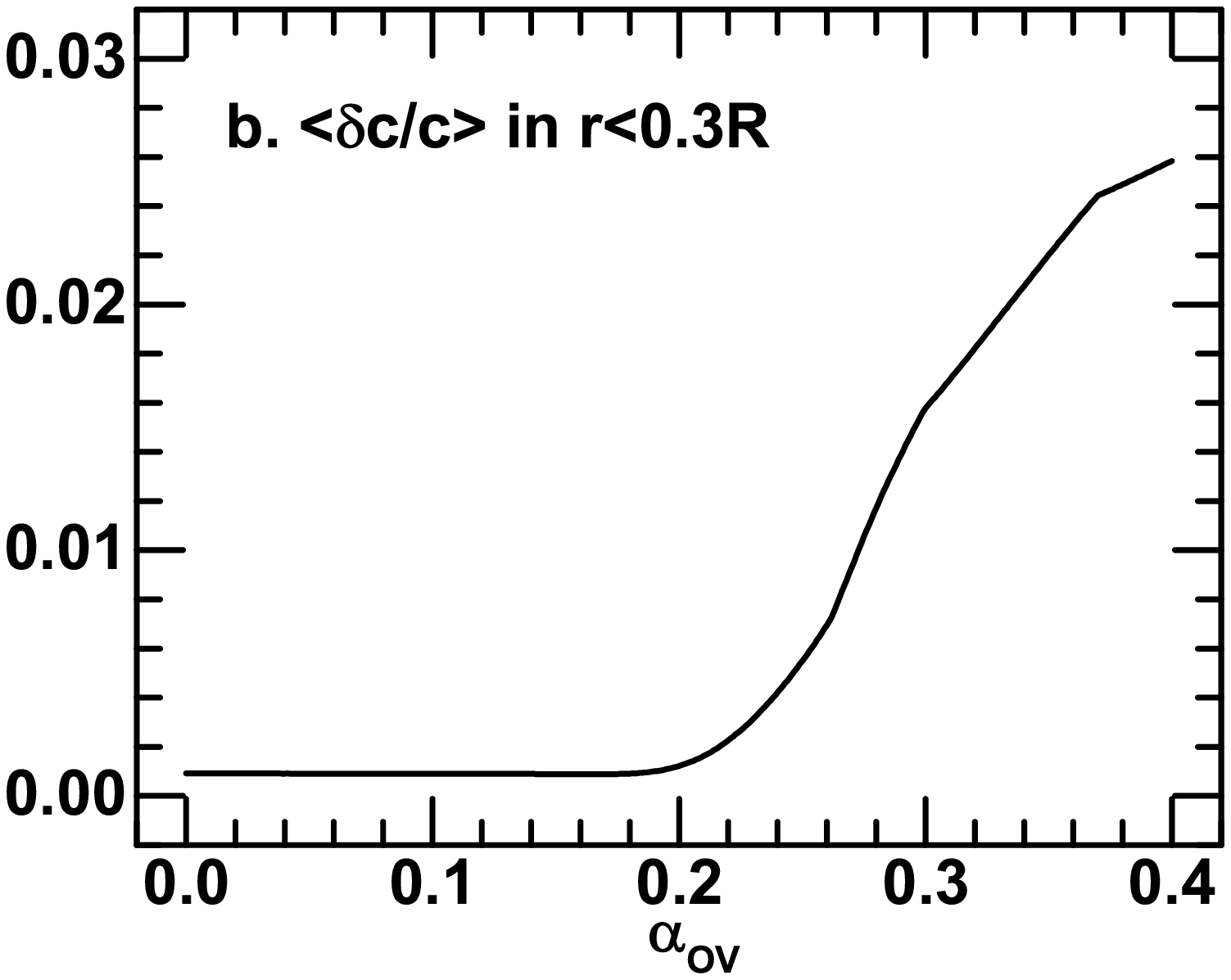}
	\includegraphics[width=0.66\columnwidth]{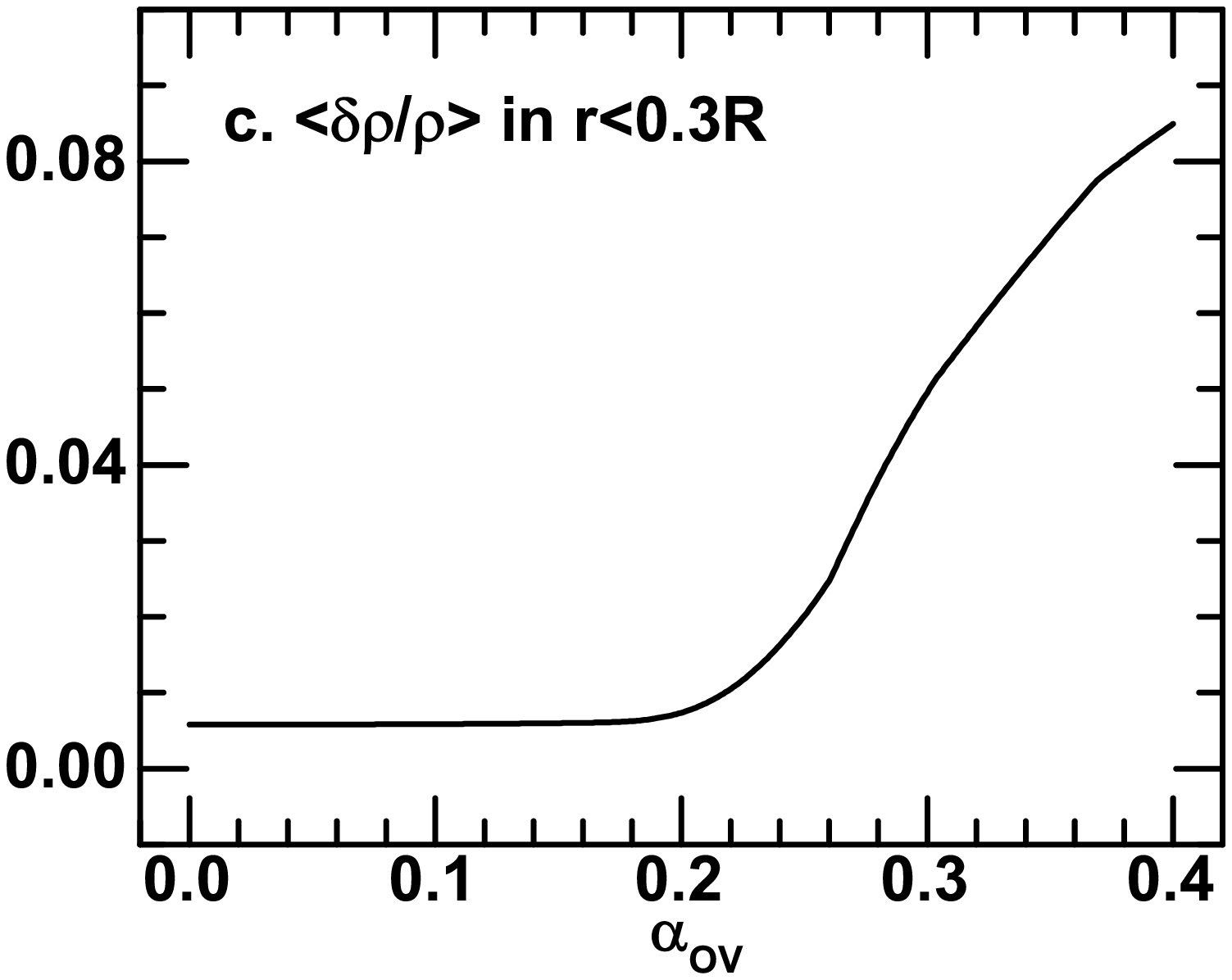}
	\includegraphics[width=0.66\columnwidth]{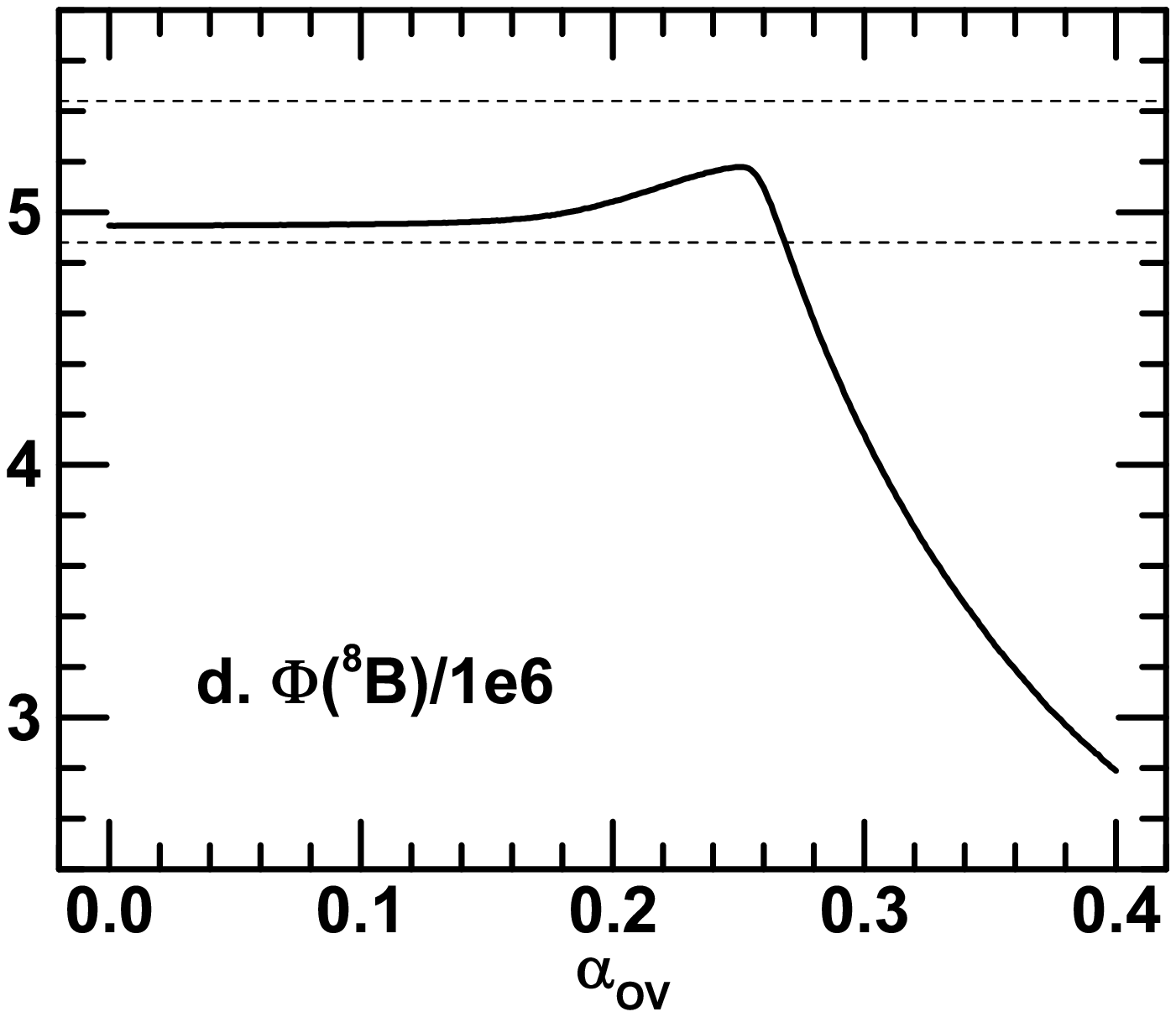}
	\includegraphics[width=0.66\columnwidth]{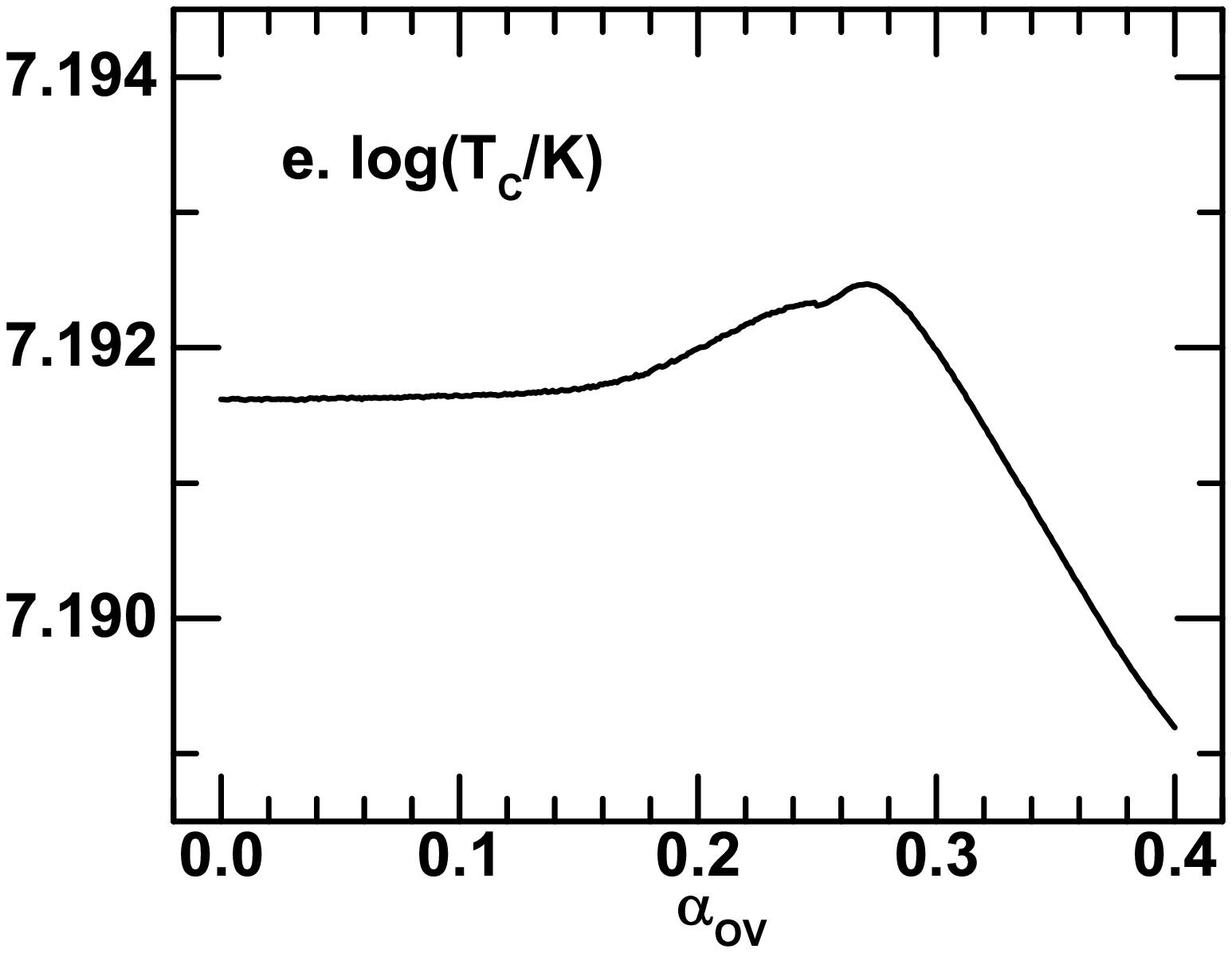}
	\includegraphics[width=0.66\columnwidth]{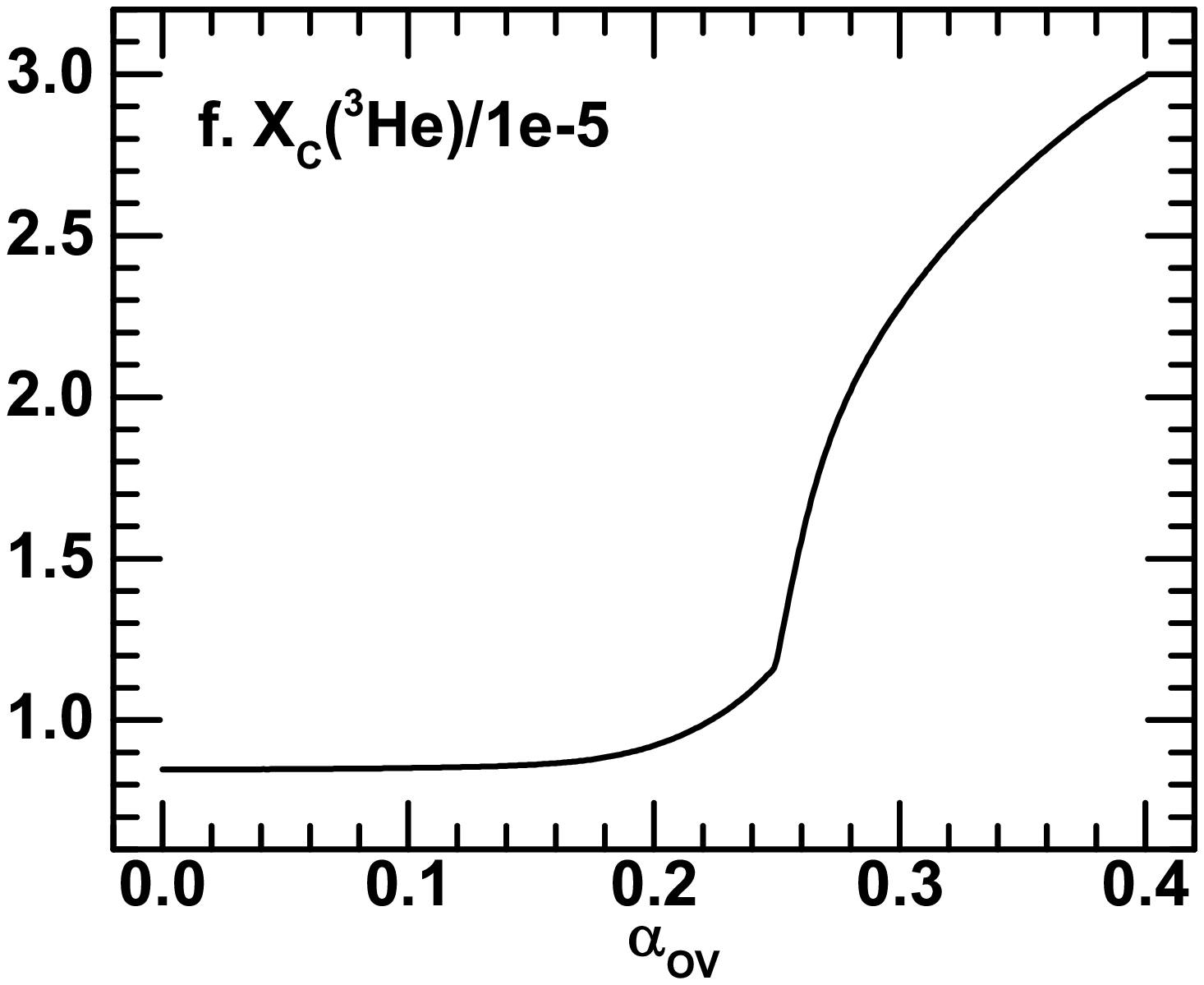}
	\includegraphics[width=0.66\columnwidth]{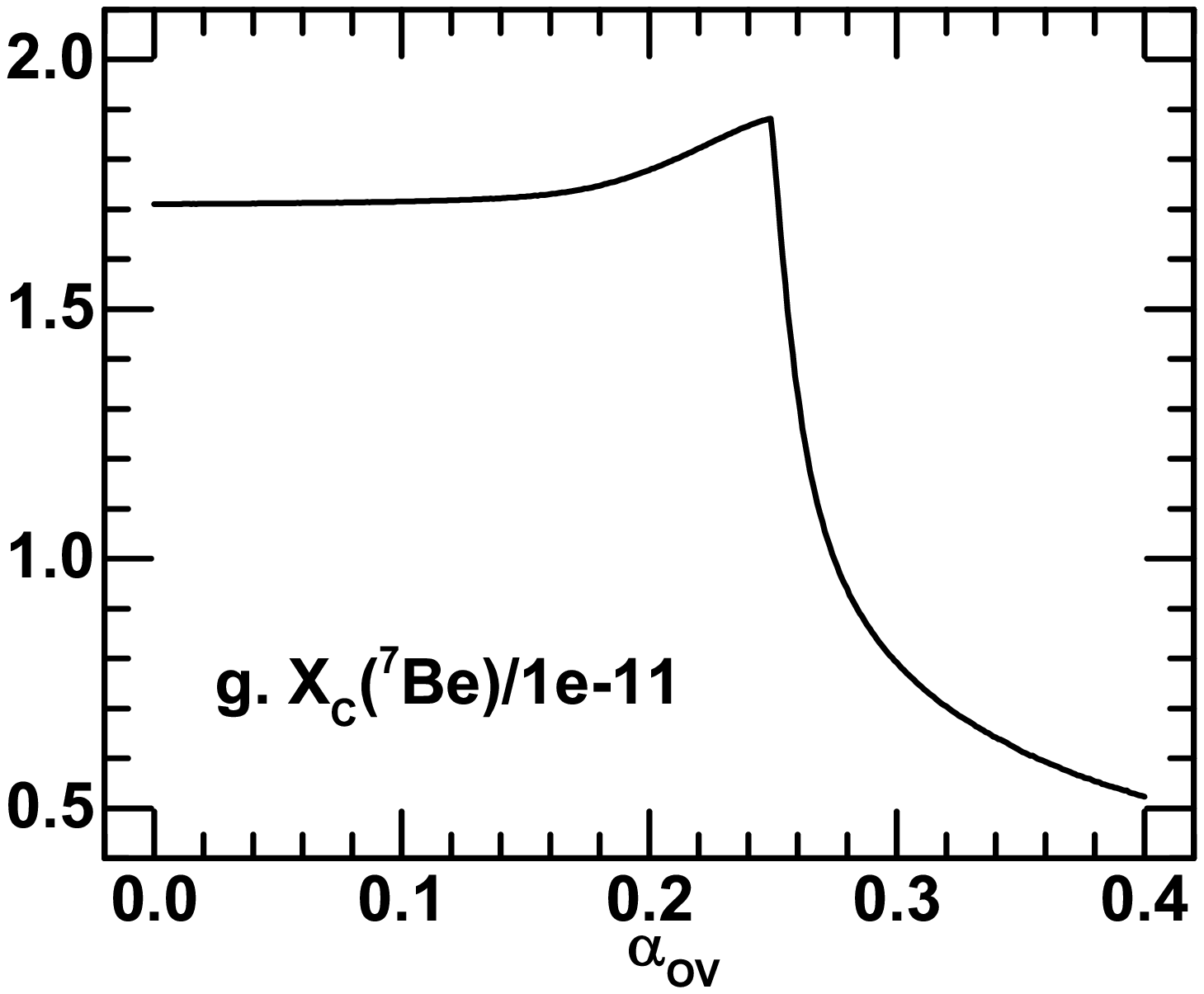}
	\includegraphics[width=0.66\columnwidth]{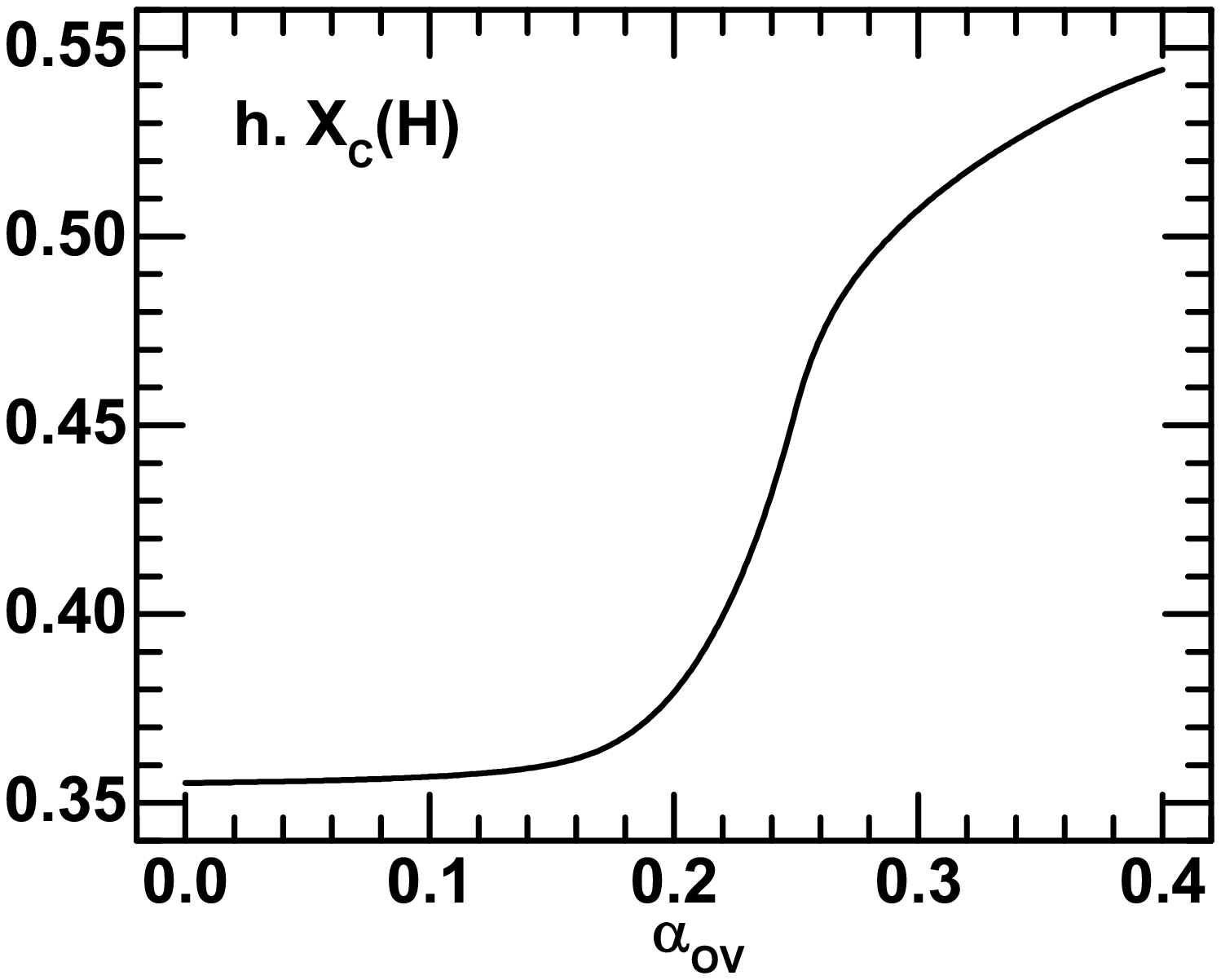}
	\caption{Properties of the cores of models of the present Sun with classical overshoot (cf.\ equation~\ref{eq:com}). The dashed lines in panel d show the discussed range of ${\rm{\Phi(^8B)}}$, in units of $10^6\,{\rm cm^{-2} s^{-1}}$, based on observation \citep{Bergstrom16} and theoretical uncertainty of $\sim$5\% (see text).}
    \label{resstd}
\end{figure*}

Solar models based on COM with $0 \leq \alpha_{\rm{ov}} \leq 0.4$ (step 0.001) have been calculated to investigate the effects of core overshoot mixing on the properties of the solar core. Figure~\ref{resstd} shows some properties of the cores of the solar models at the present solar age, i.e., the fraction of convective core mass $M_{\rm{cz}}/{\rm M_\odot}$, \textit{rms} sound-speed deviations $\langle \delta c / c \rangle$ and density deviations $\langle \delta \rho / \rho \rangle$ in the solar core with $r<0.3\,R$, the $^8$B neutrino flux ${\rm{\Phi(^8B)}}$, the central hydrogen, $^3$He and $^7$Be abundances $X_{\rm{c}}(\rm{H})$, $X_{\rm{c}}(^3\rm{He})$ and $X_{\rm{c}}(^7\rm{Be})$, and the central temperature $\log T_{\rm{c}}$. $\langle \delta c / c \rangle$ and $\langle \delta \rho / \rho \rangle$ are calculated based on \citeauthor{basu09}'s \citeyearpar{basu09} helioseismic inferences of sound speed and density.

Figures~\ref{resstd} (a-c) show that $\langle \delta c / c \rangle$ and $\langle \delta \rho / \rho \rangle$ are strongly correlated with the size of the convective core. The sound-speed and density profiles of solar models in the core with $r<0.3\,{\rm R_\odot}$ are in reasonable agreements with the helioseismic inferences when the core is not convective. However, the deviations of sound-speed and density profiles becomes significant when a convective core exists. Although the uncertainties of helioseismic inferences of sound speed and density in the core (e.g., $ \sim 0.1\%$ for sound speed and $ \sim 1\%$ for density) are significantly higher than that in the bulk of the sun, the deviations of the models with a convective core are significantly larger than the uncertainties, indicating that the structure of the core in solar models with a convective core is inconsistent. As shown in Fig.~\ref{resstd}d, ${\rm{\Phi(^8B)}}$ is also correlated with the size of the convective core. When the core is not convective ($\alpha_{\rm{ov}}\leq 0.25$), it increases with the increase of $\alpha_{\rm{ov}}$. When the core is convective, it decreases with the increase of $\alpha_{\rm{ov}}$. Observations have shown ${\rm{\Phi(^8B)}}=5.16 \times 10^6 (\pm2.2\%) \flun$ \citep{Bergstrom16}, while revision of the inferred solar surface abundances, giving rise to the so-called solar abundance problem \citep[see, e.g.,][for a review]{buldgen2019b} may lead to a model uncertainty which can be estimated as $\sim5\%$, from the range of ${\rm{\Phi(^8B)}}$ between the AGSS09Ne and GS98 SSMs compositions \citep[e.g.,][]{ZLCD2019}; thus a large convective core with $M_{\rm{cz}}>0.02 \msun$ or $\alpha_{\rm{ov}}> 0.28$ is excluded. The $^7$Be neutrino flux ${\rm{\Phi(^7Be)}}$ of those solar models changes in a small range between $4.58 \times 10^9$ and $4.74 \times 10^9 \flun$, consistent with observations \citep[e.g.][]{Bergstrom16}.

\begin{figure}
	\includegraphics[width=0.98\columnwidth]{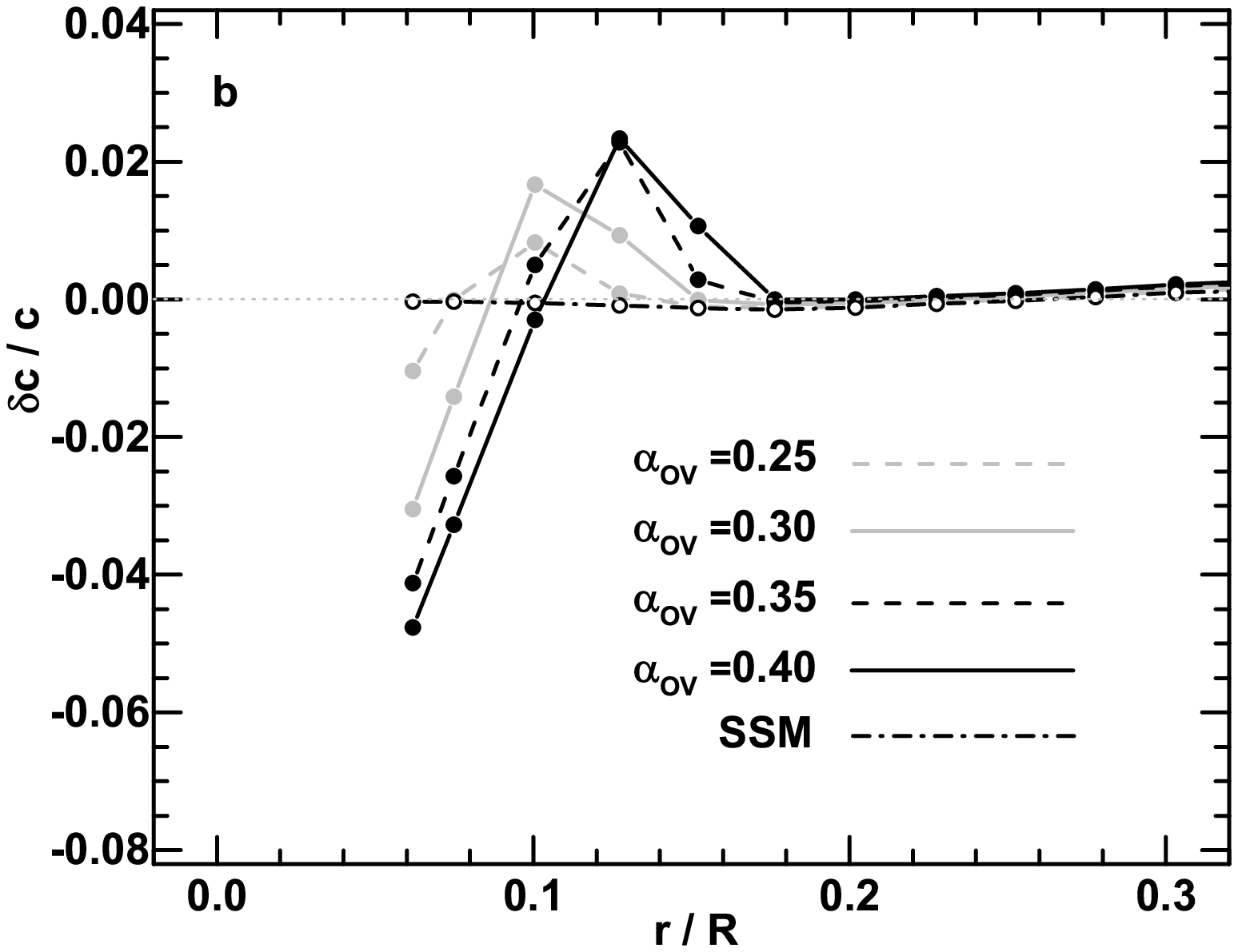}
	\includegraphics[width=0.98\columnwidth]{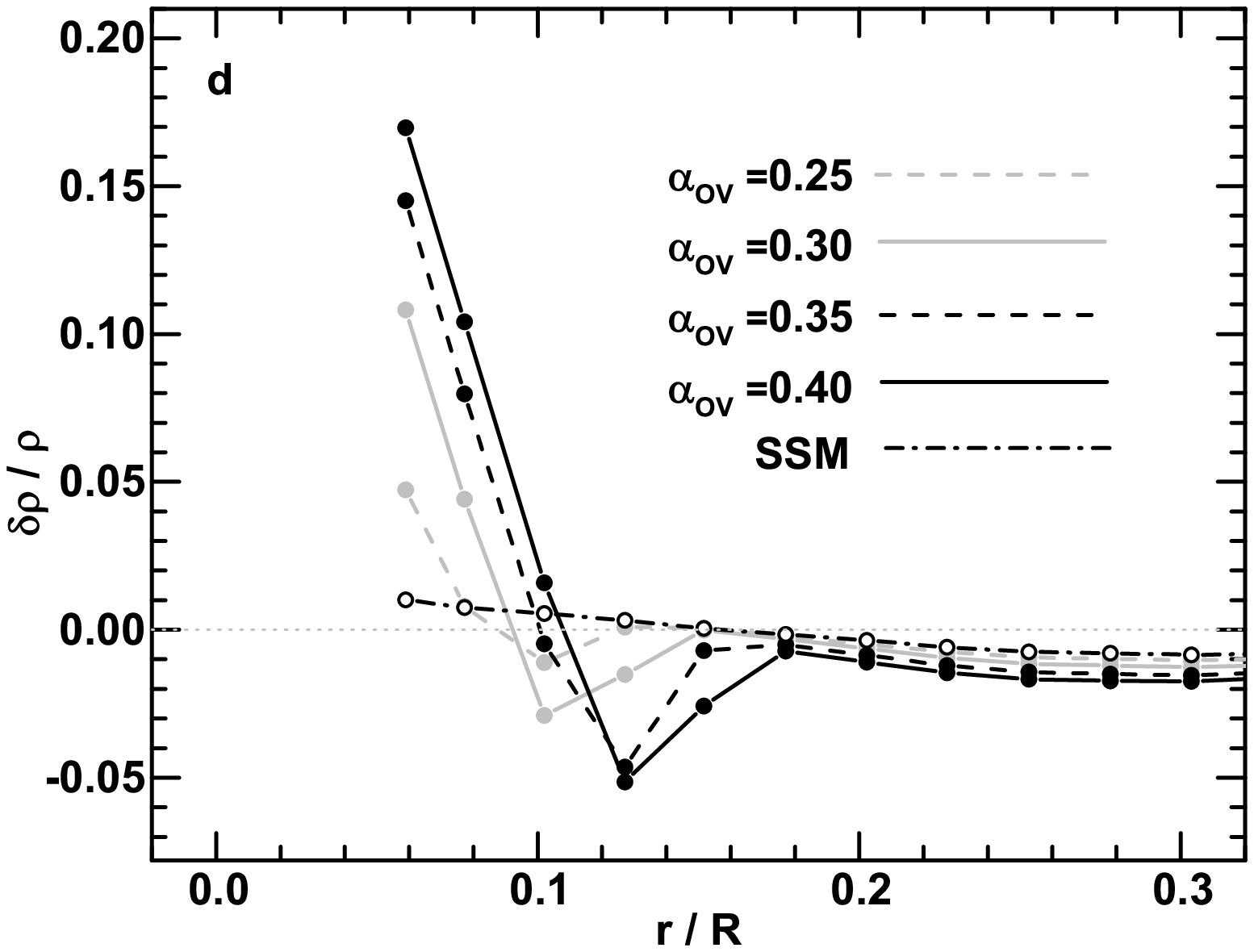}
	\caption{Sound-speed and density deviations of models of the present Sun with different COM core overshoot mixing parameter $\alpha_{\rm{ov}}$ in the sense of $1-c_{\rm model}/c_{\odot}$ and $1-\rho_{\rm model}/\rho_{\odot}$, where $c_{\odot}$ and $\rho_{\odot}$ are the helioseimic inferences from \citet{basu09}. }
    \label{dcdrhostd}
\end{figure}

From these results we conclude that the helioseismic results argue strongly against the presence of a convective core in the present Sun when the COM is used, limiting $\alpha_{\rm ov}$ to be less than 0.25. Similarly, $\Phi({}^8{\rm B})$ at most allows a tiny convective core.

\begin{figure}
	\includegraphics[width=0.98\columnwidth]{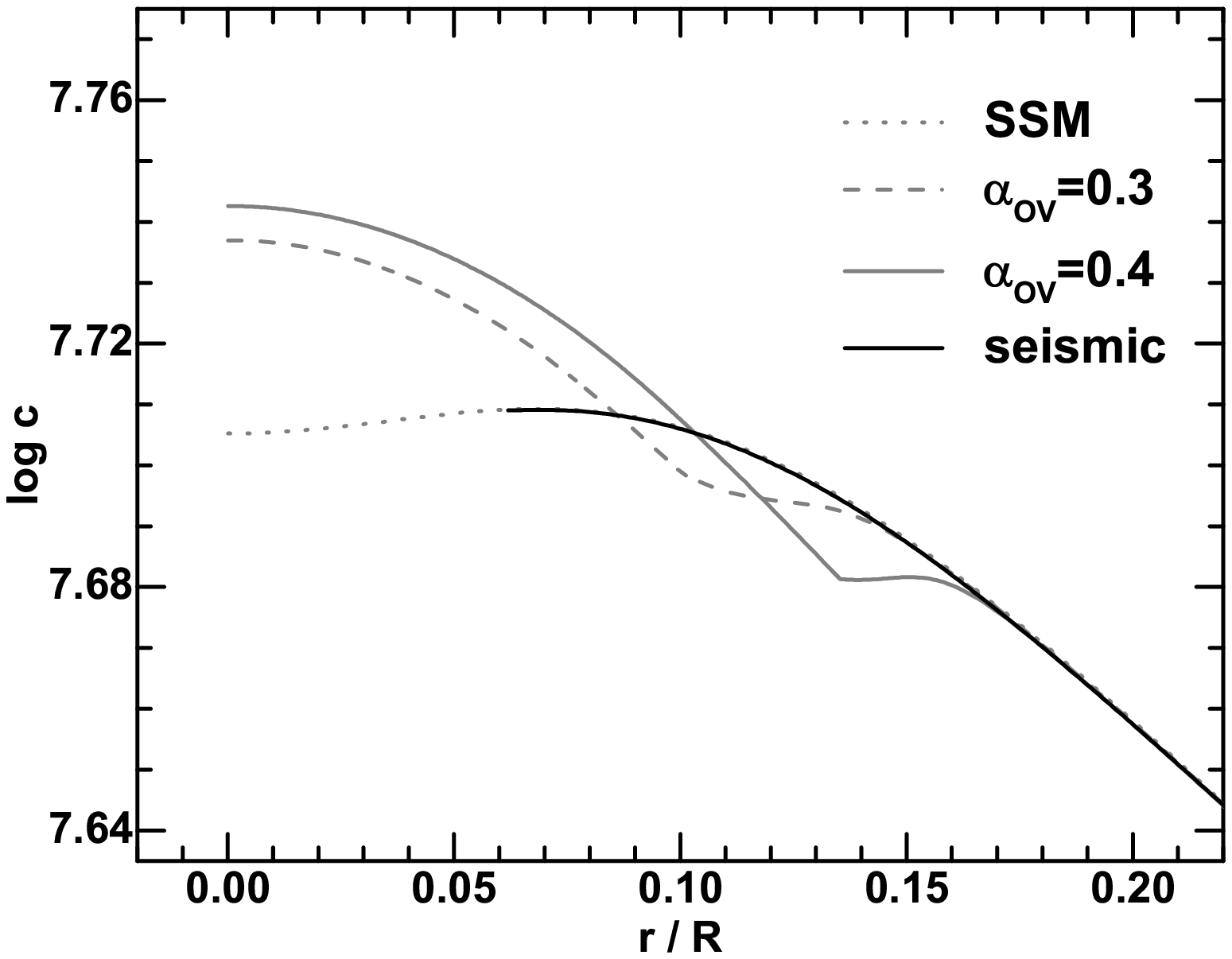}
	\includegraphics[width=0.98\columnwidth]{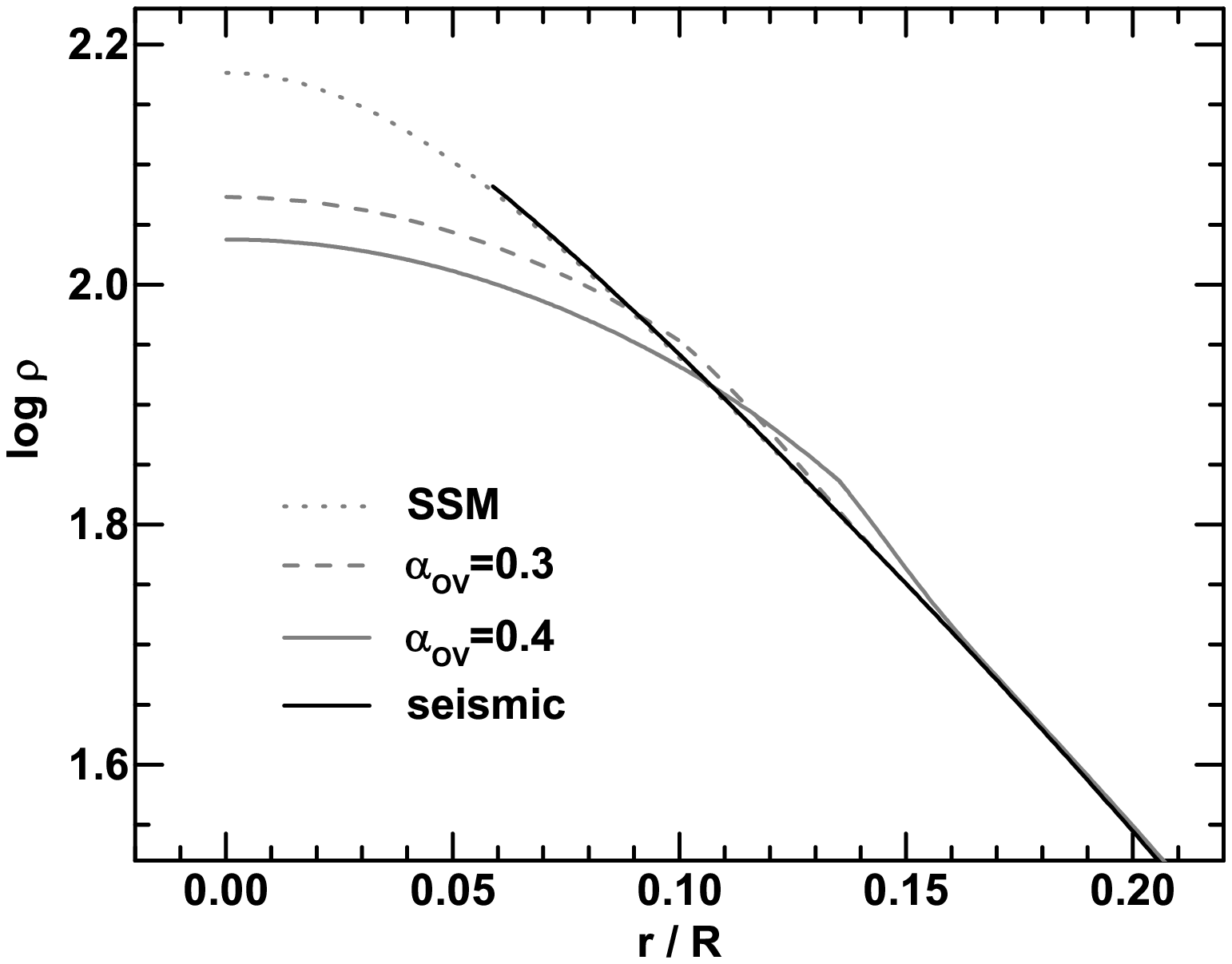}
	\includegraphics[width=0.98\columnwidth]{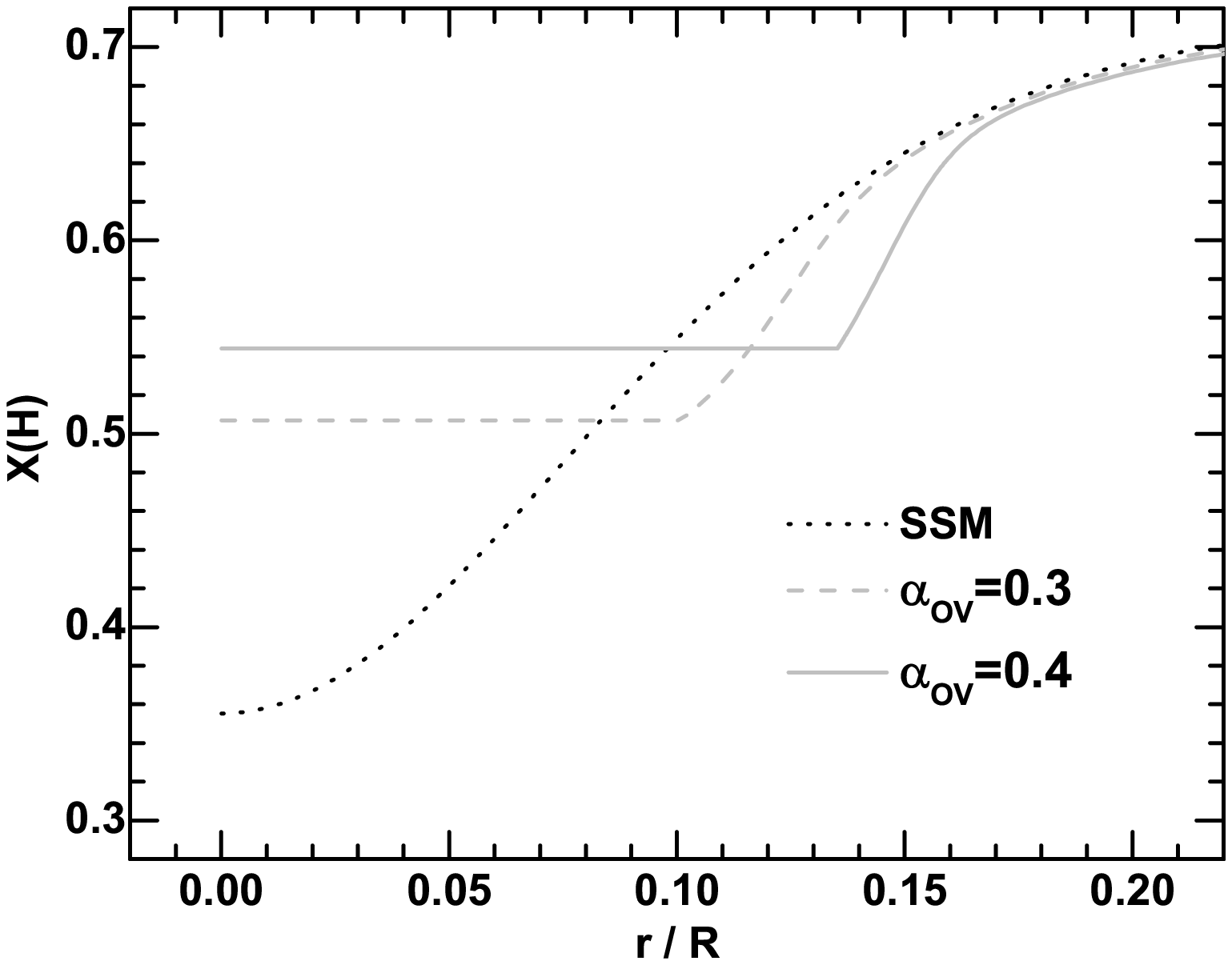}
	\caption{Sound speed, density and hydrogen abundance profiles in the core of the solar models with different value of $\alpha_{\rm{ov}}$. The black solid lines are smoothly interpolated through the helioseimic inferences from \citet{basu09}, the dotted lines are for model with $\alpha_{\rm{ov}}=0$ (the standard solar model), the grey dashed lines and the grey solid lines are for models with $\alpha_{\rm{ov}}=0.3$ and $\alpha_{\rm{ov}}=0.4$, respectively. }
    \label{crhofit}
\end{figure}

In order to understand the dependence of $\langle \delta c / c \rangle$ and $\langle \delta \rho / \rho \rangle$ on $\alpha_{\rm{ov}}$, the details of sound-speed and density deviations of those solar models are shown in Fig.~\ref{dcdrhostd}. For $r>0.3\,R$, the sound-speed and density deviations are basically identical to the SSM; thus they are not shown. It is found that a larger convective core (i.e., a high $\alpha_{\rm{ov}}$) leads to highly significant deviations of sound speed and density in the convective core. The sound-speed and density profiles in the core of the solar models with different values of $\alpha_{\rm{ov}}$ are shown in Fig.~\ref{crhofit}. It is found that, for a solar model with a convective core, the sound speed and its gradient are too high and the density and its gradient are too low in the convective core. Those features can be explained by the properties of the convective core as follows. The convective mixing leads to a higher central hydrogen abundance than that of the SSM as shown in Fig.~\ref{resstd}h and Fig.~\ref{crhofit}, thus a lower density is required to balance the pressure caused by the gravity of the star since $P \propto \rho T / \mu $, while it also leads to a higher sound speed since $c^2 \propto T / \mu$. The gradient of sound speed and density can be written as
\begin{equation} \label{eqdcdr}
\frac{{\dd \ln {c^2}}}{{\dd r}} \approx \frac{\dd }{{\dd r}}\left(\ln \frac{P}{\rho }\right) \approx \frac{\dd }{{\dd r}}\left(\ln \frac{T}{\mu }\right) = - \frac{1}{{{H_P}}}(\nabla - {\nabla _\mu })
\end{equation}
and
\begin{equation} \label{eqdrhodr}
\frac{{\dd \ln \rho }}{{\dd r}} \approx \frac{{\dd \ln P}}{{\dd r}} - \frac{{\dd \ln {c^2}}}{{\dd r}} \approx - \frac{1}{{{H_P}}}(1 - \nabla + {\nabla _\mu }).
\end{equation}
Because $\nabla < \nabla_{\rm{ad}}$ and $\nabla_{\mu}>0$ for a radiative core and $\nabla \approx \nabla_{\rm{ad}}$ and $\nabla_{\mu}=0$ for a convective core, $\nabla - {\nabla _\mu }$ in a convective core is larger than that in a radiative core and this leads to a higher gradient of sound speed and a lower gradient of density in the convective core. As mentioned above, the overshoot region is assumed to be radiatively stratified. If assuming an adiabatic stratified overshoot region, its effects on the gradients of the sound speed and density should lead to more significant deviations.

The effect of $\alpha_{\rm{ov}}$ on ${\rm{\Phi(^8B)}}$ can be understood by analyzing the status of the nuclear burning and mixing in the solar core, described by the central H, $^7$Be and $^3$He abundances, i.e., $X_{\rm{c}}({\rm{H}})$, $X_{\rm{c}}({\rm{^7Be}})$ and $X_{\rm{c}}({\rm{^3He}})$, and the central temperature $\log T_{\rm{c}}$, as shown in Fig.~\ref{resstd}. $X_{\rm{c}}({\rm{H}})$ is positively correlated with $\alpha_{\rm{ov}}$ as shown in Fig.~\ref{resstd}h, since the overshoot mixing brings hydrogen into the core. For $\alpha_{\rm{ov}}<0.25$, $X_{\rm{c}}({\rm{^3He}})$ is also positively correlated with $\alpha_{\rm{ov}}$ as shown in Fig.~\ref{resstd}f because the $^3$He abundance is positively correlated with the H abundance in nuclear equilibrium. For $\alpha_{\rm{ov}}>0.25$, the convective core survives in the solar model; thus $X_{\rm{c}}({\rm{^3He}})$ significantly increases due to the convective/overshoot mixing. The increase of $X_{\rm{c}}({\rm{^3He}})$ is more significant than the increase of $X_{\rm{c}}({\rm{H}})$ near $\alpha_{\rm{ov}}=0.25$. This is because the burning timescale of $^3$He is much shorter than that of H, and therefore $X_{\rm{c}}({\rm{H}})$ has a memory of historical overshoot but $X_{\rm{c}}({\rm{^3He}})$ does not. For $\alpha_{\rm{ov}}<0.25$, the central temperature is positively correlated with $\alpha_{\rm{ov}}$ as shown in Fig.~\ref{resstd}e. That is because opacity is positively correlated with $X_{\rm{c}}({\rm{H}})$, so that a higher $X_{\rm{c}}({\rm{H}})$ leads to a higher central temperature. For $\alpha_{\rm{ov}}>0.25$, however, because of the significant increase of the central $^3$He abundance, which direct determines the total reaction rate of the pp chains and dominates the total luminosity, the calibration of the total luminosity requires a decrease of the central temperature. For $\alpha_{\rm{ov}}<0.25$, $X_{\rm{c}}({\rm{^7Be}})$ is positively correlated with $X_{\rm{c}}({\rm{^3He}})$ and the central temperature; thus it increases with $\alpha_{\rm{ov}}$. For $\alpha_{\rm{ov}}>0.25$, the $^7$Be abundance is significantly diluted by the convective/overshoot mixing because the generation rate of $^7$Be is strongly positively correlated with the temperature; thus $^7$Be is mainly produced in the convective core. The electron-capture of $^7$Be is much less sensitive to temperature than the proton-capture of $^7$Be. Consequently $X_{\rm c}({\rm ^8B})$, and hence ${\rm{\Phi(^8B)}}$, are positively correlated with temperature and the $^7$Be abundance, and therefore it increases with $\alpha_{\rm{ov}}$ for $\alpha_{\rm{ov}}<0.25$ due to the increase of the $^7$Be abundance and $\log T_{\rm{c}}$ and quickly decreases for $\alpha_{\rm{ov}}>0.25$ because of the mixing diluting the $^7$Be abundance.

The convective core survives to the present solar age only for $\alpha_{\rm{ov}}>0.25$; in this case the $^8$B neutrino flux is not in agreement with the observations. However, the sound-speed and density deviations are more sensitive than the neutrino flux since they start to be significant when $\alpha_{\rm{ov}}>0.20$. Thus, by using the strong constraint provided by the sound-speed and density deviations, it is indicated in the COM case that the convective core should vanish before $t=2$\,Gyr which is the lifetime of the convective core of the model with $\alpha=0.20$.

\section{Core properties of solar models with EDOM} \label{SecComp}

\subsection{Abundance profiles in the overshoot region} \label{SecComp1}

The diffusion coefficient of mixing in the overshoot region in the COM is much higher than that in the EDOM. In the EDOM, on the other hand, the quickly decreasing exponential diffusion coefficient covers a range of many order of magnitude. This leads to a complexity on the abundance profile of elements involved in nuclear reactions.

In regions where there is no mixing or enhanced diffusion the evolution of the abundances is controlled by the local nuclear reactions, including local nuclear equilibrium since D, ${}^3{\rm He}$, ${}^7 {\rm Be}$ and ${}^7 {\rm Li}$ all have nuclear timescales that are much shorter compared with the evolution timescale of the Sun. In regions with diffusive mixing the behaviour of an element depends on the relative magnitude of the nuclear and the diffusion timescale. In regions where the mixing timescale is shorter than the nuclear timescale a spatially constant abundance is obtained, as a suitable average over the mixed region of the nuclear equilibrium abundance. In contrast, for elements with nuclear timescales substantially shorter than the mixing timescale, local nuclear equilibrium will apply, obviously depending on the abundance of other relevant elements. This leads to different overshoot lengths for elements with different burning timescales.

Another property of the EDOM is that the effective overshoot distance increases with the stellar age. Taking hydrogen for example, since $D$ decreases in the overshoot region, the effective overshoot distance defined as $l_{\rm{ov},dif}=r_1-r_{\rm{cz}}$ can be estimated based on the mixing timescale at $r_1$ being equal to the stellar age, i.e., $t \sim \tau_{\rm{mix}} \sim (r_1-r_{\rm{cz}})^2 / D(r_1)$. Therefore $l_{\rm{ov},dif} \propto \sqrt{t}$. In the COM, $D$ truncates at the boundary of the overshoot region so that there is no such effect.

Although there is no observational data directly relating to the abundance profiles in the core, the investigation on the abundance profiles helps to understand the interaction between the mixing and nuclear reactions in the stellar interior. A detailed analysis of the abundance profiles in the convective overshoot region is presented in the Appendix.

\subsection{Other properties of the core} \label{SecComp2}

\begin{figure*}
	\includegraphics[width=0.66\columnwidth]{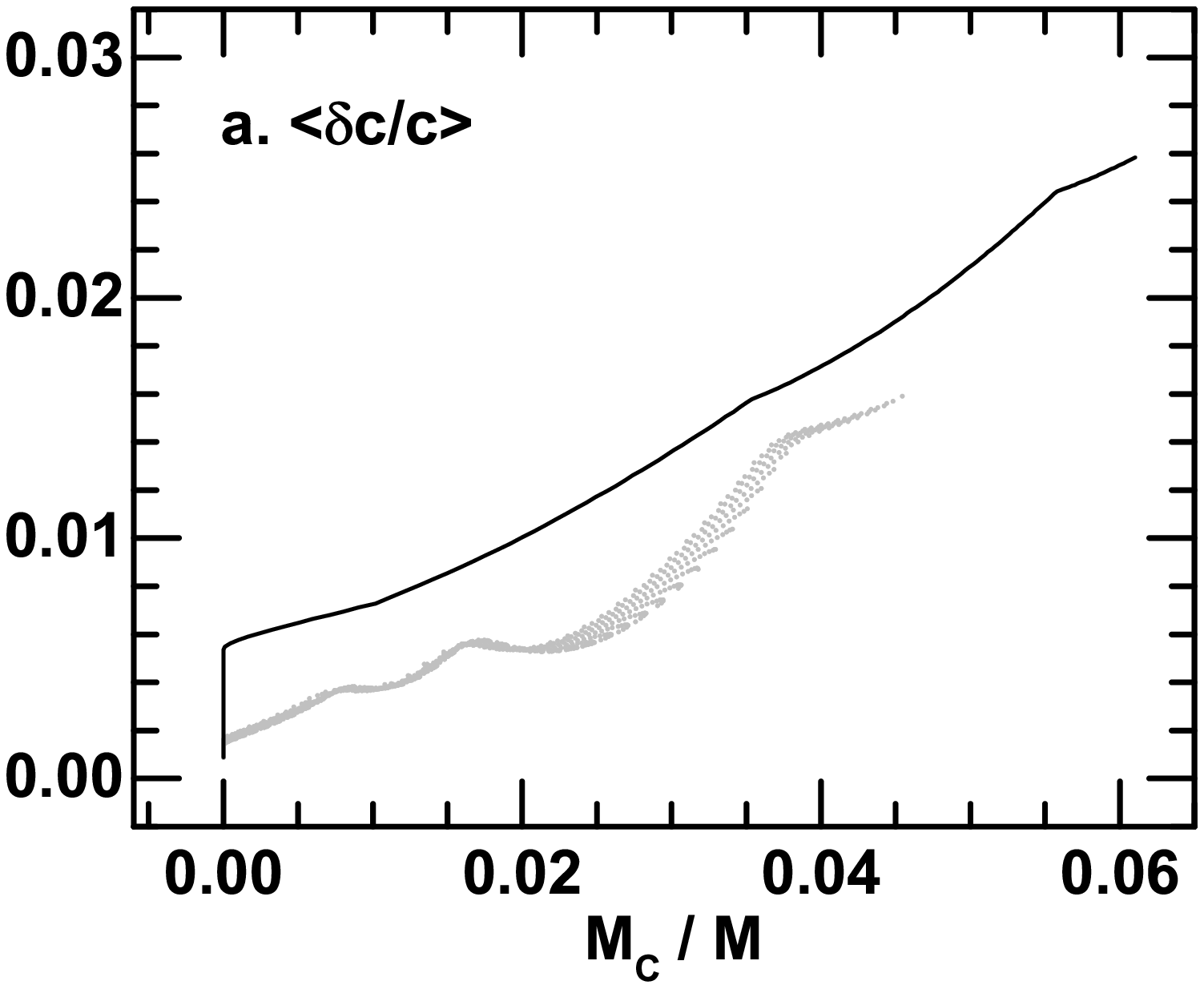}
	\includegraphics[width=0.66\columnwidth]{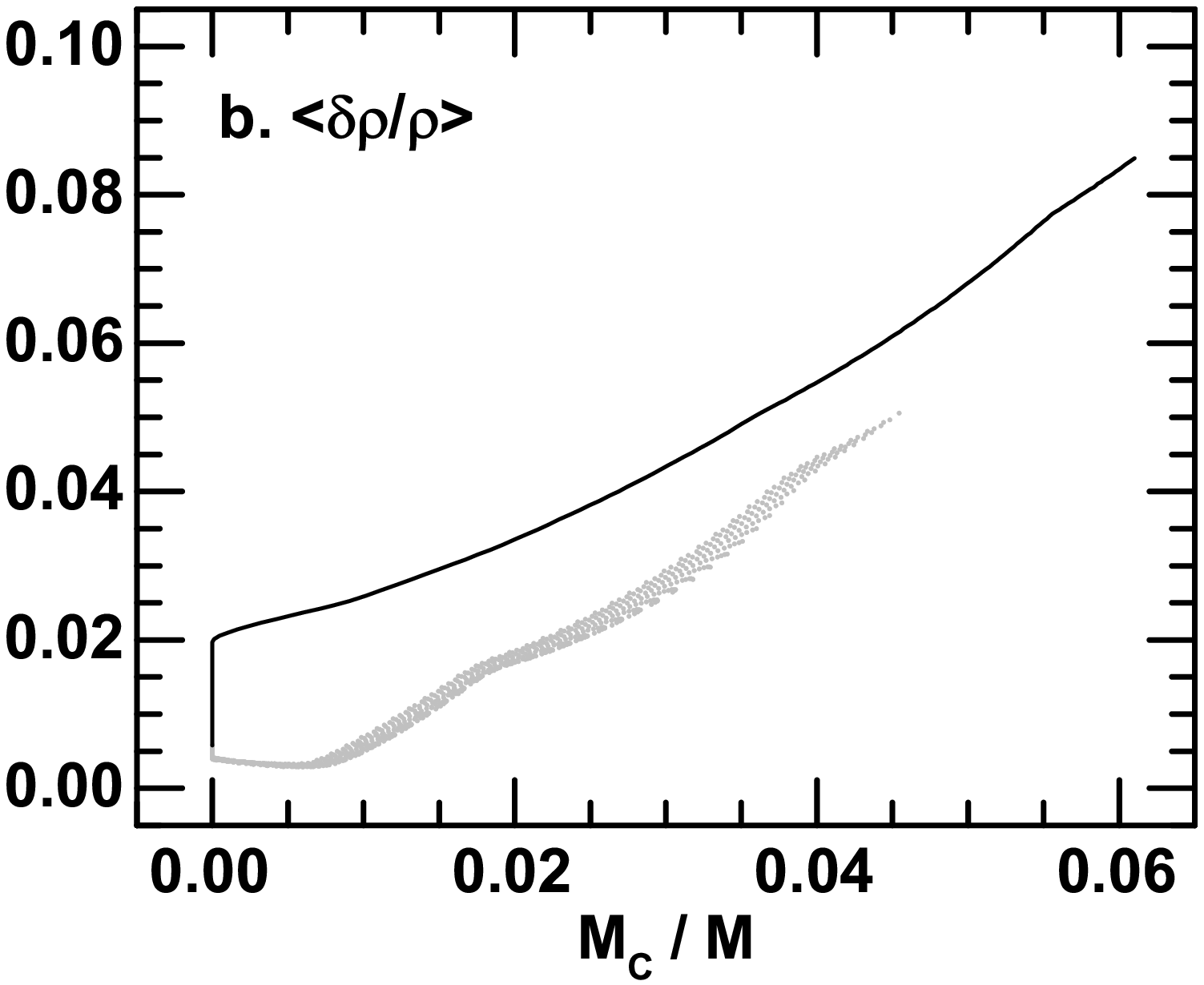}
	\includegraphics[width=0.66\columnwidth]{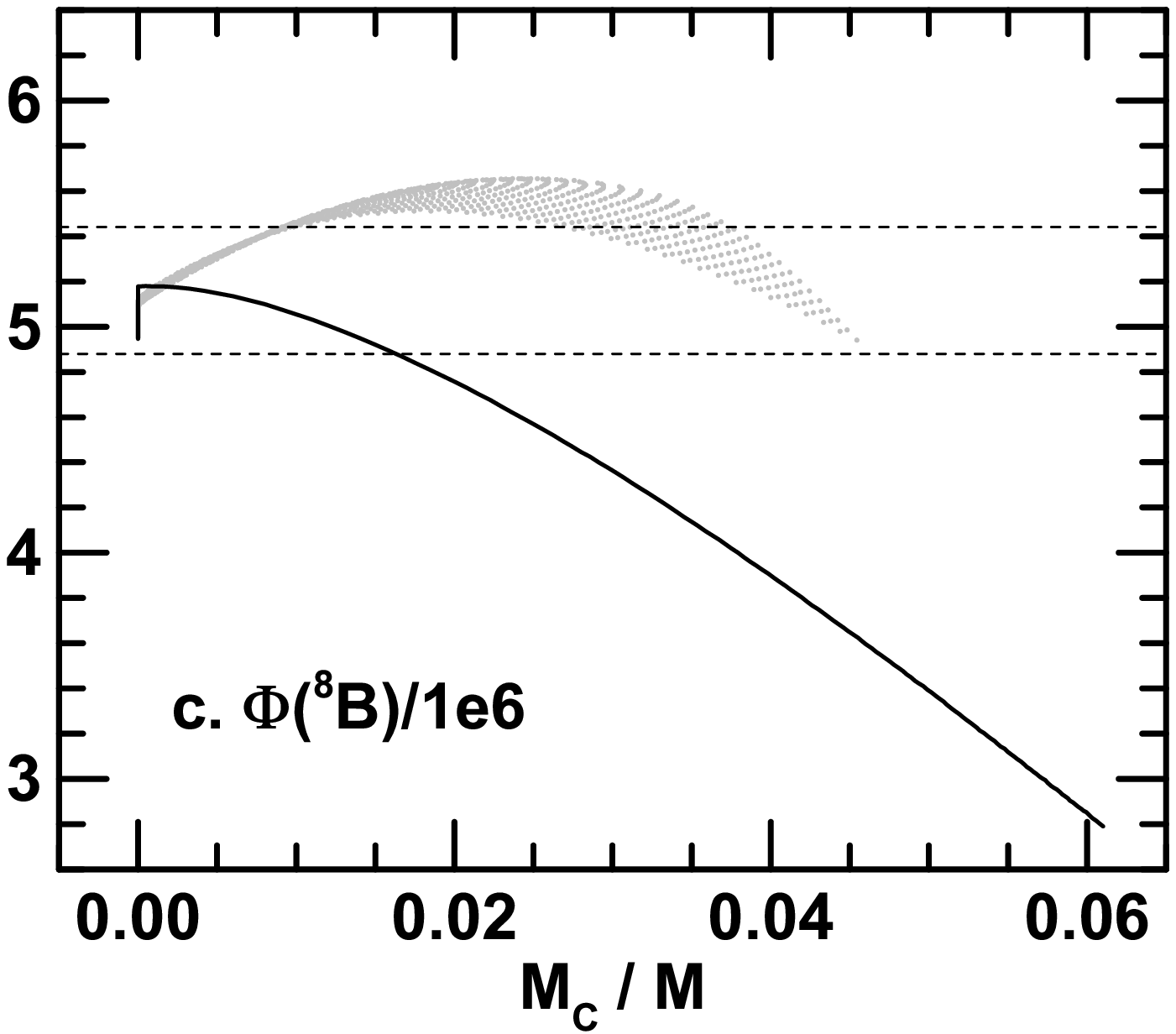}
	\includegraphics[width=0.66\columnwidth]{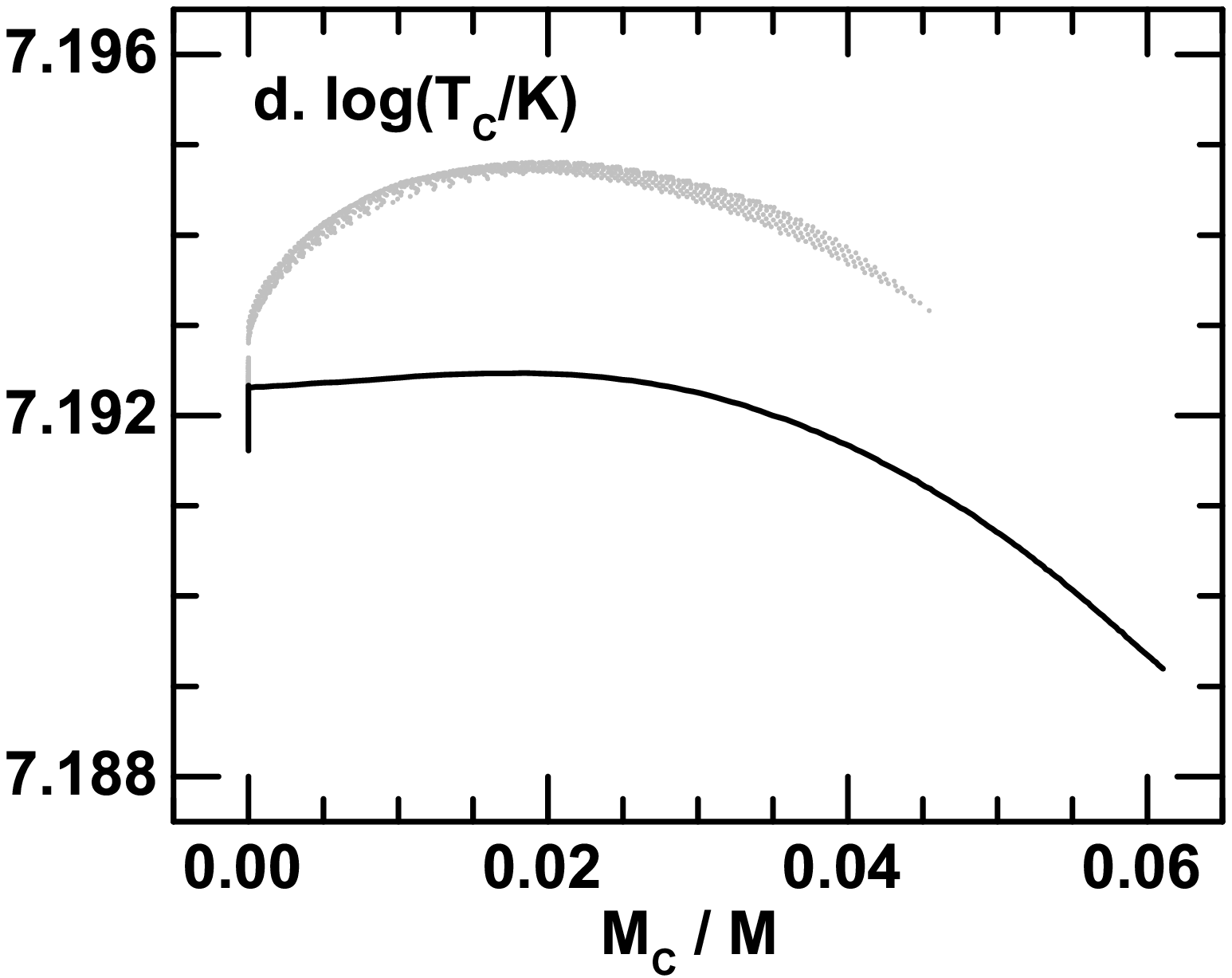}
	\includegraphics[width=0.66\columnwidth]{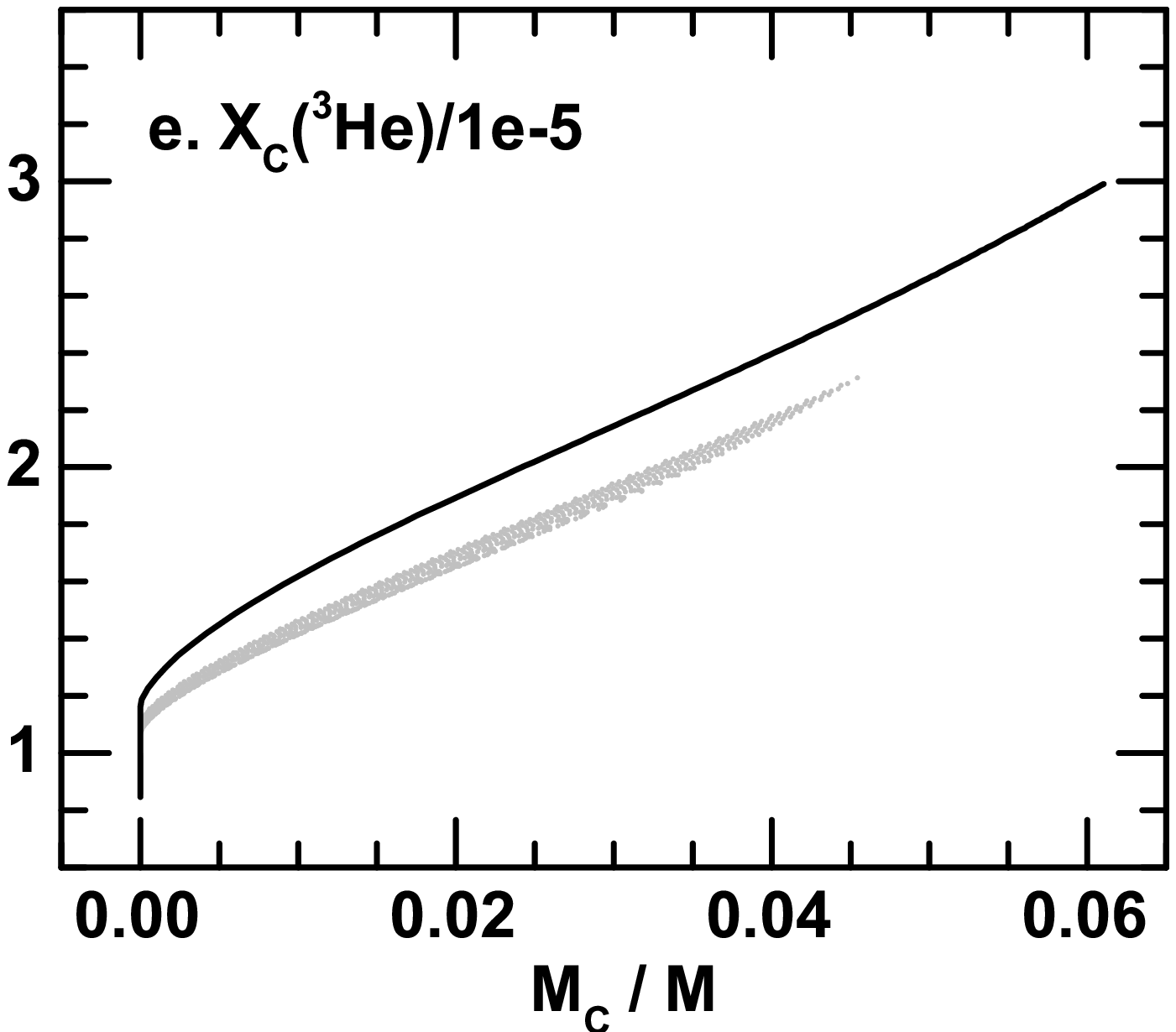}
	\includegraphics[width=0.66\columnwidth]{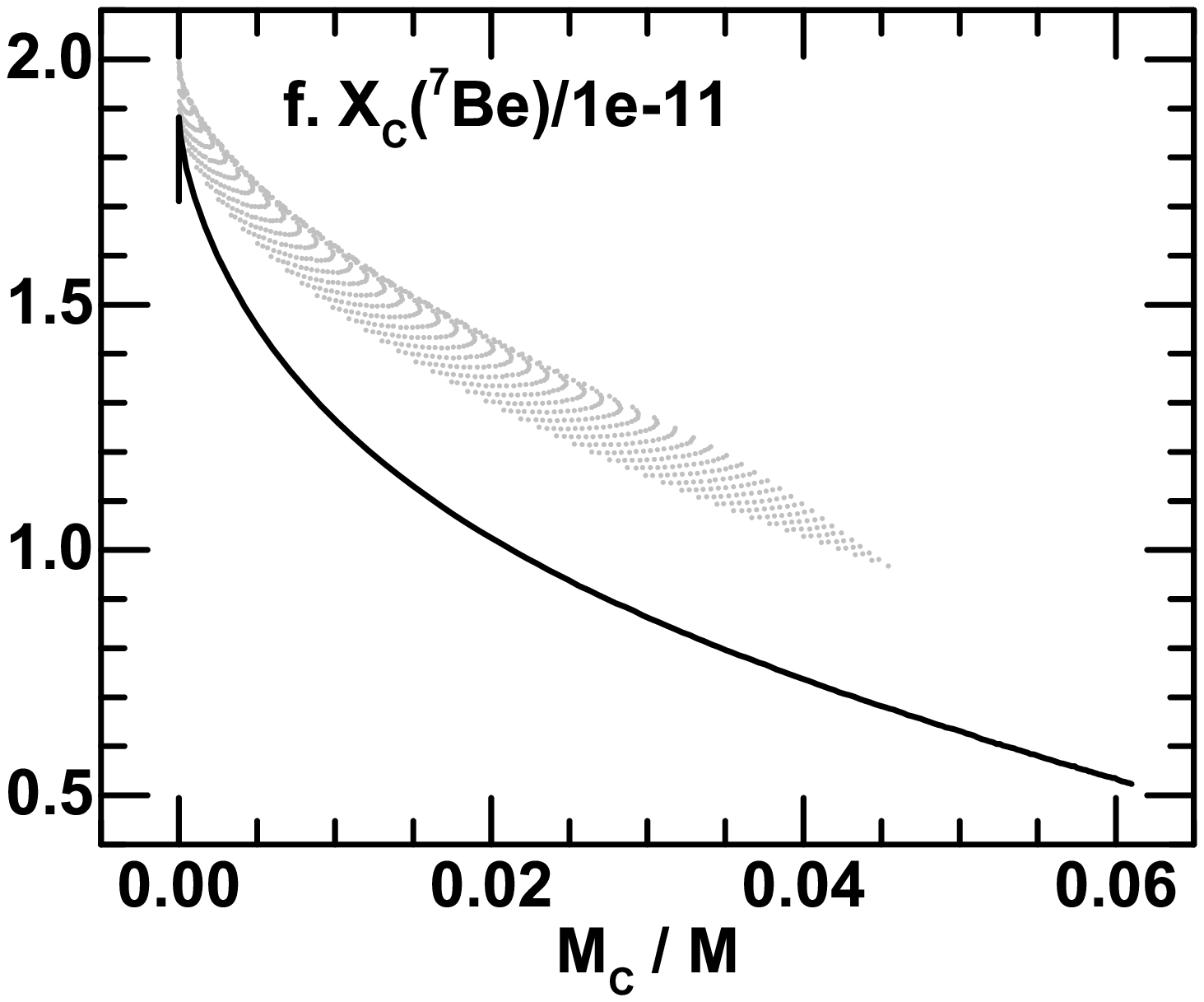}
	\includegraphics[width=0.66\columnwidth]{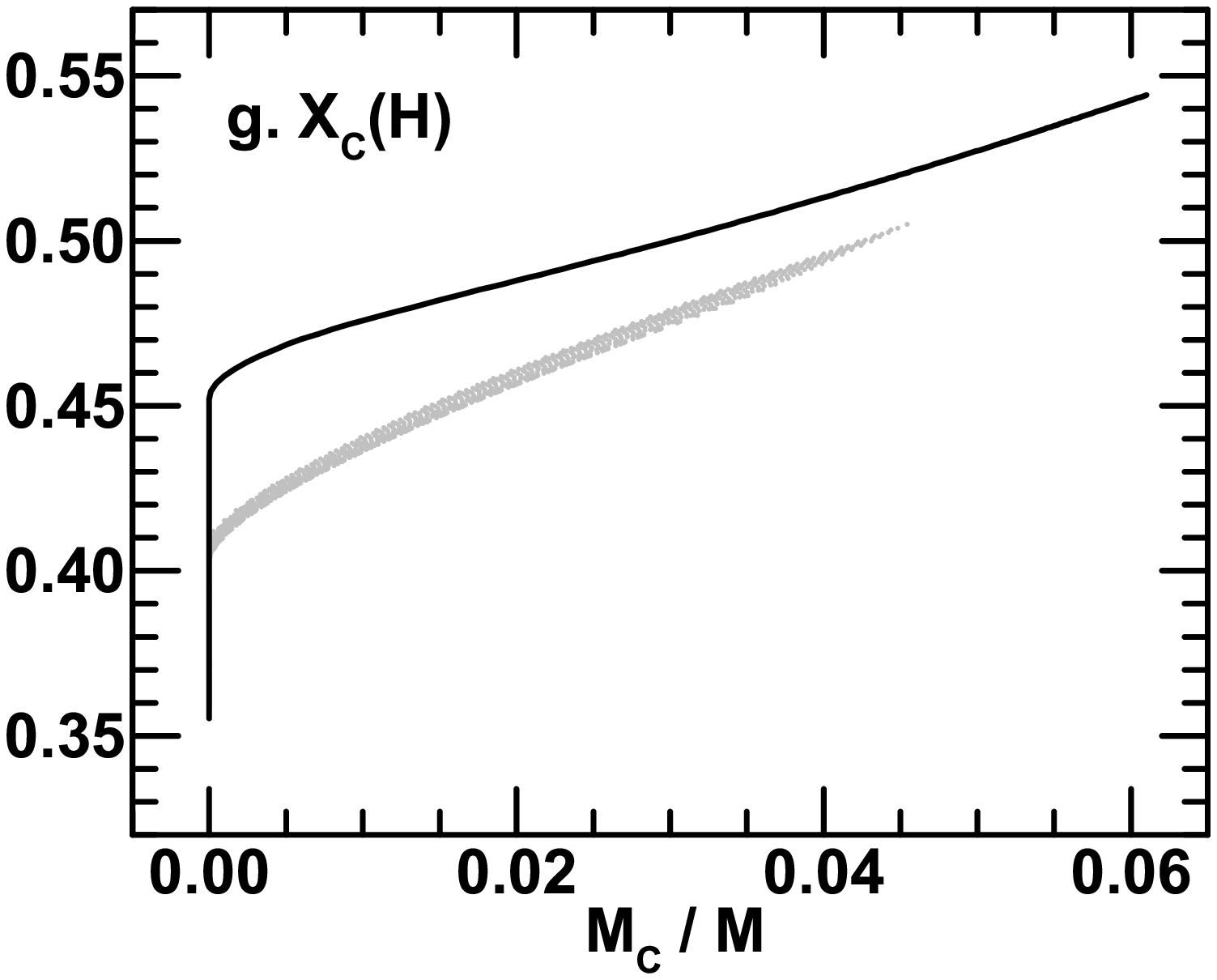}
	\caption{Properties, as a function of the mass of the convective core, of the cores of models of the present Sun with overshoot mixing. Solid lines are for models with COM and grey dots are for models with EDOM. The dashed lines in panel d shown the range of determinations of ${\rm{\Phi(^8B)}}$, in units of $10^6\,{\rm cm^{-2} s^{-1}}$. }
   \label{resdif1d}
\end{figure*}

Solar models with $-6 \leq \log C \leq 0$ (step 0.2) and $2.1 \leq \log \theta \leq 2.6$ (step 0.01) based on EDOM (equation~\ref{OVMmodel}) have been calculated to investigate the effects of the diffusion model of the core overshoot mixing on the properties of the present solar core. In order to compare them with models with COM, we consider only models with a present convective core and set the convective core mass fraction as the independent variable. Figure~\ref{resdif1d} shows some main properties of the cores of the models at the present solar age, i.e., $\langle \delta c / c \rangle$ and $\langle \delta \rho / \rho \rangle$ in the solar core with $r<0.3\,R$, ${\rm{\Phi(^8B)}}$, $X_{\rm{c}}(\rm{H})$, $X_{\rm{c}}(^3\rm{He})$ and $X_{\rm{c}}(^7\rm{Be})$, and $\log T_{\rm{c}}$. The solid line shows the solar model with COM and the grey dots show results for the solar models with EDOM. It is found that the dependence of each variable on the convective core mass fraction in the solar models with EDOM is similar to that of the solar models with COM. We note that, as a result of the range of values considered in $C$ and $\theta$ for the EDOM models their results cover a band of values; however the width of this is constrained by carrying out the analysis at fixed mass of the convective core. The qualitative analysis in Section \ref{SecFullymix} is also applicable for the models with EDOM. However, the core properties show systemic differences between EDOM and COM.

In order to understand the reason of the differences of the observable variables $\langle \delta c / c \rangle$, $\langle \delta \rho / \rho \rangle$ and ${\rm{\Phi(^8B)}}$ between EDOM and COM as shown in Fig.~\ref{resdif1d}(a-c), we have first to investigate $X_{\rm{c}}(\rm{H})$, $X_{\rm{c}}(^3\rm{He})$ and $X_{\rm{c}}(^7\rm{Be})$, and $\log T_{\rm{c}}$. Those are shown in Fig.~\ref{resdif1d}(d-g). For a given mass fraction of the convective core, the central hydrogen abundances of solar models with EDOM are less than those of COM. This is because $\dd l_{\rm{ov},dif}/\dd t>0$ for EDOM and $\dd l_{\rm{ov}}/\dd t<0$ for COM. The latter has stronger overshoot than the former before the present solar age, thus $X_{\rm{c}}(\rm{H})$ of the models with COM are higher. The same reason holds for $X_{\rm{c}}(^3\rm{He})$. As discussed above, the efficiency of $^3$He and $^7$Be mixing in EDOM is weaker than in COM. Therefore the solar model with EDOM has lower $X_{\rm{c}}(^3\rm{He})$. Concerning ${}^7{\rm Be}$ the abundance in the core, as discussed above, reflects an averaged nuclear equilibrium over the fully mixed region; since this extends further for COM than for EDOM, the average nuclear equilibrium abundance includes lower temperatures, resulting in a lower ${}^7{\rm Be}$ abundance for COM than for EDOM in the core. For a given mass fraction of the convective core, the solar model with EDOM has higher central temperature because the hydrogen and $^3$He abundances are lower so that the calibration of luminosity requires a higher central temperature.

${\rm{\Phi(^8B)}}$ of the models with EDOM is higher than those with COM as shown in Fig.~\ref{resdif1d}c. The reason is that the models with EDOM have higher $X_{\rm{c}}(^7\rm{Be})$ and central temperature. The former is because the overshoot of $^7$Be in \hbox{EDOM} is much weaker than that in COM due to its short burning timescale (see the Appendix). The models with EDOM show smaller deviations of the sound speed and density in the core relative to the Sun as illustrated in Fig.~\ref{resdif1d}a and b. This results from their lower central hydrogen abundance. The solar model with COM show higher sound speed and lower density than those of the helioseismic inferences as shown in Fig.~\ref{crhofit}. The lower central hydrogen abundance in the models with EDOM leads to a higher $\mu$ so that a higher density is required to balance the pressure and a lower sound speed is obtained because $c^2\propto\mu^{-1}$. Those reduce the deviation of sound speed and density for the solar models with EDOM, compared with the models computed using COM.

Another difference between the results of the COM and EDOM calculations is shown in Fig.\,\ref{resdif1d}, namely that the sound-speed and density deviations at $M_c=0$ can be significant and this occurs only in the COM case. As discussed above, this leads to the stronger constraint that the lifetime of the convective core should be less than 2\,Gyr in the COM case. The results of the EDOM show no degeneracy at $M_c=0$, indicating that there is no such stronger constraint. The possible reason of the difference is as follows. The historical strength of the overshoot mixing in the COM case is stronger than that in the EDOM case since $\dd l_{\rm{ov},dif}/\dd t>0$ for EDOM and $\dd l_{\rm{ov}}/\dd t<0$. Therefore it requires a much longer time to establish a high enough composition gradient to reduce the sound-speed and density deviations in the COM case.

\begin{figure*}
	\includegraphics[width=0.66\columnwidth]{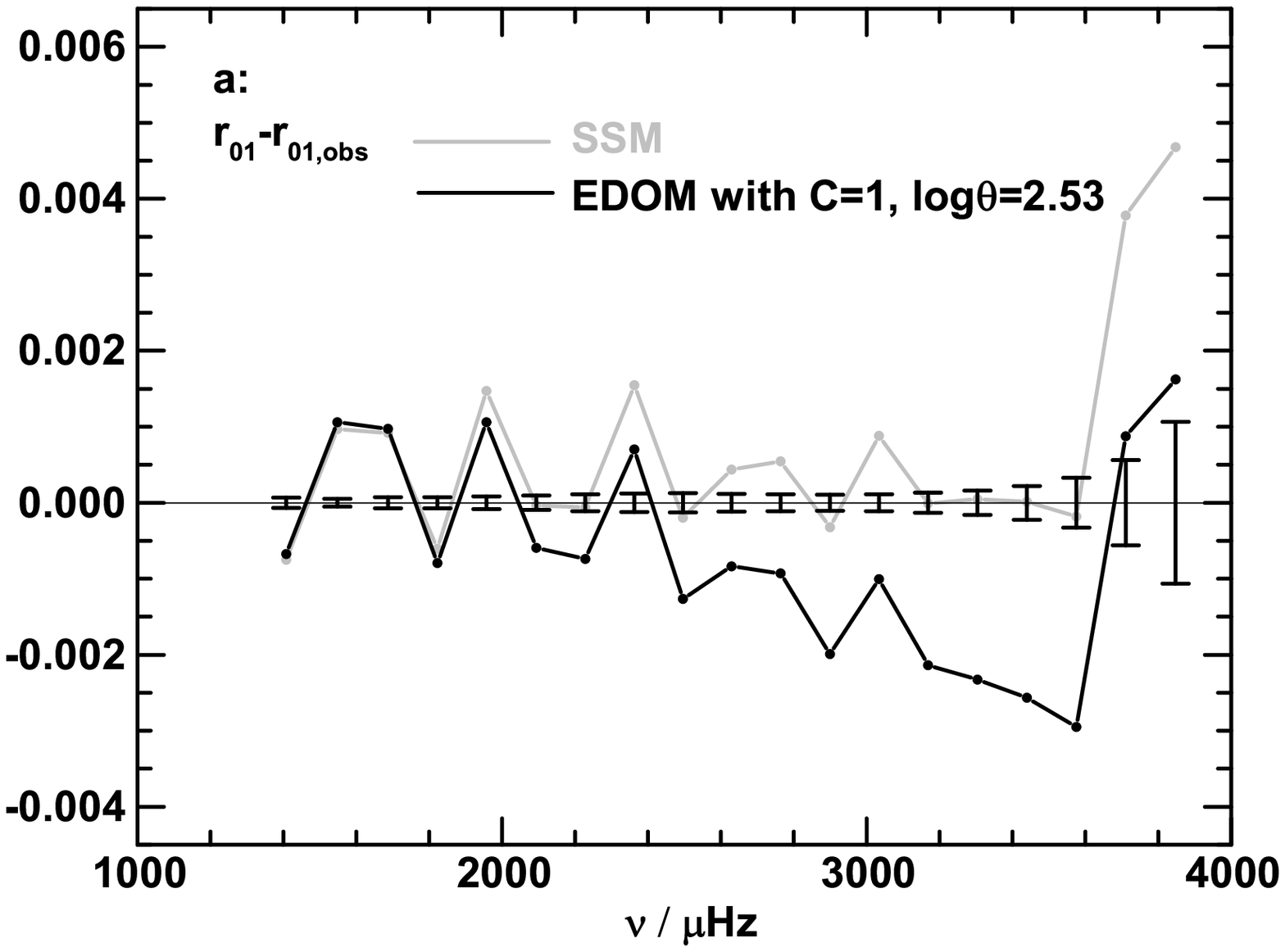}
	\includegraphics[width=0.66\columnwidth]{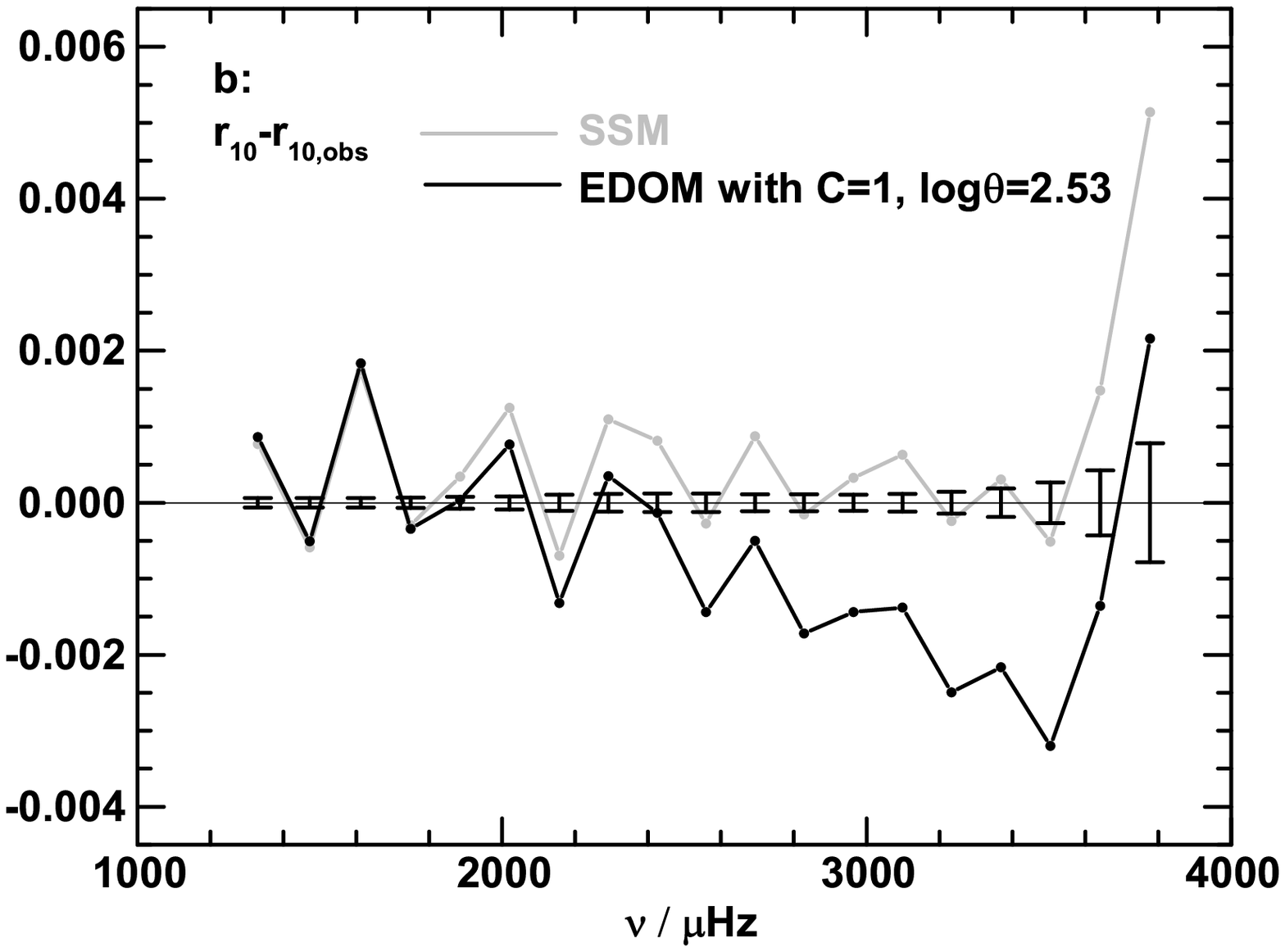}
	\includegraphics[width=0.66\columnwidth]{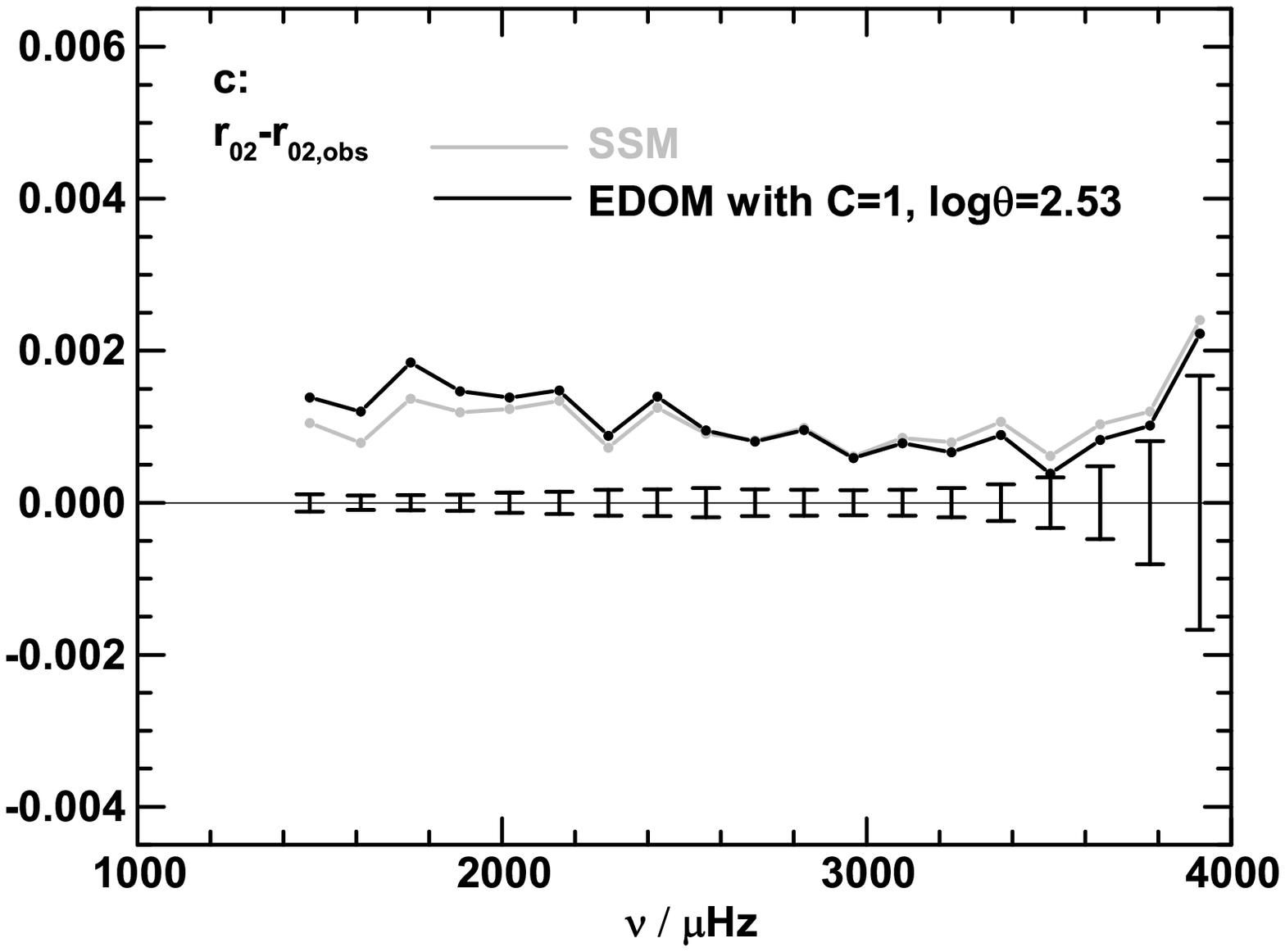}
    \caption{ Deviations of $r_{01}$, $r_{10}$ and $r_{02}$ ratios of the SSM and a typical EDOM solar model with $C=1$ and $\theta=2.53$ from observations. For the SSM and the EDOM solar models $\log \chi^2_{r}$ are about 3.4 and 3.6, respectively. }
    \label{solarratios}
\end{figure*}

\subsection{The ratios of small to large frequency separations} \label{SecComp3}

\begin{figure}
	\includegraphics[width=\columnwidth]{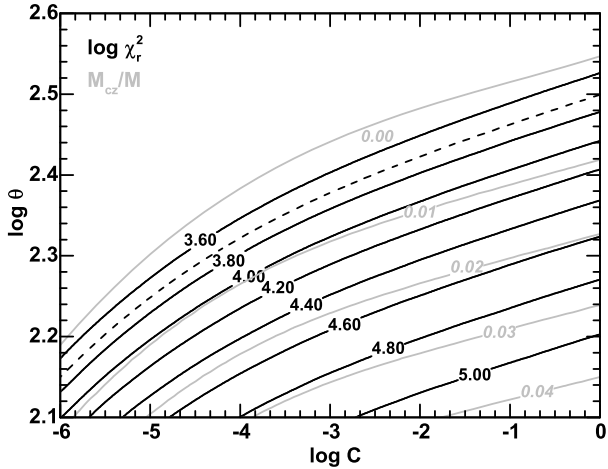}
    \caption{ $\log \chi^2_{r}$ and mass fraction of the convective core of the EDOM solar models with different values of $C$ and $\theta$ are shown by the black and grey contours, respectively. The dashed line is $\log \chi^2_{r} = 3.8$, considered as the lower limit of the parameter $\theta$ by $\chi^2_{r}$. }
    \label{chi2rsun}
\end{figure}

\citet{RV2003} demonstrated that the ratios of small to large separations are sensitive to the properties of the core for solar-like stars. Therefore the ratios could constrain the core overshoot parameters. We have calculated the frequencies of the EDOM solar models in the range of $l=0 - 2$ and $600<\nu/({\mu \rm Hz})<5000$ and then the ratios $r_{01}$, $r_{10}$ and $r_{02}$.
Observations of solar oscillation frequencies for $l=0 - 2$ are taken from \citet{Broomhall2009}.
Deviations of the ratios of the SSM and a typical EDOM solar model with $C=1$ and $\theta=2.53$ are shown in Fig.\,\ref{solarratios}.
The fit of the ratios to the observations is characterized by $\chi^2$,
calculated as
\begin{equation} \label{chi2ij}
{\chi_{ij}}^2 = \sum\limits_{i = 1}^N {{{(\frac{{{r_{ij}} - {r_{ij,obs}}}}{{\delta {r_{ij}}}})}^2}}.
\end{equation}
The uncertainty $\delta r_{ij}$ of the observed $r_{ij}$ is calculated by assuming all the individual frequencies being independent.
\citet{Roxburgh2018} demonstrated that $r_{01}$ and $r_{10}$ are correlated and suggested that only one of those ratios should be combined with $r_{02}$ in comparing the ratios with observations. Here we define the total $\chi^2_{r}$ as:
\begin{equation} \label{chi2tot}
\chi^2_{r} = \chi^2_{010} + \chi^2_{20},
\end{equation}
where
\begin{equation} \label{chi2r010}
\chi^2_{010} = \frac{\chi^2_{01} + \chi^2_{10}}{2}
\end{equation}
to avoid overfitting; meanwhile $r_{01}$ and $r_{10}$ are balanced in the total $\chi^2_{r}$. Results for the EDOM solar models are shown in Fig.\,\ref{chi2rsun}. The minimum in the calculated parameter space is $\log \chi^2_{r} \approx 3.5$. However, the AGSS09Ne SSM gives a smaller $\chi^2_{r}$, $\log \chi^2_{r}\approx 3.4$. Because $\chi^2_{r}$ decreases with increasing $\theta$ and decreasing $C$ as shown in Fig.\,\ref{chi2rsun}, while the SSM is equivalent to the case of $\theta=\infty$ and $C=0$, $\chi^2_{r}$ of the SSM can be regarded as the minimum of $\chi^2_{r}$, i.e., $\chi^2_{r,{\rm min}}=3.4$. As a suitable constraint on the range of the parameters based on the frequency separation ratios we suggest $\chi^2_{r} (C,\theta) \leq 2 \chi^2_{r,{\rm min}}$. Therefore the parameters are constrained in the range with $\log \chi^2_{r} \leq 3.7$, shown as the dashed line in Fig.\,\ref{chi2rsun}, favouring a radiative core.

\citet{Bellinger2016} developed a machine learning code to obtain stellar parameters for given observational data. They investigated the Sun by application of machine learning to the oscillation frequencies, including the ratios of small to large frequency separations, obtaining a small value of $\alpha_{\rm ov}=0.060\pm0.015$. In this case, the Sun should have no convective core, based on Fig.\,\ref{resstd}a. This is consistent with our analysis of the ratios which also favours a radiative core.

\section{Parameter constrained for EDOM of the core overshoot} \label{SecDiffusion}

\begin{figure}
	\includegraphics[width=\columnwidth]{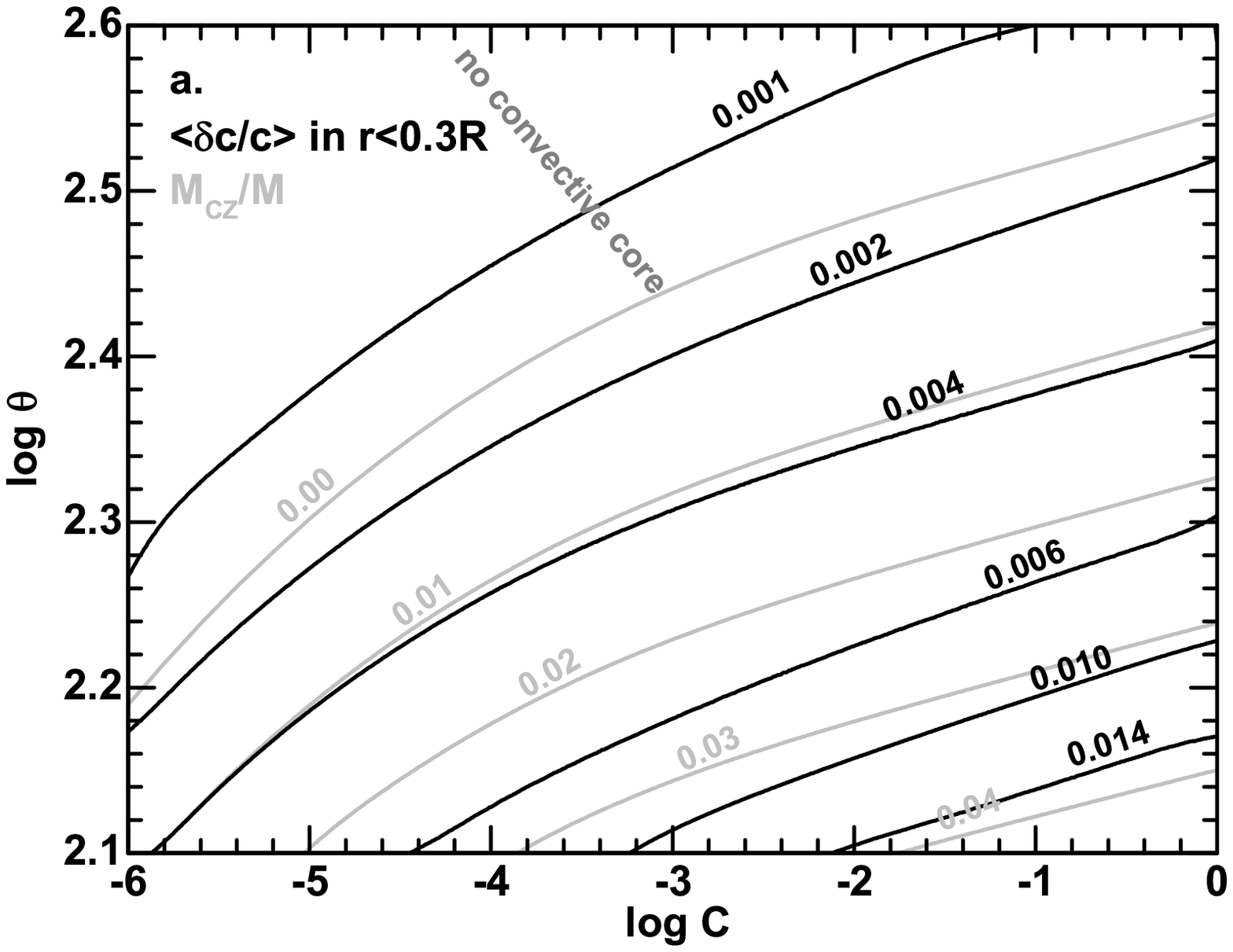}
	\includegraphics[width=\columnwidth]{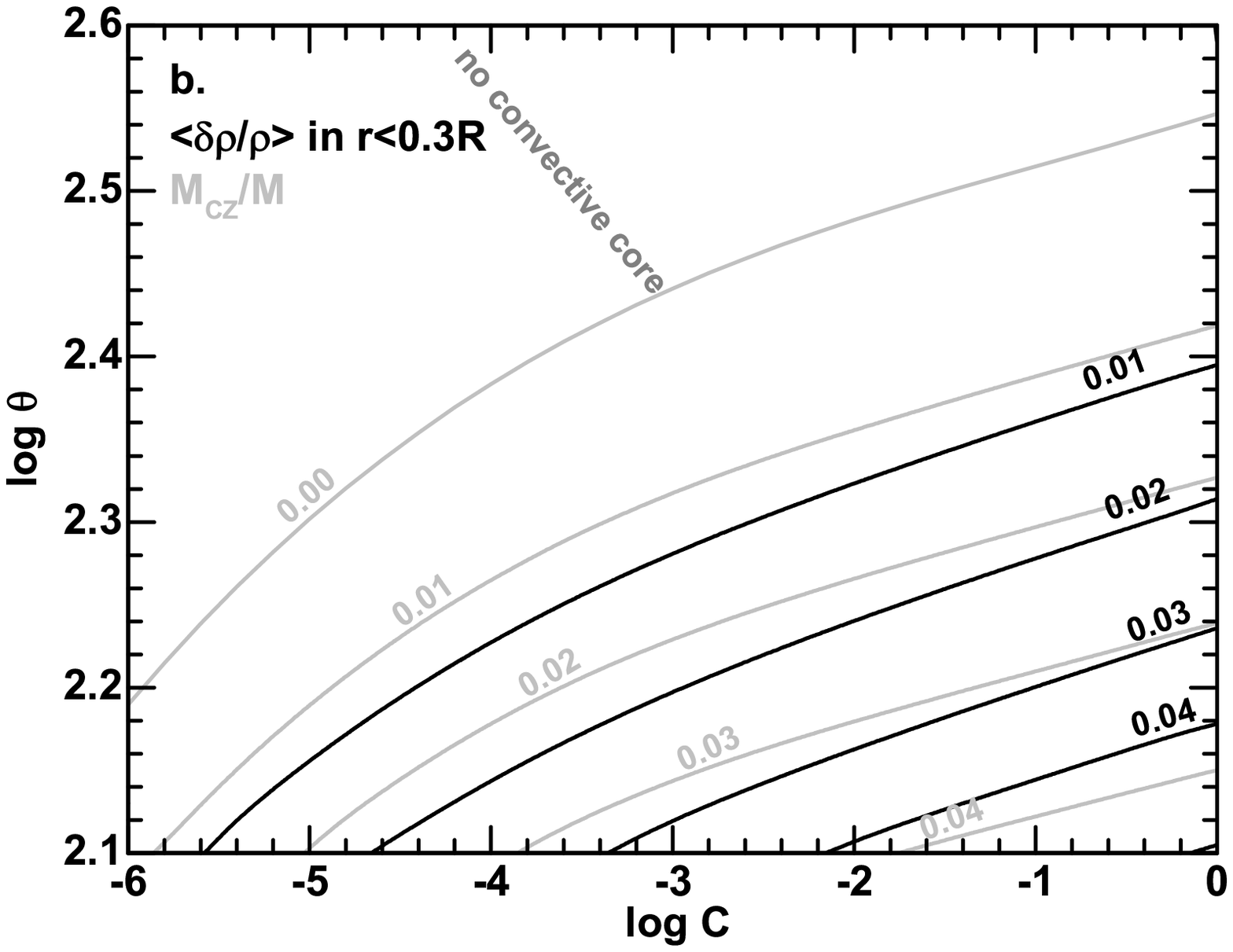}
	\includegraphics[width=\columnwidth]{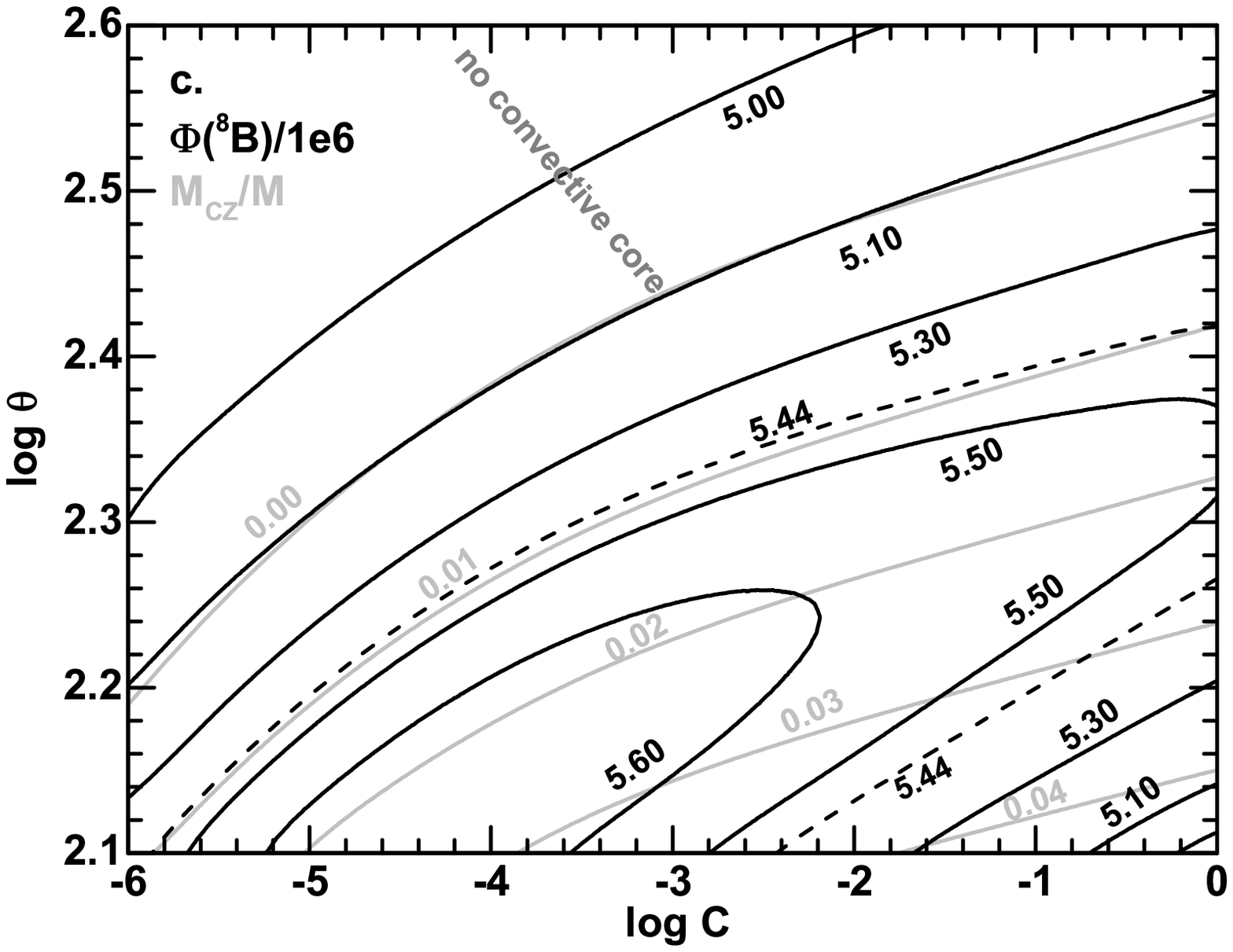}
	\caption{Properties of cores of solar models with diffusion model of overshoot mixing. Grey contours show the mass fraction of the convective core. Black contours show the corresponding values of the title of each subfigure. The black dashed contour in the figure of the $^8$B neutrino flux shows the upper limit of ${\rm{\Phi(^8B)}}$, in units of $10^6\,{\rm cm^{-2} s^{-1}}$, taking into account both observational and theoretical uncertainties. }
    \label{resdif2d}
\end{figure}

The dependence of properties of the solar core ($M_{\rm{cz}}/{\rm M_\odot}$, $\langle \delta c / c \rangle$ and $\langle \delta \rho / \rho \rangle$ in the solar core with $r<0.3\,R$, and ${\rm{\Phi(^8B)}})$ of the solar models on the parameters of EDOM are shown in Fig.~\ref{resdif2d}. It is found that the sound-speed and density deviations in the solar core quickly increase when the mass fraction of the convective core becomes larger. Therefore the helioseismic inferences of sound speed favour a radiative core. Taking into account 2.2\% observational and 5\% theoretical uncertainties, it is found that the reasonable range of ${\rm{\Phi(^8B)}}$ of the solar model is from $4.88 \times 10^6$ to $5.44 \times 10^6 \flun$. The resulting $^8$B neutrino fluxes shown in the figure also favour a radiative core or a tiny convective core with the mass fraction less than 1\% (noting that the region of ${\rm{\Phi(^8B)}}<5.44 \times 10^6 \flun$ in the large $M_{\rm{cz}}/M$ corner is strongly excluded by the helioseismic inferences).

Based on these results, it is reasonable to conclude that the present Sun should have no convective core. To characterize this, we introduce a critical $\theta$ denoted $\theta_{\rm{cr}}$, such that $\theta<\theta_{\rm{cr}}$ will leads to a convective core in the solar model at the present solar age.
From our analysis we find that
\begin{equation} \label{critheta}
\log \theta_{\rm cr} = 2.53 + 0.007 {\log}C - 0.008 {\log ^2}C \pm 0.02,
\end{equation}
which is derived from a quadratic polynomial fitting of the data in Fig.~\ref{resdif2d}. This formula is consistent with the constraint given by the ratios of small to large separations since $\log \chi^2_{r} \leq 3.7$ is satisfied if $\theta > \theta_{\rm{cr}}$.

\begin{figure}
	\includegraphics[width=\columnwidth]{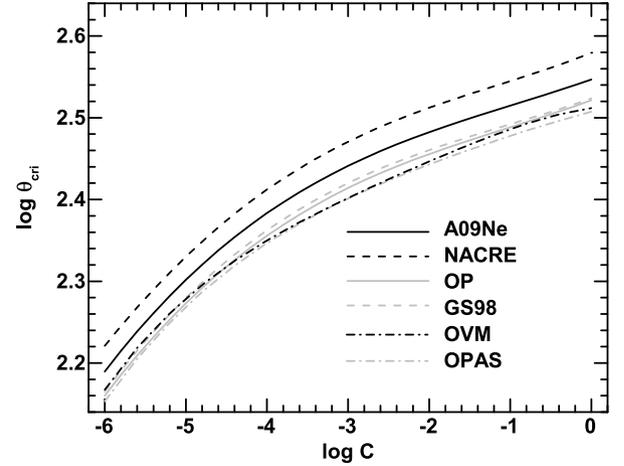}
	\caption{Critical value of $\theta$ with different $C$. A convective core exists in the solar model when $\theta<\theta_{\rm{cr}}$. Four cases of different input physics are compared. The black solid line denoted as 'A09Ne' is the standard case. The black dashed line denoted 'NACRE' is the case where the SFII nuclear reaction rates are replaced by the NACRE rates.
For the grey solid line denoted 'OP' and the grey dashed-dotted line denoted 'OPAS' the OPAL opacity tables are replaced by, respectively, the OP \citep{OP} and the OPAS \citep{OPAS1,OPAS2} tables, while for the grey dashed line denoted 'GS98' the AGSS09 solar composition is replaced by the GS98 composition. Finally, for the black dashed-dotted line denoted 'OVM' convective envelope overshoot \citep{ZLCD2019} is taken into account. }
    \label{crisolar}
\end{figure}

For solar models, especially for the structure of the solar core, there are some crucial input physics, e.g., the solar composition, opacity and nuclear reaction rates, whose current uncertainties could significantly affect the solar structure \citep{serenelli2016,vinyoles2017,buldgen2019c,cd2021}. The uncertainties of composition and opacity lead to global variation of the solar model, as reflected in larger deviations from the helioseismic inferences, known as the solar abundance problem \citep{serenelli2009}. The uncertainties of the nuclear reaction rates could impact the structure of the core. We have tested different input physics to investigate their effects. The critical $\theta-C$ relations in those cases are shown in Fig.~\ref{crisolar}.
In the standard case denoted as A09Ne, the solar composition is the A09Ne composition, the opacity is based on the OPAL tables, and the nuclear reaction rates are from SFII. The difference between the NACRE case and the standard case is that the SFII rates have been replaced by the NACRE rates \citep{NACRE}. The difference between the OP or OPAS case and the standard case is that the OPAL tables have been replaced by the OP \citep{OP} or OPAS \citep{OPAS1,OPAS2} tables to obtain the opacities in solar interior. The difference between the GS98 case and the standard case is that the A09Ne composition has been replaced by the GS98 composition \citep{GS98}. We have also tested the effect of the extra mixing below the base of the convection envelope, which is based on the model of the convective overshoot mixing and turbulent kinetic energy flux \citep{ZLCD2019}, denoted as the OVM case. It is found that the critical $\theta-C$ relation changes little ($\Delta\theta_{\rm{cr}}<0.06$) when those crucial input physics change.

\section{Link to the ``solar spoon''} \label{SecSS}

\citet{Dilke1972} proposed a ``solar spoon'' mechanism that the $\epsilon$-mechanism of H and $^3$He burning on $g-$modes could excite an instability leading to a mixing in the solar core. The mixing reduces the composition gradient and then restrains the instability. The next time of the instability occurs when a sufficient composition gradient is built up. \citet{Dilke1972} suggested that the mixing occurs in the solar core with enclosed mass $0.25 \msun$ every 250\,Myrs \citep[see also][]{cd1974},
the mixing leading to a slight reduction of the solar luminosity ($\sim5\%$) and a significant reduction of the solar neutrino fluxes. The former relates to the Earth's ice ages and the latter relates to the low observed solar neutrino fluxes at a time where the neutrino oscillations had not been confirmed yet.

\begin{figure}
	\includegraphics[width=\columnwidth]{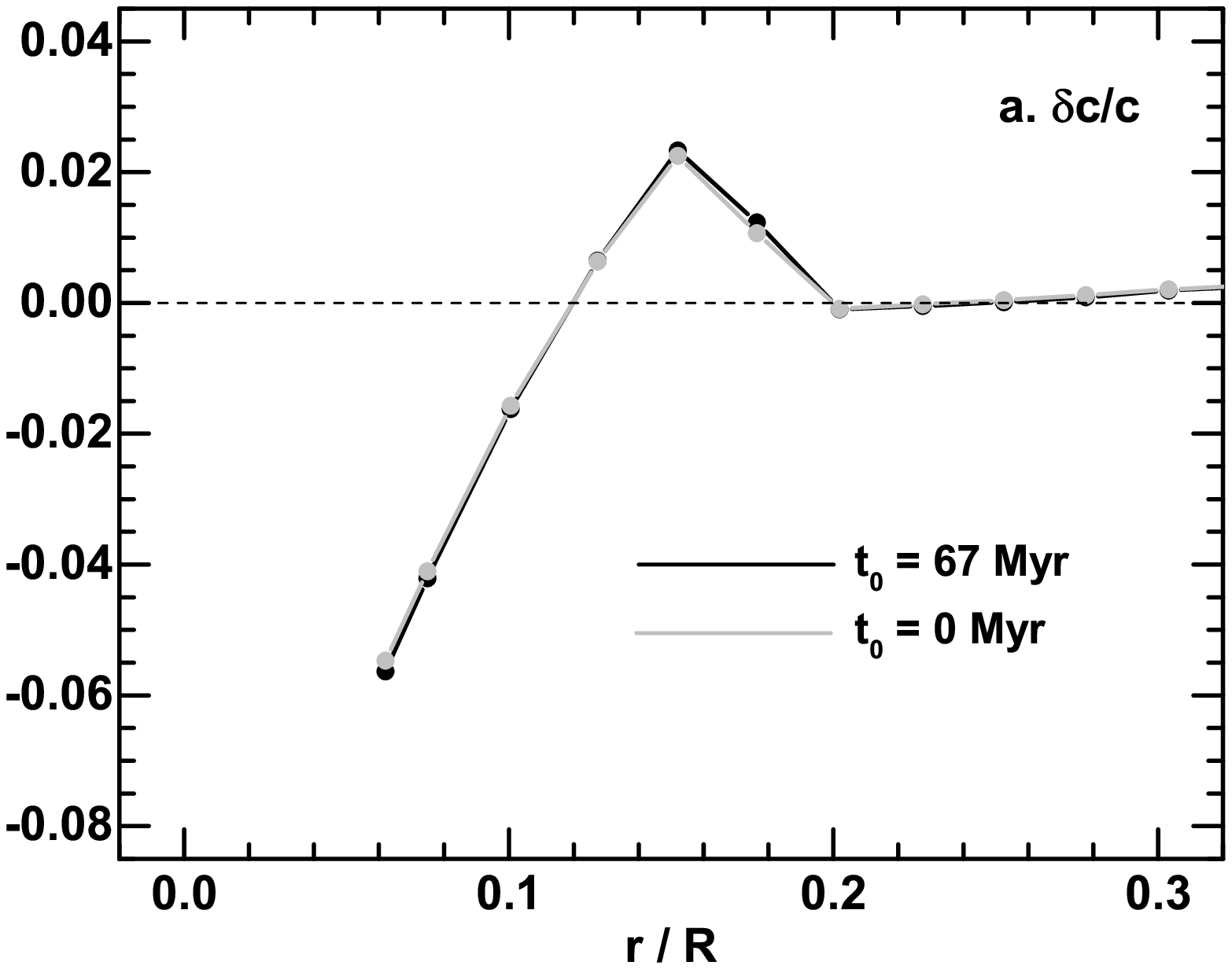}
	\includegraphics[width=\columnwidth]{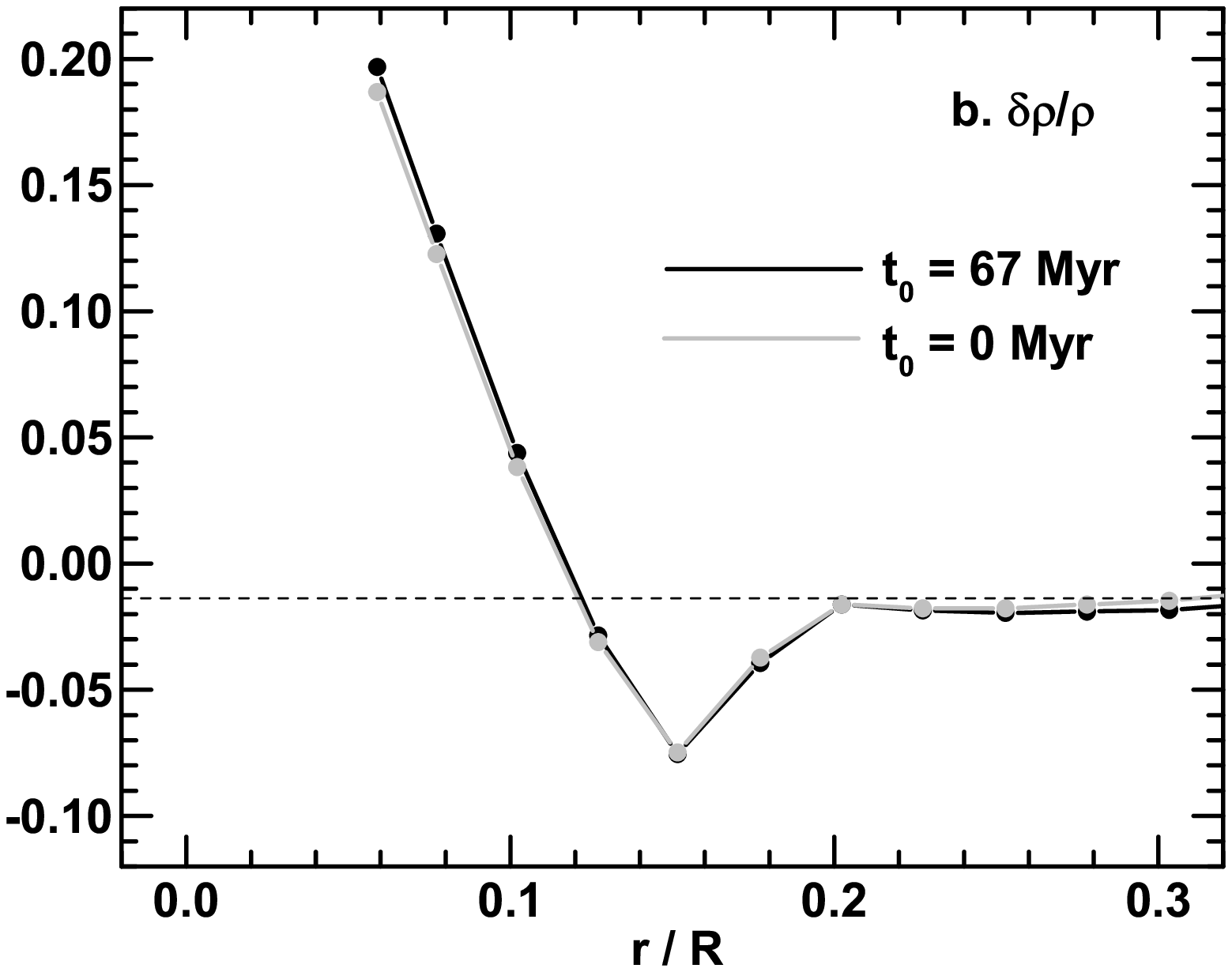}
    \caption{The sound-speed and density deviations of the two solar models with the ``solar spoon'' mixing. The black lines correspond to the case of $t_0=67$Myr and the grey lines to the case of $t_0=0$Myr. }
    \label{spooncrho}
\end{figure}

This mixing leads to a variation of the properties of the solar core that could be tested. We have calculated solar models based on the input physics of the AGSS09Ne standard solar model with an additional instantaneous mixing in the solar core with $M<0.25 \msun$ at the age $t=t_0+250n$\,Myrs where $n=1,2,3,...$ is integer. Two cases are calculated, i.e., $t_0=67$\,Myr and $t_0=0$. In the former case, the last ``solar spoon'' mixing occurs 3\,Myr before the current solar age which is consistent with the start of the quaternary glaciation on Earth. In the latter case, the last ``solar spoon'' mixing occurs 70\,Myr before the current solar age. The solar model with $t_0=67$\,Myr is directly impacted by the mixing and the solar model with $t_0=0$ is in equilibrium. Therefore the two models include the two possible cases that the stellar model is in equilibrium and not in equilibrium. The sound-speed and density deviations of the two solar models are shown in Fig.\,\ref{spooncrho}. The deviations are quite large because of the low composition gradient in the core due to the ``solar spoon'' mixing. The $^7$Be and $^8$B neutrino fluxes of the two solar models are about $3.4 \times 10^9$ and $3.0 \times 10^6 \, {\rm cm^{-2} \, s^{-1}}$, respectively. Both are significantly lower than the observations. Another consequence is to result in a longer main-sequence stage lifetime as 14.4\,Gyr, higher than the age of the universe, raising an issue of the origin of the solar-mass red giants. Because of the above problems, we regard the possibility of the ``solar spoon'' mechanism with a fast mixing in the core with 0.25 solar mass occurring at ~250\,Myr intervals to be unlikely.

\section{Parameter constraints based on other solar-mass stars} \label{SecHD}

\citet{deheuvels2016} (in the following D16) have investigated some solar-mass \textsl{Kepler} target stars and constrained the properties of their cores, i.e., convective or not and the mass fraction of te convective core if it exists. The investigation is based on the analysis of the $r_{010}$ ratios of the small to large frequency separations of $l=0$ and $l=1$ modes \citep{RV2003}, which have been shown to be sensitive to the structure of the stellar core \citep[e.g.,][]{RV2003,provost2005,hd203608}. We use those stars as samples to constrain the overshoot parameters. The following information is taken from D16: the large separations $\Delta\nu$ is taken from Table~1 in D16, the effective temperature $T_{\rm eff}$ is the mean value of \citet{bruntt2012} and \citet{pinsonneault2012}, the stellar masses for stars without convective core are taken from Table~1 in D16, the ratio of the metallicity to hydrogen abundance at the stellar surface $(Z/X)_{\rm{D16}}$ is taken from Tables~1 and 2 in D16. $Z/X$ listed in the last column shows the range of the metallicity to hydrogen abundance at the stellar surface of the satisfactory stellar models calculated in this paper.

\begin{table*}
	\caption{ Information of some solar-mass stars investigated by \citet{deheuvels2016}. $(Z/X)_{\rm D16}$ is the value obtained in that paper, while $Z/X$ is the value required in our calculation to calibrate $T_{\rm eff}$. }\label{KIC_inf}
\begin{tabular}{lccccc}
\hline\noalign{\smallskip}
	KIC ID    &$\Delta\nu$($\mu$Hz)& $T_{\rm eff}$(K) & $M/{\rm M_{\odot}}$   & $(Z/X)_{\rm{D16}}$ & $Z/X$ \\
\hline\noalign{\smallskip}
   5184732    &       95.64        &   5840(60)   &    1.20(1)      &  0.057(1)  &  0.046(1)  \\
   6106415    &      104.20        &   5990(60)   &    1.15(6)      &  0.020(3)  &  0.032(1)  \\
   6116048    &      100.72        &   6000(90)   &    1.07(5)      &  0.014(2)  &  0.022(1)  \\
   6225718    &      106.00        &   6230(60)   &    1.26(3)      &  0.019(1)  &  0.034(1)  \\
   6933899    &       72.26        &   5850(60)   &    1.14(4)      &  0.026(3)  &  0.030(1)  \\
   7206837    &       79.10        &   6350(80)   &    1.44(4)      &  0.035(2)  &  0.043(1)  \\
   7510397    &       62.43        &   6160(80)   &    1.36(4)      &  0.017(1)  &  0.025(2)  \\
   8228742    &       62.29        &   6090(70)   &    1.33(5)      &  0.018(1)  &  0.023(2)  \\
   8394589    &      109.44        &   6180(90)   &    1.18(8)      &  0.011(1)  &  0.026(1)  \\
  10454113    &      105.55        &   6160(70)   &    1.27(2)      &  0.023(2)  &  0.039(1)  \\
  10516096    &       84.43        &   6030(110)  &    1.11(4)      &  0.021(3)  &  0.019(1)  \\
  12009504    &       88.38        &   6170(120)  &    1.20(1)      &  0.021(2)  &  0.024(1)  \\
  12258514    &       74.96        &   5990(60)   &    1.24(2)      &  0.028(2)  &  0.027(2)  \\
\hline
\end{tabular}
\end{table*}

For each star, we have calculated stellar models with different values of $C$ ($\log C$ from $-6$ to 0 with a step of 0.5) and $\theta$ ($\log \theta$ from about 1.5 to 3, a little different for each star, with a step of 0.1) to constrain their range based on the ratios of small to large frequency separations $r_{01}$, $r_{10}$ and $r_{02}$ (see Section~\ref{SecComp3}), while the mixing-length parameter $\alpha=1.8$ is the same as in the solar case, and the initial helium abundance is set to $Y=0.26$. The reference solar composition is based on the GN93 composition \citep{GN93}, following the MESA stellar models in D16. \footnote{\citet{deheuvels2016} have compared the results of GN93 composition with those of AGSS09 composition and found little difference.} The suggested stellar mass in Table~1 is adopted. The metallicity is iteratively adjusted to calibrate the effective temperature $T_{\rm{eff}}$ of the stellar model whose large frequency separation $\Delta\nu$ is consistent with the observations (values of $T_{\rm{eff}}$ and $\Delta\nu$ for each star are listed in Table~1). Oscillation frequencies with $l=0 - 2$ in a range covering the observed frequencies are calculated for the calibrated stellar models whose $T_{\rm{eff}}$ and $\Delta\nu$ are consistent with the values in Table~1. The ratios $r_{01}$, $r_{10}$ and $r_{02}$ and then $\chi^2_{r}$ are calculated. We adopt the parameter space in which $\chi^2_{r} (C,\theta) \leq 2\chi^2_{r,min}$ as the recommended range of the overshoot parameters. For each star, three cases of stellar mass are calculated, i.e., the centre value, the lower limit and the upper limit. For example, for KIC 8394589, we calculated three cases with $M=1.18 \msun$, $M=1.10 \msun$ and $M=1.26 \msun$. The suggested parameter space defined by $\chi^2_{r} (C,\theta) \leq 2\chi^2_{r,min}$ in each case of the stellar mass are combined as the final recommended parameter space of the overshoot mixing for each star.

Molecular diffusion was not considered for the models in Table~1 of D16 or in our calculations. It is well known that the molecular diffusion results in strong depletion of helium and heavier elements in the outer layers, and eventually to a pure hydrogen envelope, in stars of mass higher than about $1.2 \msun$ (depending on metallicity). Because the opacity is sensitive to heavy elements, complete depletion of helium and heavy elements impacts the stellar radius and then $\Delta\nu$. Excessive settling of helium and heavy elements is in contrast to the observation of the surface element abundances \citep[e.g.,][]{VM1999} and the measurement of helium by using its glitch signature in the observed oscillation frequencies \citep{verma2017,verma2019} of A- and F- type stars. Physical processes missing in our calculations, such as radiative levitation \citep[e.g.,][]{TRM1998,dotter2017,deal2018,deal2020}, or extra dynamical mixing \citep{verma2017,verma2019,deal2020} are required to compensate for the quick downward settling of helium and heavy elements in the stellar envelope. \citet{deal2020} showed that the combination of molecular diffusion, radiative levitation and rotational mixing can produce reasonable results of surface abundance and stellar parameters.

For each star, the ratio $Z/X$ of the stellar models satisfying the condition of the convective core suggested by D16 for models with different $\log C$ and $\log \theta$ are listed in the last column in Table~1. The resulting $Z/X$ are generally largely consistent with the suggested value by D16 except for some stars showing significant deviations: KIC8394589, KIC6225718, and KIC10454113. The possible reason for the deviation of metallicity is the difference in the calculations between D16 and this paper. D16 assessed the best-fitting models by using the minimum of $\chi^2$, with the differences of $T_{\rm{eff}}$ between models and observations contributing to the total $\chi^2$. In this paper, we have calibrated $T_{\rm{eff}}$ for the stellar model with given $\Delta\nu$. This calibration of $T_{\rm{eff}}$ corresponds to applying a very small uncertainty (much less than the uncertainty suggested by the observations) in $T_{\rm{eff}}$ in assessing the best-fitting model by using the minimum of $\chi^2$. This adds extra weight to $T_{\rm{eff}}$ and could reduce and change the range of metallicity.

\begin{figure}
	\includegraphics[width=\columnwidth]{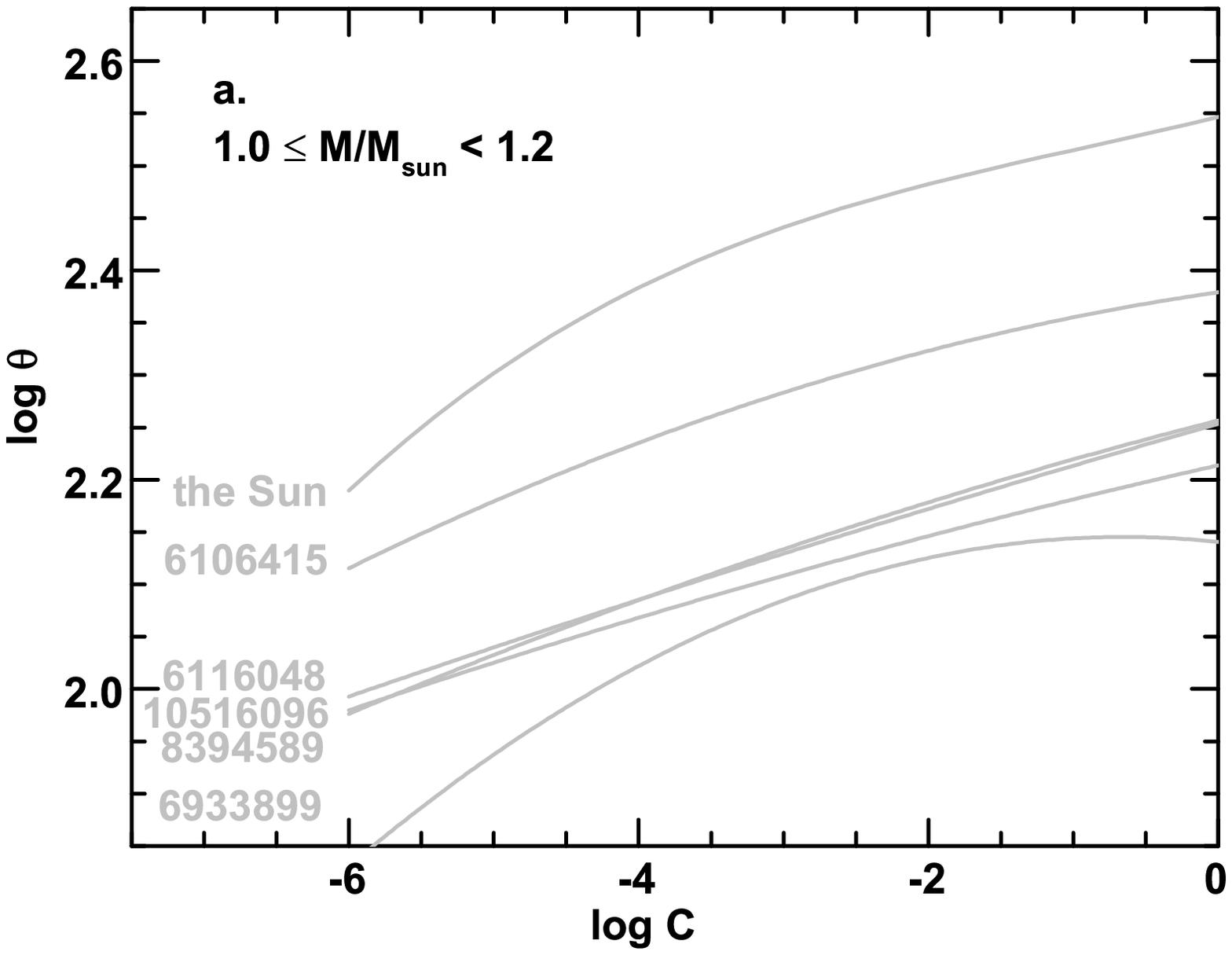}
	\includegraphics[width=\columnwidth]{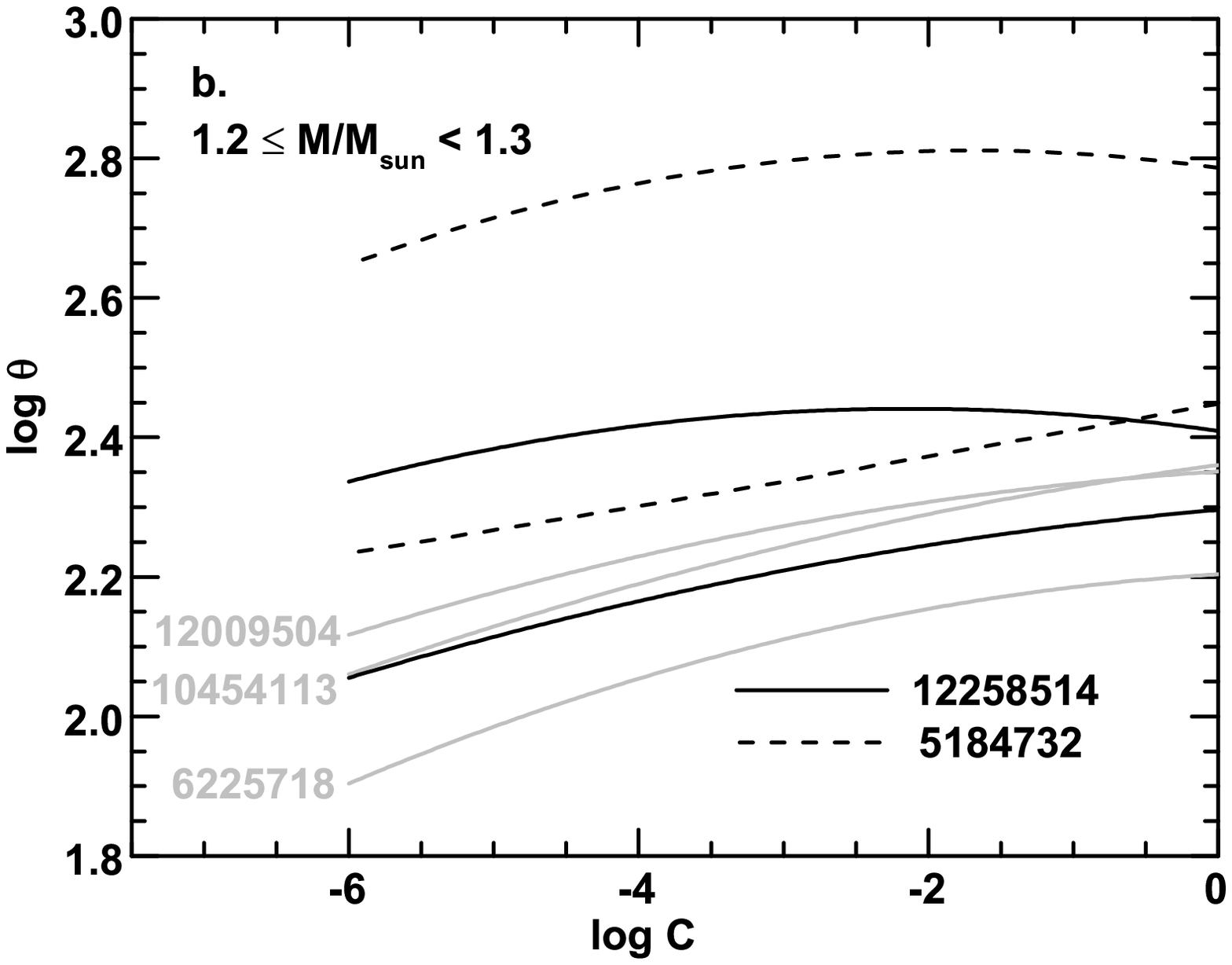}
	\includegraphics[width=\columnwidth]{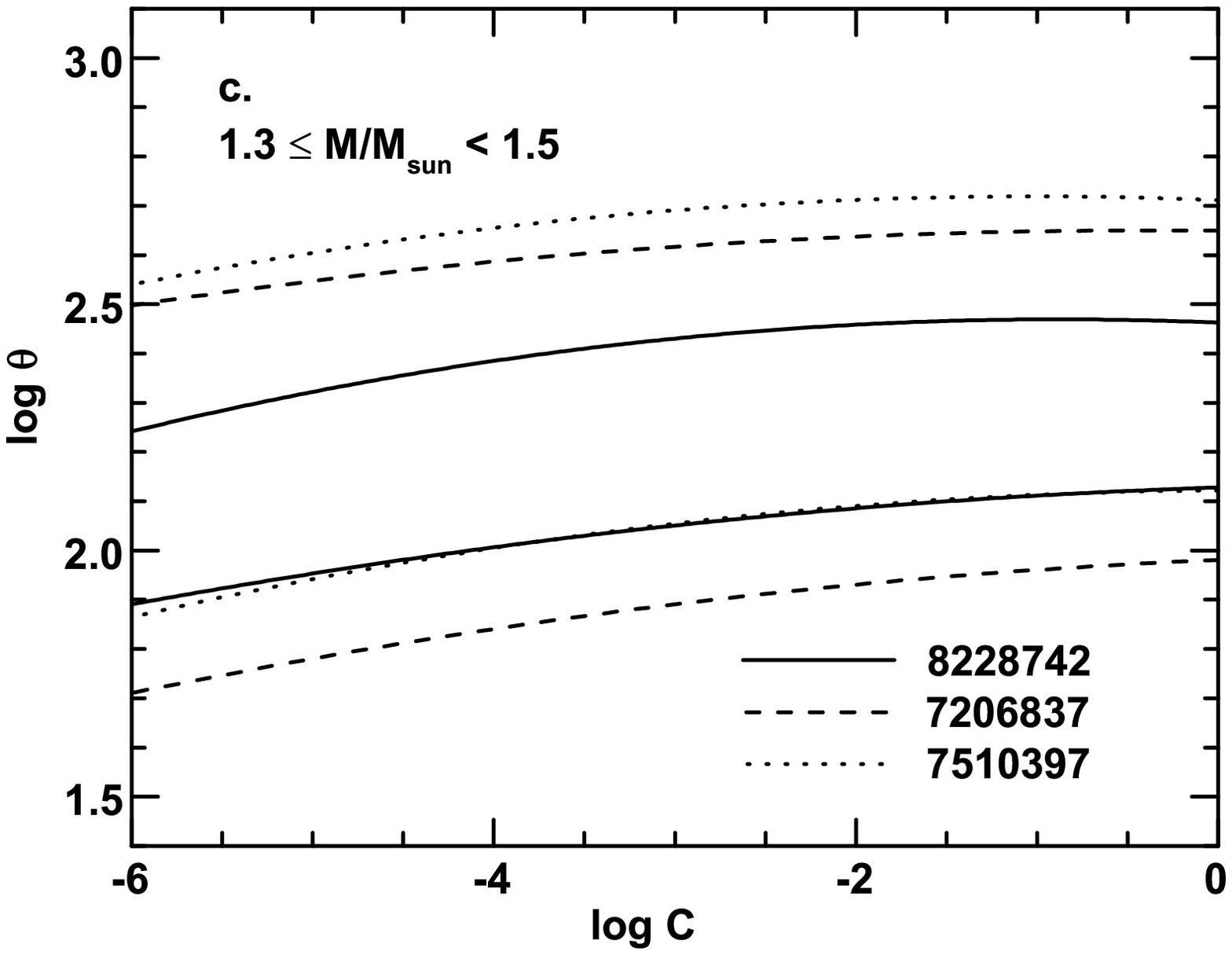}
    \caption{ Parameter constraints based on the FWHM of $\chi^2_{r}$ with fitting by quadratic functions. The fitting errors are typically less than 0.05. The grey lines show the lower limits of $\theta$ provided by the star with the KIC ID denoted to the left of the line. The black lines (solid, dashed and dotted) are shown in pairs, present the recommended parameter range provided by the star with the KIC ID denoted in the key. }
    \label{KICtheta}
\end{figure}

The lower limits of $\theta$ for the stars with mass less than $1.20 \msun$, i.e., KIC8394589, KIC6106415, KIC6116048, KIC10516096 and KIC6933899 are shown in Fig.\,\ref{KICtheta}a. The parameter constraints provided by the stars with mass 1.20$\leq M/{\rm M_\odot} <$1.30, i.e., KIC6225718, KIC5184732, KIC10454113, KIC12009504 and KIC12258514 are shown in Fig.\,\ref{KICtheta}b. KIC6225718, KIC10454113, KIC12009504 only provide lower limits of $\theta$. KIC5184732/KIC12258514 constrain $\theta$ in a range between the two dashed/solid lines. For the three stars with mass higher than $1.30 \msun$, i.e., KIC7206837, KIC7510397 and KIC8228742, corresponding parameter ranges are shown in Fig.\,\ref{KICtheta}c. No constraint for the parameter $C$ has been found in those stars.

\begin{figure}
	\includegraphics[width=\columnwidth]{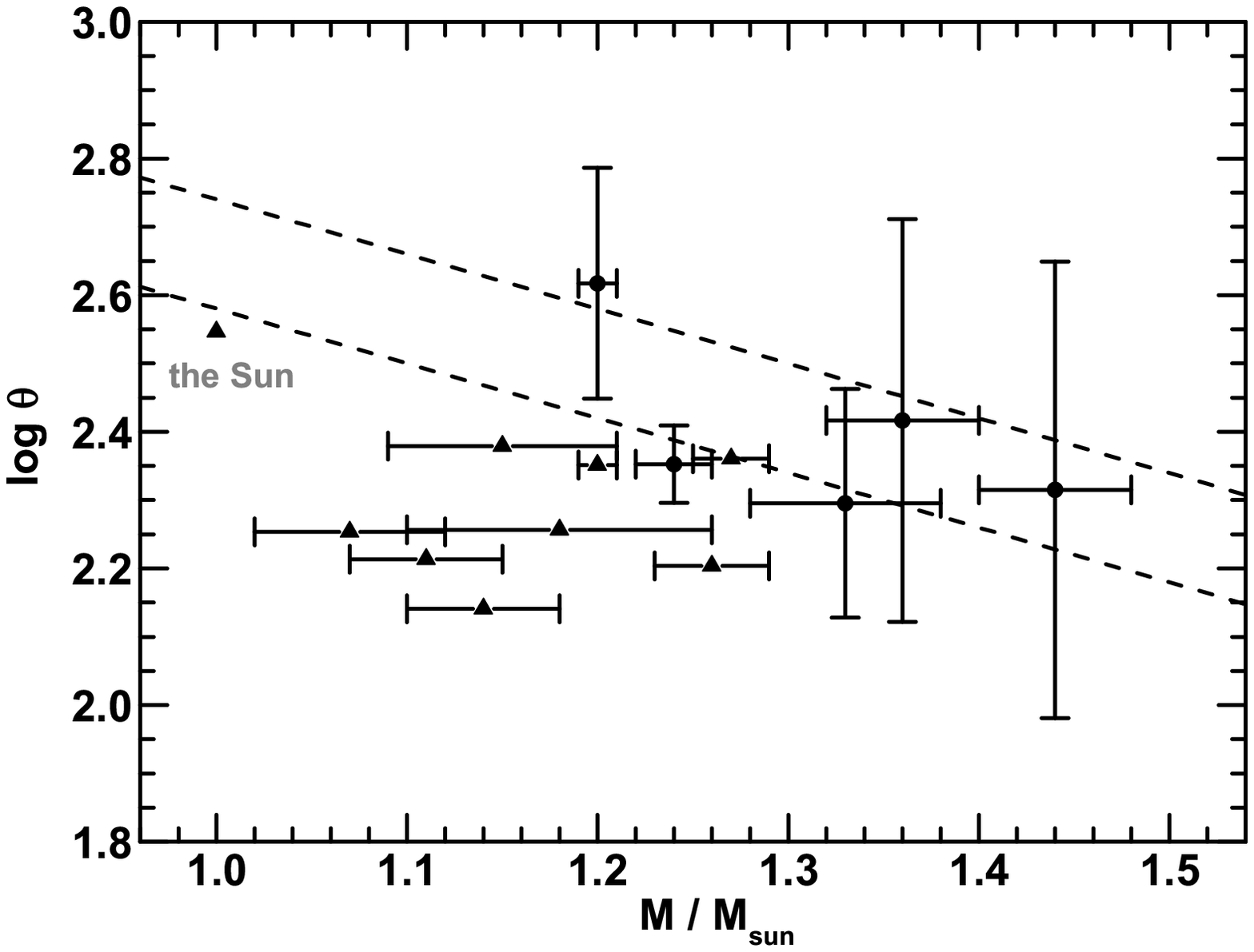}
	\includegraphics[width=\columnwidth]{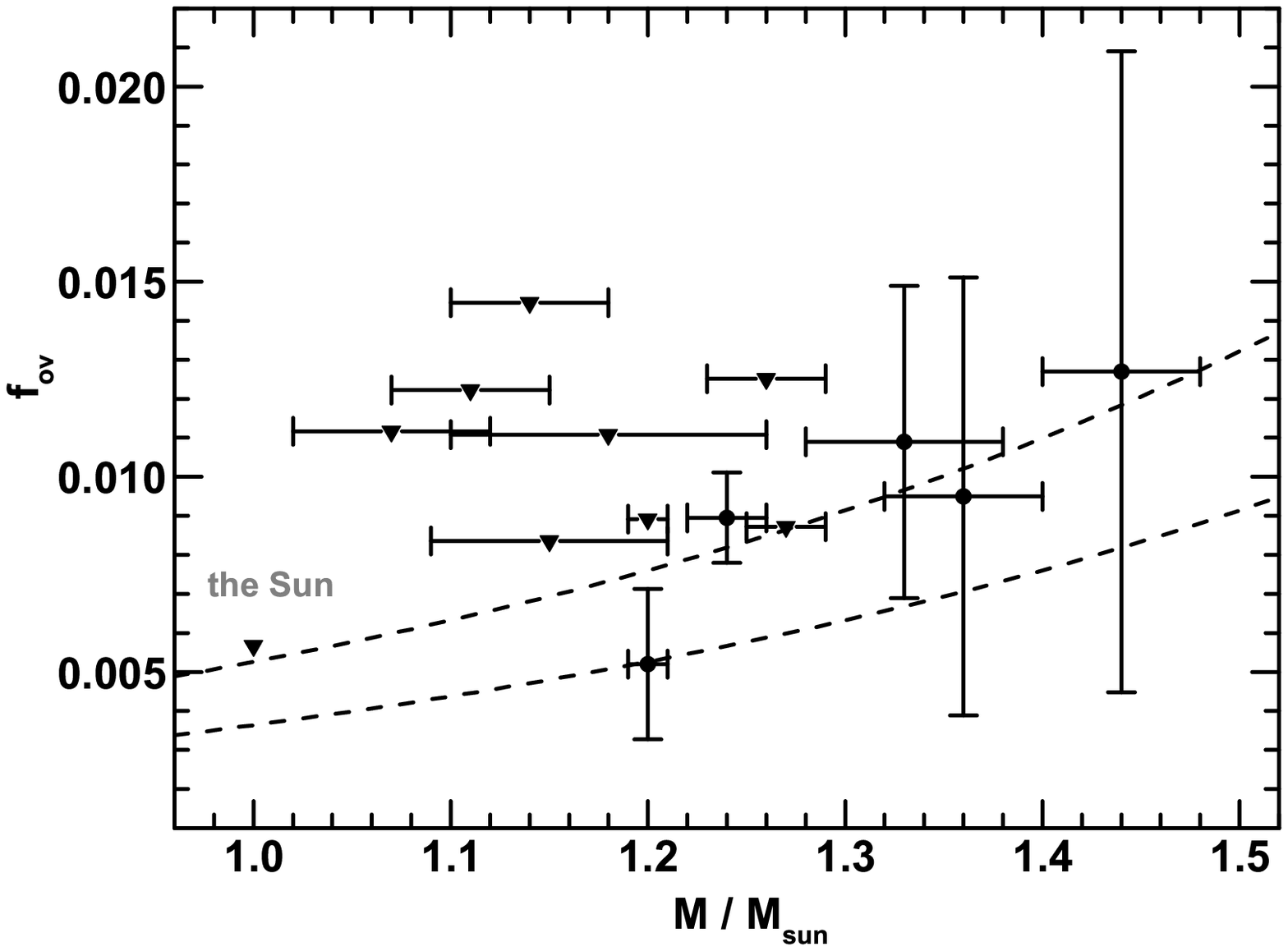}
	\caption{Relation between $\theta$ (top) and $f_{\rm{ov}}$ (bottom) and stellar mass in the case of $C=1$. The arrows mean the lower limit for $\theta$ and the upper limit for $f_{\rm{ov}}$. The dashed lines present the recommend parameter ranges. }
    \label{KIC_sum}
\end{figure}

A special case is $C=1$, when the overshoot model equation~(\ref{OVMmodel}) reduces to the widely used \citeauthor{herwig2000}'s \citeyearpar{herwig2000} model, equation~(\ref{herwigovm}), with $\theta=2/f_{\rm ov}$. The parameter constraints provided by all the 13 KIC stars and also the Sun in the case of $C=1$ are shown in Fig.\,\ref{KIC_sum}a. The strongest constraints are from the Sun, KIC10454113 and KIC12258514. The dashed lines present the recommended parameter range based on the linear fitting with ensuring all lower limits being satisfied:
\begin{equation} \label{crithetaKIC13}
\log \theta = 3.46 - 0.80 \frac{M}{\rm M_\odot} \pm 0.08.
\end{equation}
The results and the recommended parameter range of $f_{\rm ov}$ for \citeauthor{herwig2000}'s \citeyearpar{herwig2000} model are shown in Fig.\,\ref{KIC_sum}b. It should be pointed out here that there is a little difference between the $f_{\rm ov}$ in \citeauthor{herwig2000}'s \citeyearpar{herwig2000} model and the value of $2/\theta$ since the constant $H_P$ in that model has been replaced by local $H_P$ as mentioned above. The local $H_P$ decreases with $r$ near the boundary of the convective core. Using $2/\theta$ could underestimate $f_{\rm ov}$. However, the difference should be tiny because the diffusion coefficient decreases with $r$ much faster than $H_P$ does and and this quantity varies little in the efficient mixing region. It is shown that $f_{\rm ov}$ increases with stellar mass, qualitatively similar to the results of the calibration of eclipsing binaries \citep[e.g.,][]{claret2017,claret2018,claret2019}. Our results show a slightly higher value of $f_{\rm ov}$ than they did. Their results indicated no overshoot for $M < 1.2 \msun$. Our results of KIC 5184732 and KIC 1258514 indicate the existence of overshoot mixing for $1.2 \msun$ stars. It should emphasized here that the above investigation is not a full asteroseismic investigation because the stellar mass, the large separation and the effective temperature are fixed for the stellar models whose small to large separation ratios have been assessed. We shall carry out a full asteroseismic investigation with free stellar parameters in the near future.

\begin{figure}
	\includegraphics[width=\columnwidth]{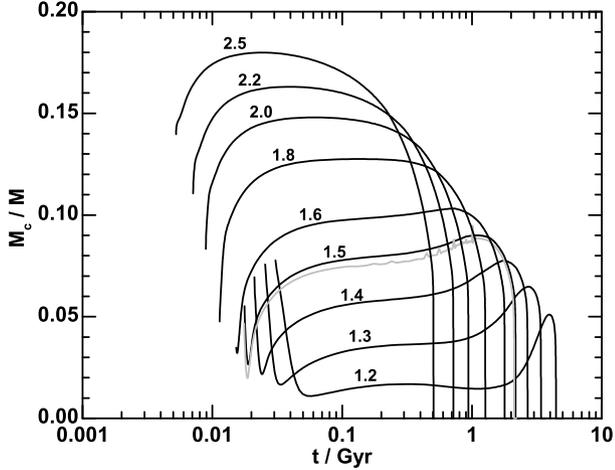}
	\caption{Evolution of the mass fraction of the convective core for stellar mass from $1.2-2.5 \msun$. The black lines are for the stellar models with the EDOM with $\log C=-6$ and $\log \theta=2.8$. The numbers denote the stellar mass. The grey line is the case of a $1.5 \msun$ star without overshoot mixing, showing significant instability of the size of the convective core due to the semi-convection. The abundances for all models are $X=0.7$ and $Z=0.02$ with the GN93 solar composition of heavy elements, as for \citet{SilvaAguirre2020}. The stellar age includes the PMS stage. Only the main sequence stage (defined by $X_0-X_c>0.01$ in the YNEV code) is shown. }
    \label{mcovm}
\end{figure}

For the calculations of the stellar evolution of the stars with mass from about $1.2-2.5 \msun$, it is well known that the size of the convective core is sensitive to the details of the numerical calculations since the semi-convection leads to an instability of the location of the boundary of the convective core \citep[see, e.g.,][]{SilvaAguirre2020}. In the YNEV code, we always set dense mesh points (by a factor of $\sim$10) near all convective boundaries to alleviate the uncertainty of the size of convection zones caused by meshing. On the other hand, it has been found that an strong enough overshoot mixing could remove the semi-convection \citep[see, e.g.,][]{xiong86,MengZhang2014}. We have tested the weakest case of the EDOM in our parameter space, i.e., $\log C=-6$ and $\log \theta=2.8$. The mass fractions of the convective core for stellar mass between $1.2-2.5 \msun$ as functions of stellar age are shown in Fig.\,\ref{mcovm}. It is found that the mass fractions of the convective core are smooth and the instability of the location of the boundary of the convective core is removed. The semi-convection problem for the low mass stars is eliminated by taking into account core overshooting mixing. Therefore our results are not affected by semi-convection.

\section{Discussion and Conclusions} \label{SecConclusion}

The convective overshoot mixing is a significant uncertainty in stellar physics. Classical overshoot modelling (COM) is implemented by assuming full mixing in a region whose extent is parametrized by $\alpha_{\rm ov}$, in units of the pressure scale height $H_P$ or the extent of the convective core. However, numerical simulations and stellar turbulent convection models have shown that the overshoot mixing should be described as a diffusion process with an exponentially decreasing diffusion coefficient.

In this paper, we consider both the classical and the exponential-diffusion models. The diffusion overshoot mixing is modelled with a diffusion coefficient with a general exponential behaviour, i.e., equation~(\ref{OVMmodel}), including two parameters $\theta$ and $C$. The exponent $\theta$ represents the e-folding length of the diffusion coefficient and the parameter $C$ represents the possible decrease of the diffusion coefficient near the convective boundary due to variation of the characteristic length. The formula recovers to the widely used \citeauthor{herwig2000}'s \citeyearpar{herwig2000} model when the parameters are set as $C=1$ and $\theta=2/f_{{\rm{ov}}}$. In order to investigate the effects of the exponential diffusion overshoot model and the range of the parameters for low-mass stars, we have investigated the properties of the core in solar models with convective core overshoot mixing.

For ZAMS solar-mass stars, there is a convective core because of the high temperature sensitivity of the $^{12}$C proton capture reaction and $^3$He fusion. The convective core vanishes in a short time ($\sim0.1$ Gyr) when the primordial $^{12}$C and $^3$He are depleted. It is found that the convective core overshoot mixing prolongs the lifetime of the convection in the core because of a cyclic mechanism of $^3$He driven by overshoot mixing and nuclear reactions, i.e., overshoot mixing transports $^3$He from the overshoot region into the core, $^3$He is consumed in the core, and $^3$He is reproduced in the overshoot region. The temperature sensitivity of the energy release from out-of-equilibrium $^3$He burning is sufficient to drive convection in the core if the overshoot mixing is sufficiently efficient to keep a high $^3$He abundance in the core.

If the core overshoot mixing maintains a convective core in the solar models until the present solar age, the sound-speed and density profiles in the solar core are not consistent with the helioseismic inferences because the mixing enlarges the hydrogen abundance leading to a lower density and a higher sound speed and also the gradients of sound speed and density are determined by the temperature gradient and $\mu$ gradient which are affected by the existence of convection as shown by equations~(\ref{eqdcdr})-(\ref{eqdrhodr}). The presence of a convective core will lead to a strong signal in the gradients of sound speed and density in the core, resulting in significant deviations in sound speed and density as shown in Fig.~\ref{dcdrhostd}. The solar $^8$B neutrino flux also favours a radiative core because a convective core leads to high core temperature, thus resulting in a too high $^8$B neutrino flux. Those results provide a constraint on the parameters of the general diffusion model of the core overshoot mixing for solar-mass stars, e.g., equation~(\ref{critheta}). Analysis of the small to large frequency separation ratios leads to a similar upper limit on the strength of the overshoot mixing. If the classic overshoot model is adopted, the length of the overshoot region should be less than $0.25\,\min(H_P, r_{\rm cz})$. We have investigated the consequences of some uncertainties in the input physics for solar models, e.g., replacing the composition, opacity tables, or nuclear reaction rates, and have found that their effects are small on our conclusions.

We analyzed the overshoot mixing process of elements involved in the pp chains. The exponential diffusion overshoot model leads to different effective overshoot mixing lengths for elements with different nuclear equilibrium timescale. The effective duration of mixing of an elements is constrained by its nuclear equilibrium timescale. An element with a shorter nuclear equilibrium timescale requires a higher diffusion coefficient to show a given strength of mixing. Therefore the exponential diffusion overshoot model predicts a shorter effective overshoot mixing length for that element. This is a significant difference from the classical overshoot model which predicts an uniform overshoot length for all elements.

The exponential overshoot model, equation~(\ref{OVMmodel}), was applied to some \textsl{Kepler} solar-mass stars investigated by \citet{deheuvels2016} who constrained the core status (convective or not, mass fraction of the convective core if convective) of those stars. Based on the frequencies and stellar parameters of those stars, the overshoot parameter $\theta$ can be further constrained by using a least squares deviation of the $r_{010}$ and $r_{02}$ ratios. A recommended parameter range of $\theta$ is obtained for stars with $1<M/{\rm M_\odot}<1.5$. A tendency of decreasing $\theta$ with increasing stellar mass is revealed. On the other hand, the overshoot parameter $C$ cannot be constrained in this investigation; $\theta$ is the most influential parameter for the strength of the overshoot mixing because it determines the e-folding length of the overshoot region. Therefore the effect of the variation of $C$ is relatively weaker than that of $\theta$. It requires more samples and a more accurate investigation to probe the value of $C$.

The resulting tendency of decreasing $\theta$ with increasing stellar mass for low-mass stars is not surprising. It is equivalent to the well-known result that $f_{\rm ov}$ or $\alpha_{\rm ov}$ should increase with the increasing stellar mass \citep[e.g.,][]{WooDemarque2001,Demarque2004,Pietrinferni2004,VandenBerg2006,claret2007,claret2016,claret2017,claret2018,claret2019,bressan2012,deheuvels2016,Hidalgo2018}.
On the other hand, as the stellar mass decreases the size $r_{\rm{cz}}$ of the convective core becomes smaller and $H_P$ at the convective boundary becomes larger; thus if the characteristic length of the overshoot mixing is measured in $H_P$ a constant $f_{\rm ov}$ or $\alpha_{\rm ov}$ leads to a rather strong overshoot mixing for the low-mass stars. In this case, rather than $H_P$, the radius of the convective core $r_{\rm{cz}}$ could be a better unit to measure the characteristic length in overshoot region. This is because the size of the core constrains the characteristic length in the convective core which is correlated with that in the overshoot region. For intermediate mass stars with $2.5<M/{\rm M_{\odot}}<5$, the characteristic length in the convective core is smaller than the size of the core so that it is insensitive to the size of the core and then a constant $f_{\rm ov}$ or $\alpha_{\rm ov}$ is expected \citep[see, e.g.,][]{claret2017,claret2018,claret2019}. However, for massive stars with $M/{\rm M_{\odot}}>8$, the width of the main sequence indicates a mass-dependent overshoot again \citep{castro2014,scott2021}. The possible reason is that the strength of the convective boundary mixing caused by the turbulent entrainment depends on conditions and varies with stellar mass \citep{scott2021}. In this paper, extra mixing including the turbulent entrainment and rotational mixing are not taken into account. If they are present, the constraints of the strength of the overshoot mixing should be regarded as the constraints of total strength of all kinds of mixing in the mixing layer above the convective core.

For both overshoot models in this paper, we assumed that the temperature gradient is purely radiative in the overshoot region. However, it should be noticed that the convective entropy flux and kinetic energy flux change the temperature gradient in the overshoot region. The convective entropy flux should increase the temperature gradient a little and make it closer to the adiabatic temperature gradient \citep[e.g.,][]{xiong02,zqsly12a,ZDXCD2012,baraffe2022}. The kinetic energy flux should decrease the temperature gradient and reduce the size of the convective core \citep{zqs14}. However, the kinetic energy flux in the core overshoot region should be low due to the low turbulent velocity in the core caused by the high density. Therefore taking into those fluxes should not significantly change the results.

\section*{Acknowledgements}

We thank the anonymous referee for providing valuable comments improving the original version.
Fruitful discussions with Dr. Yu-Feng Li, Professor Douglas Gough and Dr. Tao Wu are highly appreciated. Funding for Yunnan Observatories is co-sponsored by the Strategic Priority Research Program of the Chinese Academy of Sciences (grant No. XDB 41000000), the National Natural Science Foundation of China (grant No. 11773064 \& 12133011), the foundation of Chinese Academy of Sciences (Light of West China Program and Youth Innovation Promotion Association), and the Yunnan Ten Thousand Talents Plan Young \& Elite Talents Project. Funding for the Stellar Astrophysics Centre is provided by The Danish National Research Foundation (Grant DNRF106).

\section*{Data Availability}

The data underlying this article will be shared on reasonable request to the corresponding author.












\appendix

\section{Abundance profiles near the convective core boundary} \label{AppendixSecComp1}

\begin{figure*}
	\includegraphics[width=0.66\columnwidth]{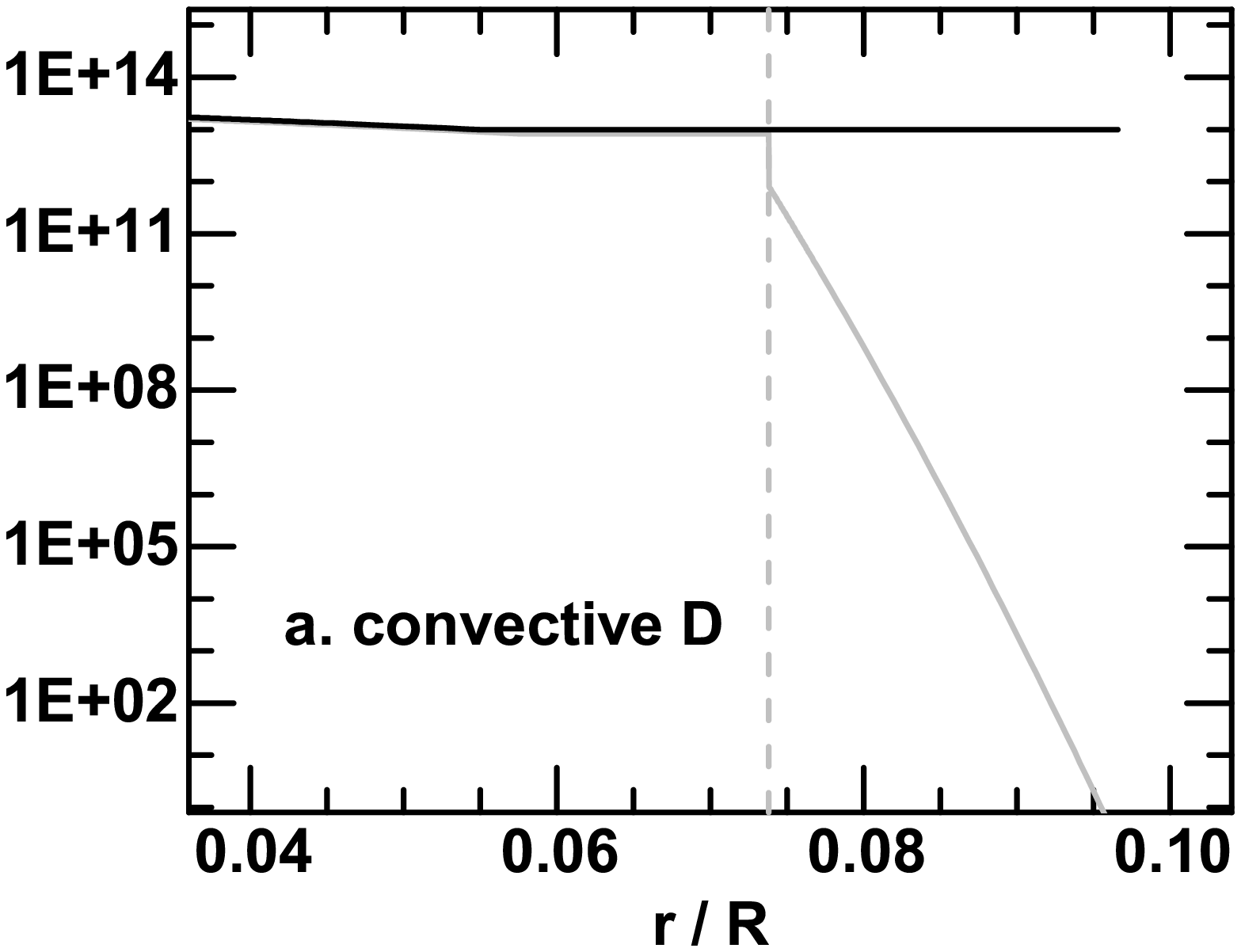}
	\includegraphics[width=0.66\columnwidth]{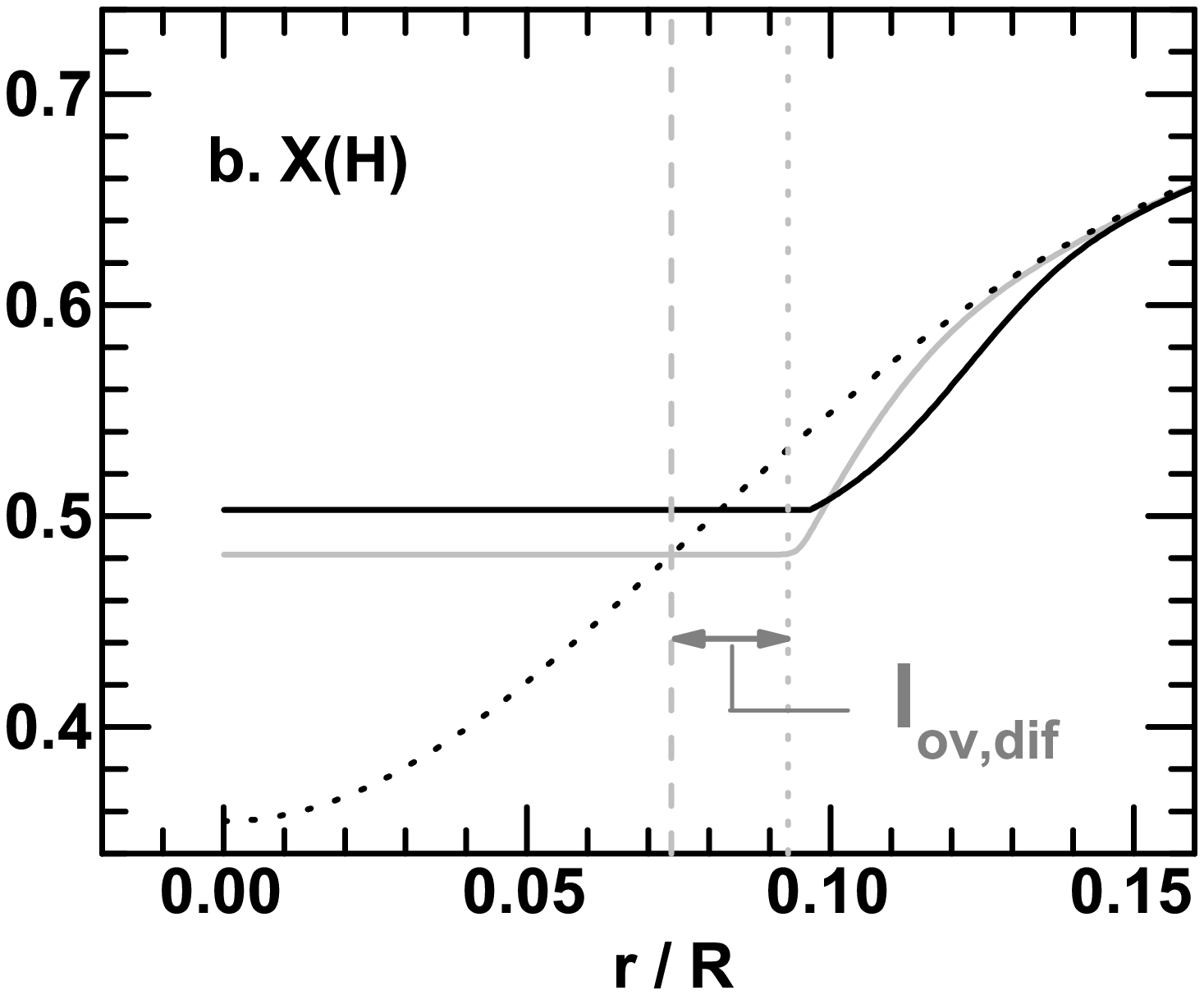}
	\includegraphics[width=0.66\columnwidth]{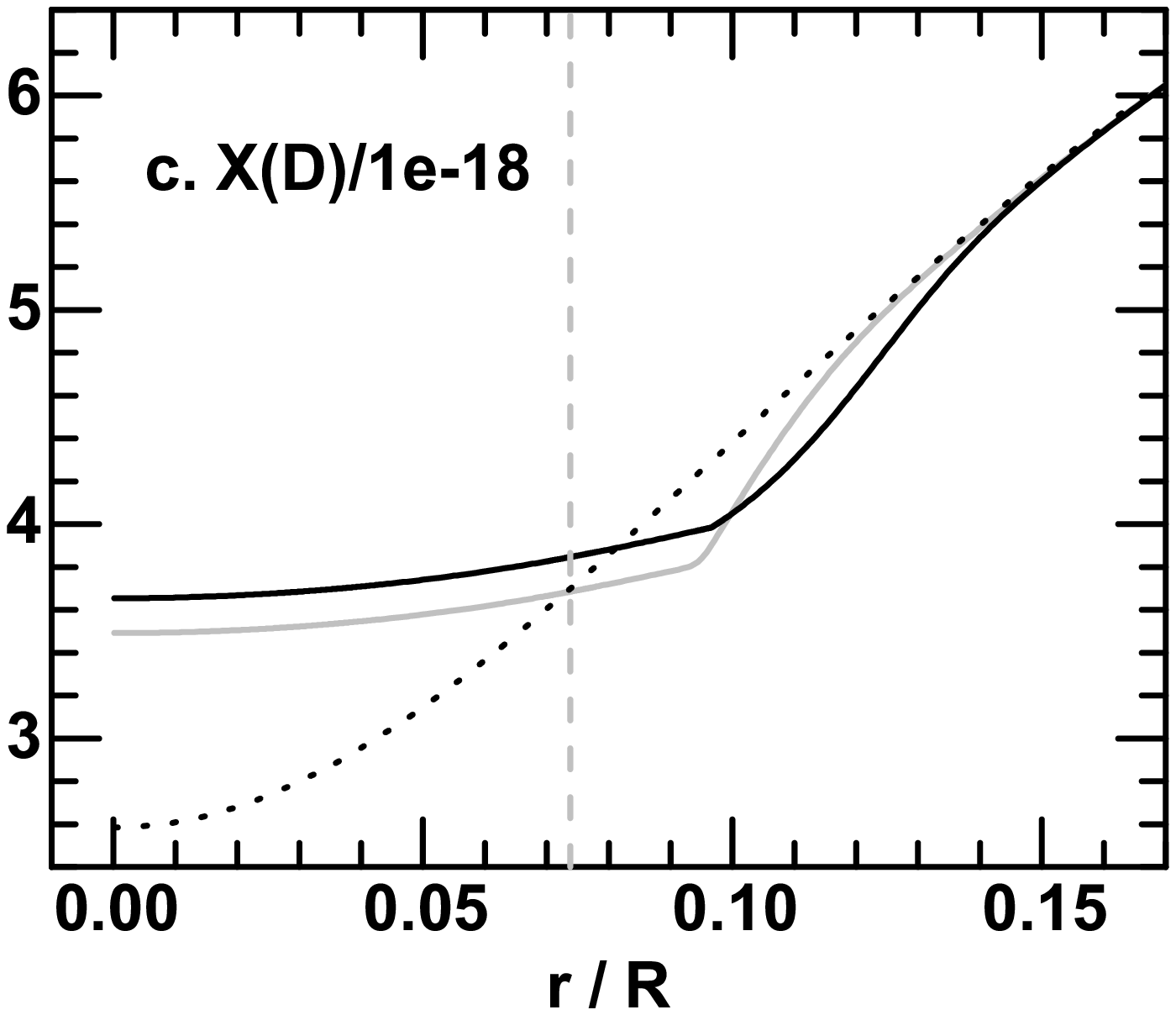}
	\includegraphics[width=0.66\columnwidth]{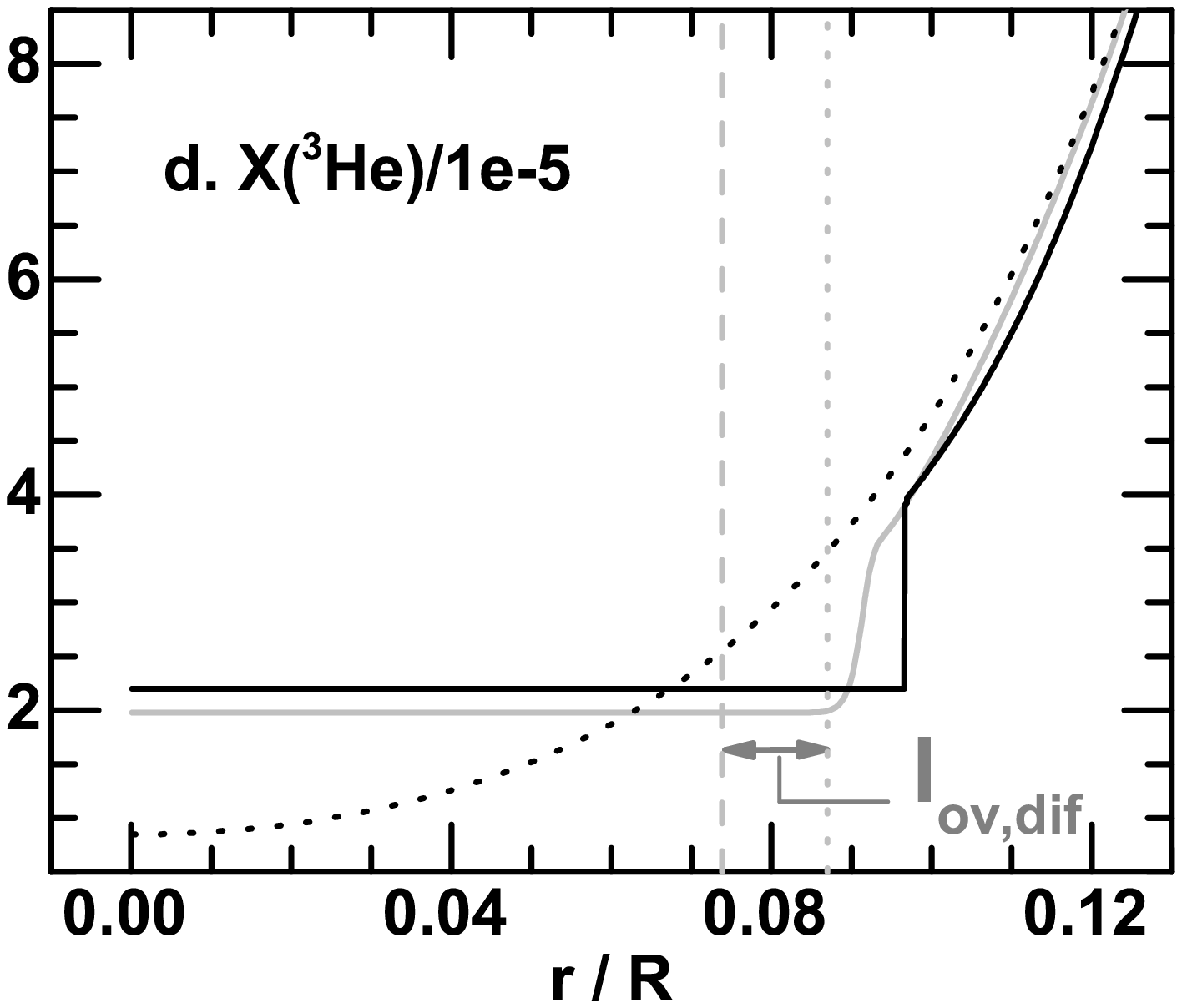}
	\includegraphics[width=0.66\columnwidth]{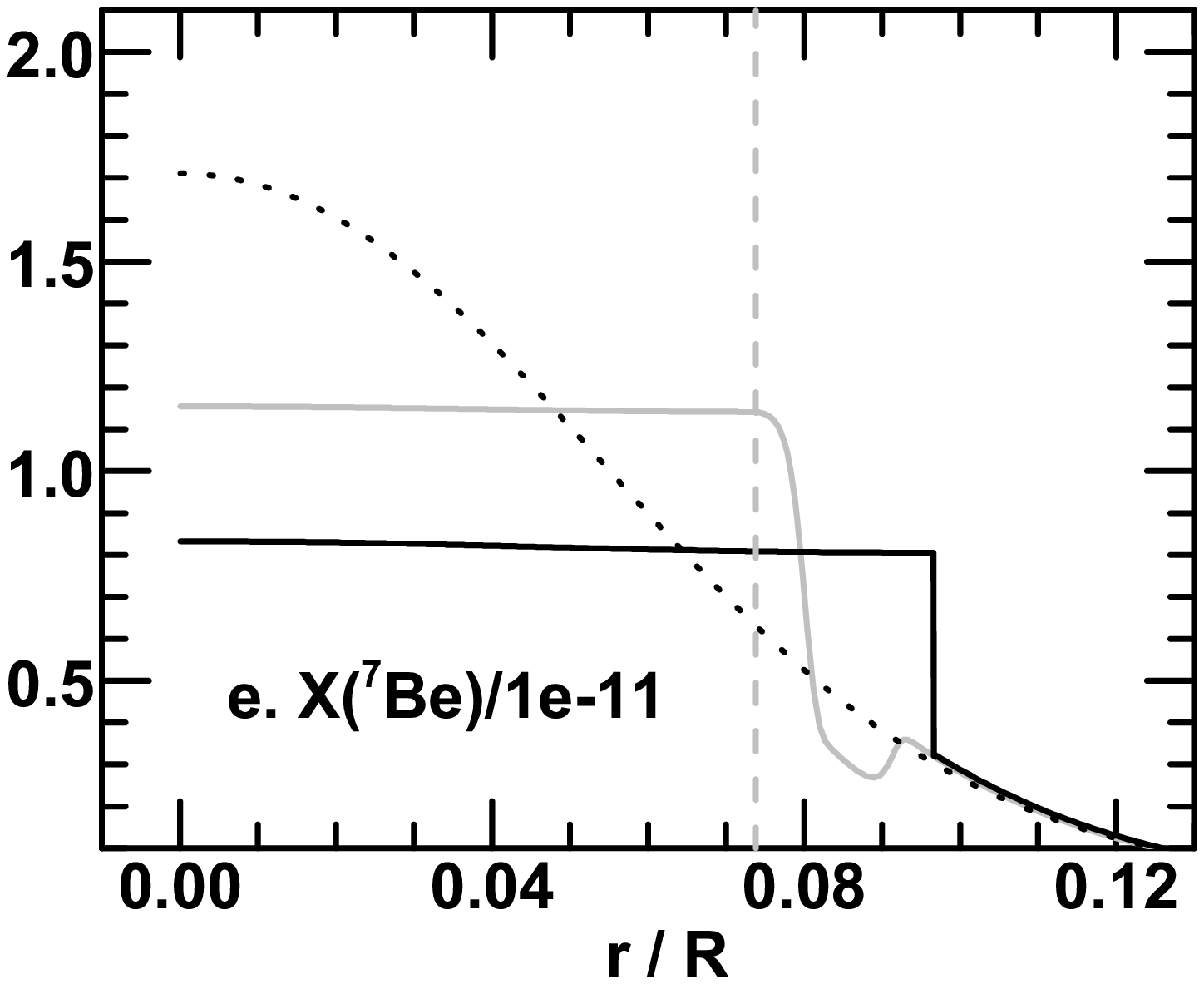}
	\includegraphics[width=0.66\columnwidth]{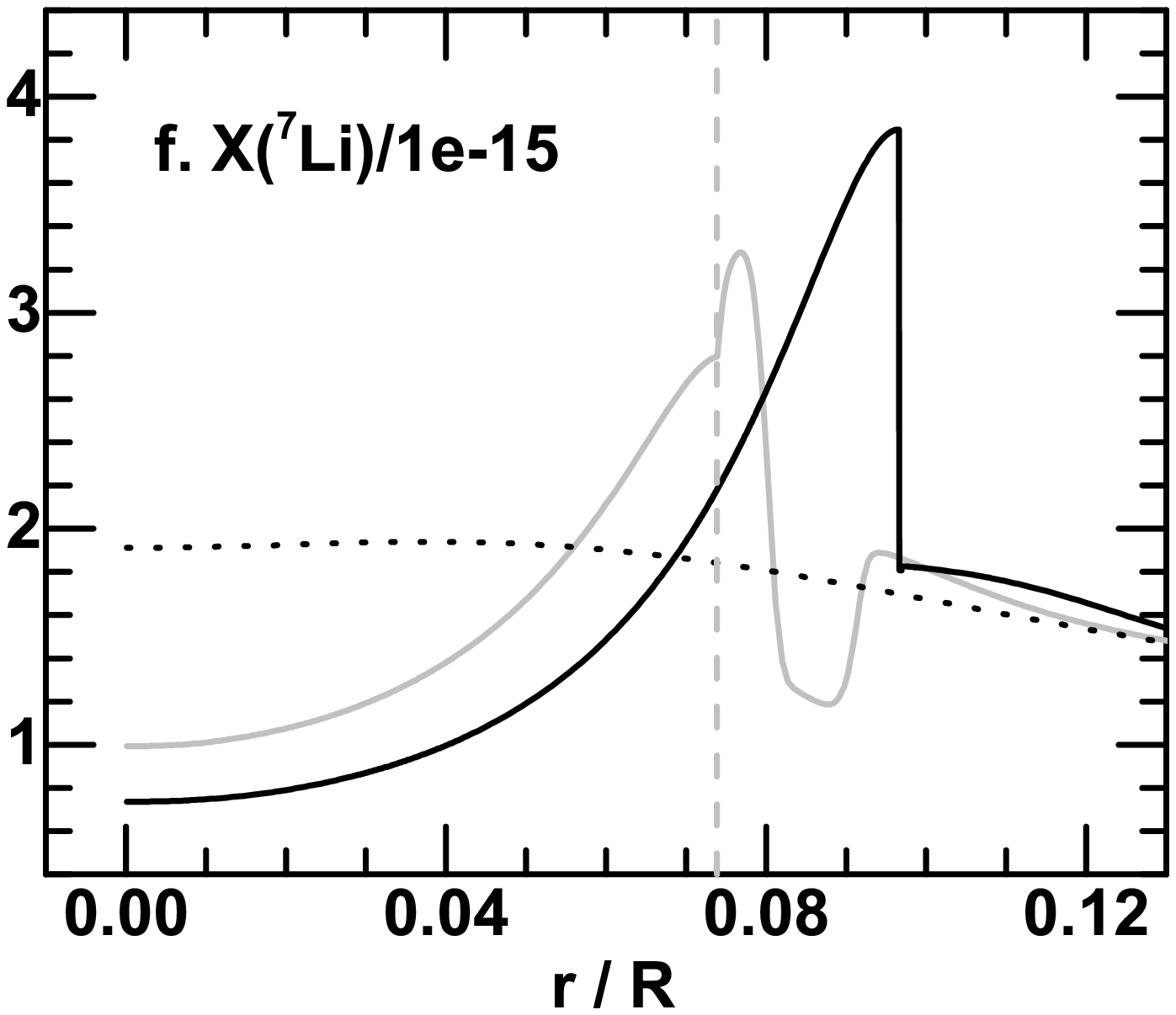}
    \caption{The diffusion coefficient and the abundance of the elements participating the pp chains.
Black dotted lines are for the case of the SSM with $\alpha_{\rm{ov}}=0$,
black solid lines are for the solar model with COM with $\alpha_{\rm{ov}}=0.293$,
	and the grey lines are for the solar model with EDOM with $C=0.1$ and ${\rm{log}}\theta=2.19$. The two solar models with overshoot have the same value of the mass fraction of the convective core $M_{\rm{cz}}/{\rm M_{\odot}}=0.0322$ and their convective boundaries are at $r_{\rm{cz}}\approx0.074$. The grey dashed line in each panel shows the location of the convective boundary of the solar model with EDOM, i.e., $r_{\rm{cz}}=0.0738\,R$. In panels b and d, the grey dotted lines and the arrows denote the effective overshoot length for H and $^3$He.}
    \label{compdx}
\end{figure*}

Here we present a detailed analysis of the abundance profiles in the stellar core affected by overshoot mixing. The diffusion coefficient of convective/overshoot mixing and the abundance of the elements participating the pp chains, i.e., H, D, $^3$He, $^7$Be, $^7$Li, of a solar model with EDOM with $C=0.1$ and ${\rm{log}}\theta=2.19$, a solar model with COM with $\alpha_{\rm{ov}}=0.293$, and the SSM are shown in Fig.~\ref{compdx}. The two models with overshoot have the same value of the mass fraction of the convective core $M_{\rm{cz}}/{\rm M_{\odot}}=0.0322$ and their convective boundary are at $r_{\rm{cz}}\approx0.074$.

The diffusion coefficient of the convective/overshoot mixing is shown in Fig.~\ref{compdx}a. For the COM (the black line), the diffusion coefficient $D$ in the overshoot region is the same as the typical value of $D$ in the convective core. For the EDOM (the grey line), $D$ drops an order of magnitude at the convective boundary because its value of the parameter $C$ is 0.1, then it decreases exponentially in the overshoot region.

D, ${}^3{\rm He}$, ${}^7 {\rm Be}$ and ${}^7 {\rm Li}$ all have nuclear timescales much shorter than the evolution timescale of the Sun, so that they are in local nuclear equilibrium where there is no mixing or overshoot diffusion. This is reflected in Fig.~\ref{compdx} in the behaviour for the SSM. The equilibrium abundances depend mainly on temperature (but also on $X({\rm H})$), such that $X({}^3{\rm He})$ increases with decreasing temperature and $X({}^7{\rm Be})$ and $X({}^7{\rm Li})$ decrease with decreasing temperature.

In the COM case, mixing is efficient throughout the convective core and the overshoot region, as reflected in Fig.~\ref{compdx}, with a diffusive timescale $r^2_{\rm{cz}}/D\sim 2.5 \times 10^6 \, {\rm s}$. This is shorter than the nuclear timescales of ${}^3{\rm He}$ ($\tau_{\rm n} \sim 10^{13} \, {\rm s}$) and ${}^7{\rm Be}$ ($\tau_{\rm n} \sim 10^{7} \, {\rm s}$); consequently the abundances of these elements as well as, obviously, hydrogen, are uniform in the mixed region. However, for D ($\tau_{\rm n} \sim 1 \, {\rm s}$) and ${}^7{\rm Li}$ ($\tau_{\rm n} \sim 10^3 \, {\rm s}$) local nuclear equilibrium applies, with $X({\rm D})$ increasing slowly, with decreasing temperature and distance to the centre, closely related to $X({\rm H})$. The behaviour of $X({}^7{\rm Li})$ is a little more complex. ${}^7{\rm Li}$ is produced from ${}^7{\rm Be}$ by electron capture, which depends little on conditions, and is destroyed by proton capture. Thus $X({}^7{\rm Li})$ is closely linked to $X({}^7{\rm Be})$. In the fully mixed region $X({}^7{\rm Be})$ is constant, and the variation in $X({}^7{\rm Li})$ reflects the variation in the proton capture rate which decreases rapidly with decreasing temperature, leading to an increase in $X({}^7{\rm Li})$. However, outside the mixed region $X({}^7{\rm Be})$ is in local nuclear equilibrium, decreasing rapidly with decreasing temperature, and resulting in the overall decrease in $X({}^7{\rm Li})$.

To analyze the more complex properties of EDOM we introduce an effective overshoot length (denoted $l_{\rm{ov},dif}$), such that the mixing is efficient over that length, making the abundance of a specific element nearly completely mixed. Setting the location of the outer point of the efficient mixed region to be $r_1$, we have $l_{\rm{ov},dif}=r_1-r_{\rm{cz}}$. Because the mixing efficiency is high enough in the convective core, $r<r_{\rm{cz}}$, for the great majority of elements, the condition of forming an efficient mixed region for $r<r_1$ is that the overshoot region $r_{\rm{cz}}<r<r_1$ is efficiently mixed. That requires $D_1 \sim l^2_{\rm{ov},dif}/\tau$ where $D_1$ is the typical diffusion coefficient in the overshoot region and $\tau$ is an effective duration of mixing. Because the diffusion coefficient decreases exponentially in the overshoot region, mixing efficiency is dominated by the minimum of the diffusion coefficient. For $\tau$, it is generally the duration of the mixing $\tau_{\rm{mix}}$, which is approximately the stellar age $t$ for the convective core overshoot of main-sequence stars. However, according to the abundance evolutionary equations, if an element participates in nuclear reactions, its abundance variation also depends on its nuclear depletion timescale $\tau_{\rm{n}}$. A reasonable viewpoint is that the abundance is dominated by the nuclear equilibrium when $\tau_{\rm{n}}<\tau_{\rm{mix}}$ and the convective/overshoot mixing when $\tau_{\rm{n}}>\tau_{\rm{mix}}$. Therefore the effective duration of mixing cannot exceed the nuclear depletion timescale. Based on the discussion above, an equation for the effective overshoot length $l_{\rm{ov},dif}$ for EDOM can be estimated from
\begin{eqnarray} \label{diflov}
D(r_1) \sim \frac{l^2_{\rm{ov},dif}}{\tau},
\end{eqnarray}
where
\begin{eqnarray} \label{diflovdef}
l_{\rm{ov},dif}=r_1-r_{\rm{cz}} \; , \quad
\tau=\min(\tau_{\rm{mix}},\tau_{\rm{n}}).
\end{eqnarray}
Defining two new dimensionless variables $x$ and $q$ as
\begin{eqnarray} \label{diflovxq}
&&x = \frac{{\theta {l_{{\rm{ov,dif}}}}}}{{{H_P}}}, \\ \nonumber
&&q = \frac{{{H_P}^2}}{{{\theta ^2}{CD_0} \tau }},
\end{eqnarray}
equation~(\ref{diflov}) can be rewritten as
\begin{eqnarray} \label{diflov2}
\exp ( - x) = q{x^2},
\end{eqnarray}
while
\begin{eqnarray} \label{diflov2d}
D({r_1}) = {CD_0}\exp ( - x).
\end{eqnarray}

In the EDOM model shown as the grey lines in Fig.~\ref{compdx}, $H_P\approx 10^{10} \, {\rm cm}$, $\theta\approx 155$ and $CD_0\approx 10^{12} \, {\rm cm^2\,s^{-1}}$, thus $q\approx4000/\tau$ where the effective duration of mixing $\tau$ is different for each element. Calculating $q$ by using equation~(\ref{diflovxq}), the effective overshoot length can be worked out by solving equation~(\ref{diflov2}). Those equations indicate that the effective overshoot length is different for each element/isotope, which is an intrinsic difference between the classical overshoot model and the exponential diffusion overshoot model.

For a main-sequence star $\tau_{\rm{n}}$ for hydrogen is the main-sequence life time that is obviously larger than its age $t$; thus $\tau=\tau_{\rm{mix}}=t\sim 10^{17} \, {\rm s}$, $q\approx 4 \times 10^{-14}$, $x\approx 24.5$ and $D(r_1)\sim 20 \, {\rm cm^2\,s^{-1}}$. The corresponding $r_1$ is about $0.093\,R$ based on Fig.~\ref{compdx}a. This is consistent with the results shown in Fig.~\ref{compdx}b. For $r>0.093\,R$ in the solar model with the diffusion overshoot model, the mixing is weak and the abundance is dominated by the preceding local burning of hydrogen. As shown in Fig.~\ref{compdx}b, for $r>r_1$, the grey line is located left of the black line, indicating that $l_{\rm{ov},dif}$, for EDOM, is shorter than $l_{\rm{ov}}$ for COM in the earlier stage. This is because $\dd l_{\rm{ov},dif}/\dd t>0$ as shown by equation~(\ref{diflov}), while $l_{\rm{ov}}\propto r_{\rm{cz}}$ in the classical overshoot model and $r_{\rm{cz}}$ decreases with age due to the retreat of the convective core such that $\dd l_{\rm{ov}}/\dd t<0$; by assumption $l_{\rm{ov},dif}=l_{\rm{ov}}$ at the present solar age. The same reason leads to the difference of the mass of the convective core between COM with $\alpha_{\rm ov}=0.25$ and EDOM with $\log C=0$ and $\log \theta=2.53$ as shown in Fig.\,\ref{resstdmc}.

For $^3$He, $\tau_{\rm{n}}\sim 10^{13} {\rm s}$ is less than the stellar age, thus $\tau=\tau_{\rm{n}}\sim 10^{13} \, {\rm s}$, $q\approx 4 \times 10^{-10}$, $x\approx 16$, $D(r_1)\sim 10^5 \, {\rm cm^2\,s^{-1}}$, and $r_1$ is about $0.087\,R$. This is consistent with the results shown in Fig.~\ref{compdx}d. In the region $0.087\,R<r<0.093\,R$ in the solar model with the EDOM, the mixing is weak and the abundance gradually changes to its nuclear equilibrium abundance, at the given hydrogen abundance.

For $^7$Be, $\tau=\tau_{\rm{n}}\sim 10^{7} \, {\rm s}$, $q\approx 4 \times 10^{-4}$, $x\approx 4.7$, $D(r_1)\sim 10^{10} \, {\rm cm^2\,s^{-1}}$, and $r_1$ is about $0.078\,R$. This is a very short $l_{\rm{ov},dif}$ as shown in Fig.~\ref{compdx}d. For $r>0.08\,R$, the ${}^7{\rm Be}$ abundance transitions to nuclear equilibrium with the $^3$He abundance,
decreasing with decreasing temperature, except at the increase in $X({}^3{\rm He})$ in the region $0.09\,R<r<0.093\,R$. An interesting result is that the effect of overshoot on $^7$Be with EDOM is small so that the $^8$B neutrino fluxes of the EDOM solar models are systematically higher than those of the COM solar models, as shown in Section \ref{SecComp2} (see Fig.~\ref{resdif1d}).

The deuterium and $^7$Li abundances are shown in Fig.~\ref{compdx}c and f. Since their very short burning timescales, they are in nuclear equilibrium. The deuterium abundance related to the hydrogen abundance and negatively correlated with temperature. The ${}^7{\rm Li}$ abundance closely tied to the abundance of ${}^7{\rm Be}$ with a scaling function that increases with decreasing temperature. Consequently, it follows the dip in the ${}^7{\rm Be}$ abundance for $0.08\,R < r < 0.09\,R$ in the EDOM case.

We note that in our treatment, the COM is different from the model with an artificially fully mixed convective core and overshoot region. The latter leads to a complete mixing for all element even deuterium and $^7$Li. However, that is unreasonable. For $^7$Li, since its burning time scale is $10^3 \, {\rm s}$ and the size of the core is $5 \times 10^9 \, {\rm cm}$, an efficient mixing requires a characteristic convective speed $5 \times 10^6 \, {\rm cm\,s^{-1}}$, i.e., Mach 0.1. Owing to the high density of the convective core, only a very small superadiabatic gradient is required for convective energy transport, such that the weak buoyancy cannot accelerate the fluid to a speed comparable with the sound speed. For deuterium, the characteristic convective speed required for efficient mixing is Mach 100 or 16\% of the speed of light. This is obviously unreasonable.


\bsp	
\label{lastpage}
\end{document}